\documentclass[11pt]{article}

\usepackage[english]{babel}

\usepackage[margin=1in]{geometry}
   
\usepackage{amsmath,amssymb,amsthm}
\usepackage{graphicx}
\usepackage[colorlinks=true, allcolors=blue]{hyperref}
\usepackage{enumerate} 
\usepackage{multirow} 
\usepackage{mathrsfs}
\usepackage{float} 
\usepackage{booktabs,comment} 
\usepackage{caption} 
\usepackage{subcaption} 
\usepackage{url} 
\usepackage{listings} 
\usepackage{mathtools} 
\usepackage{bm} 
\usepackage{longtable}

\newif\ifshow 
\showfalse 
\ifshow
\includecomment{wrap}
\else
\excludecomment{wrap} 
\fi

\newtheorem{remark}{Remark}

\usepackage{hyperref} 
\hypersetup{
    colorlinks=true,
    linkcolor=blue,
    filecolor=magenta,      
    urlcolor=blue,
    pdftitle={Overleaf Example},
    pdfpagemode=FullScreen,
    }

\usepackage[dvipsnames]{xcolor}
\usepackage[underline=true,rounded corners=false]{pgf-umlsd}
\definecolor{economic-var}{HTML}{000000}
\definecolor{geographic-var}{HTML}{000000}
\definecolor{demographic-var}{HTML}{000000}

\definecolor{cor-very-weak}{HTML}{999999}
\definecolor{cor-weak}{HTML}{000000}
\definecolor{cor-moderate}{HTML}{000000}
\definecolor{cor-strong}{HTML}{000000}
\definecolor{cor-very-strong}{HTML}{8B0000}

\DeclareMathOperator{\diag}{diag}

\newcommand{\gp}{\mathcal{GP}}

\title{Analyzing State-Level Longevity Trends with the U.S.~Mortality Database}
\author{Mike Ludkovski and Doris Padilla
\thanks{Department of Statistics \& Applied Probability, University of California, Santa Barbara, CA 93106-3110}}
\date{}

\begin{document}
\maketitle

\begin{abstract}
We investigate state-level age-specific mortality trends based on the United States Mortality Database (USMDB) published by the Human Mortality Database. In tandem with looking at the longevity experience across the 51 states, we also consider a collection of socio-demographic, economic and educational covariates that correlate with mortality trends. To obtain smoothed mortality surfaces for each state, we implement the machine learning framework of Multi-Output Gaussian Process regression (Huynh \& Ludkovski 2021) on targeted groupings of 3--6 states. Our detailed exploratory analysis shows that the mortality experience is highly inhomogeneous across states in terms of respective Age structures. We moreover document multiple divergent trends between best and worst states, between Females and Males, and between younger and older Ages. The comparisons across the 50+ fitted models  offer opportunities for rich insights about drivers of mortality in the U.S. and are visualized through numerous figures and an online interactive dashboard.

\end{abstract}

\section{Introduction}\label{sec:intro}

The United States Mortality Database \cite{usmdb} provides a high-quality complete dataset regarding Age- and Year-indexed mortality experience across the 50 U.S.~states and the District of Columbia. First published in 2019, USMDB offers a novel opportunity for actuarial and statistical insights at sub-national granularity. Given the natural similarities among the states that all share broadly similar economic, cultural, health and to some extent demographic characteristics, the main insights to be gleaned from USMDB are about the various facets of \emph{heterogeneity} across states. For example, we document that there is  a wide range of mortality improvement factors (i.e.~annual changes in mortality rates) among states, with some improving and others experiencing rising mortality. Similarly, we document a diversity of Age structures of state mortality relative to the U.S.~average, and a spectrum of Age patterns in yearly mortality changes. Such observations require a deep dive into the data and necessitate statistical tools in order to see through the intrinsic observation noise. Indeed, many of the U.S.~states have only 1-3 million inhabitants, meaning that yearly deceased counts for a single Age are in the low hundreds. As such, raw year-over-year changes in mortality rates are much too volatile to draw any conclusions and smoothing techniques are imperative. 

With the above in mind, we design a custom two-step statistical procedure for studying the USMDB. First, we advocate for data pooling, namely combining several states into a joint dataset (treating State as a factor level) in order to fuse information.  This improves smoothing thanks to the high correlation of mortality experience among states. To do so we use a (multi-output) Gaussian Process-driven stochastic mortality model. Second, we develop \emph{grouping} techniques to create the aforementioned groups of similar states to be modeled together. Targeted grouping not only maximizes goodness of fit, but also provides for additional interpretation through analysis of which states end up grouped together. On the latter point, we employ a wide range of auxiliary state-level covariates in order to investigate the relationship between economic, demographic and geographical state characteristics and the respective mortality experience. 

After constructing state groupings and fitting a stochastic mortality model, we embark on an detailed exploratory analysis via more than a dozen figures, supplemented by an online interactive RShiny dashboard \cite{Shiny}. Among others, we visualize and discuss (i) rankings of states in terms of their Age-linked mortality evolution in time; (ii) their recent mortality improvement factors;  (iii) their Age-structure of mortality rates and (iv) Age-structure of mortality improvement factors; (v) pattern of above vis-a-vis state characteristics.  We highlight several take-aways that are likely to be new to the actuarial audience in terms of the aggregate behavior of the 51 states and respective ``outliers''. To our knowledge, this is one of the first papers to present a detailed side-by-side comparison of so many inter-related mortality models. It is also one of the analyses few fully dedicated to the USMDB state dataset.

\begin{remark}
The impact of the COVID19 pandemic on U.S.~mortality has been severe. At the writing of the article the latest available data is from 2021, with all indications that 2022 is not ``back to normal'' either. Inclusion of such dramatic outliers into a stochastic mortality model is fraught since the underlying assumption is of a statistically stationary behavior across the training dataset. Moreover, it is still to be determined whether COVID19 was a 2-3 year ``interruption'' or a structural break (for example due to long-COVID health after-effects) that will permanently impact U.S.~mortality. Consequently, it is not clear whether (or how) to upweigh or downweigh the latest experience when smoothing for mortality trends. For these reasons, we choose to exclude latest data and concentrate on  pre-pandemic exploratory analysis.    
\end{remark}

\textbf{Literature review:} Our analysis of variations within U.S.~mortality connects to several non-actuarial strands of extant literature. Demographers have been highlighting growing geographical disparities in mortality within U.S.~since the late 20th century, see \cite{ezzati2008reversal,wilmoth2011geographic,currie2016mortality}, primarily focusing on standard deviation of life expectancy (LE at birth, $e_0$). These works highlight the emergence of a geographic ``belt'' so that ``the 13 worst-off states were
geographically contiguous in 2004'' \cite{Fenelon2013}. Within the economics literature, the starting point for investigating  mortality disparities in the U.S.~originated with Chetty et al.~\cite{chetty2016association} who documented strong correlation between income, geography, and LE at Age 40, $e_{40}$. Chetty et al.~primarily worked with county-level data, aggregated into commuting zones, and documented gaps of 10--14 years in LE across different parts of the U.S.. A further boost to such investigations came from the study of ``deaths of despair'' \cite{CaseDeatonBook} that are rising in mid-life U.S.~adults in the past 10+ years. Since then, many researchers have  documented growing gaps in adult mortality and life expectancies across both education and income \cite{Bosworth2018,harris2021high,risinginequLEsocio,CaseDeaton21}. We especially highlight the recent SOA-sponsored report by Barbieri \cite{Barbieri2020} which ranks counties by socioeconomic index scores (SIS) and then compares mortality by SIS deciles.

These macro effects translate into state-level differences through several channels. First, states intrinsically vary in their socio-demographics, for example in the proportions of  racial sub-groups and poverty levels. Second, states have enacted over time different policies that impact mortality. This includes anti-smoking campaigns (\cite{risinggeodis} find correlation between tobacco taxes and mortality and \cite{Fenelon2013} attributes cigarette smoking prevalence as explaining more than half of geographic differences in mortality), public health policies such as Medicare expansions, and environmental policies such as pollution mitigation. Third, migration patterns, for example increased concentration of college-educated persons along the coasts, have been amplifying regional differences by making states more heterogeneous over the past few decades. Moreover, migrants tend to have better underlying health characteristics, which improves the observed mortality profile of receiving states
and leads to higher observed mortality in sending states \cite{ezzati2008reversal}.

Beyond the above factors that create dependence between state characteristics and mortality rates, there is also ``a portmanteau of `place' effects'', cf.~Cuillard et al.~\cite{risinggeodis}. These include the rural/urban differences in mortality, climate effects, and other, yet to be pinpointed spatial factors.  The analysis in \cite{risinggeodis} parallels in several ways our work below, with three critical differences. First, Cuillard et al.~are looking at aggregate mortality at mid-life, studying aggregate mortality for ages 25--64. In contrast, we provide a much more granular, Age-specific analysis. Second, Cuillard et al.~concentrate on the working adults, where effects such as \emph{deaths of despair} (which have seen marked state-level differences \cite{CaseDeaton21}) are important; in contrast we focus on the older ages 60--84. Third, \cite{risinggeodis}~take raw mortality data as-is, without imposing any statistical analysis; this is sufficient for their age-aggregated approach but is inadequate for our deeper investigation, especially for mortality improvement factors. In sum, our work complements \cite{risinggeodis} by providing an updated, age-specific, statistically smoothed analysis of state-level mortality.

Another recent analysis on the spatial disparities in U.S.~mortality is by Vierboom et al.~\cite{vierboom2017recent,vierboom2020life,vierboom2019rising}. They use cause-specific mortality (working with 9 top-level mutually exclusive and exhaustive cause categories) across metropolitan areas and geographic regions, binning into 5-year Age intervals. The main finding in \cite{vierboom2017recent} is that spatial inequality rose between 2002 and 2016, and that areas that had lower mortality enjoyed larger gains. Such divergent trends were especially noticeable between large coastal metropolitan areas and rural Appalachia and South, and within lung cancer/respiratory diseases, as well as drug/alcohol abuse. Vierboom et al.~focus on studying LE at birth $e_0$ for ages 30--85 and work with county-level data aggregated into 40 spatial units (the 9 census divisions plus Appalachia, each broken out into Metro, Suburb, Small metro, and Rural strata).  
The companion study \cite{vierboom2020life} investigates spatial inequality in $e_{65}$ (LE at 65) for the same 40 spatial units; see also an earlier analysis in \cite{dwyer2016us} using 5-year Age bins.

Li and Hyndman \cite{li2021assessing} investigate state-level mortality through a two-level forecast reconciliation method: building single-state and national level Lee-Carter models, and then adjusting the results so that the sum of the state projections adds up to the national estimate. They primarily focus on out-of-sample performance, examining projections as much as 10 years into the future.

Several articles have recently introduced explicit consideration of spatial mortality patterns in order to handle the sparsity of death counts in small spatial units. Gibbs et al.~\cite{geo-rur-mort} use conditional auto-regressive priors  with a county-specific linear Age trend in order to borrow information across neighboring U.S.~counties. 
In a follow-up publication Hartman et al.~\cite{countyUS} construct a single multivariate spatio-temporal model to fuse data across both space and Age bins; parameter inference is done via Integrated Nested Laplace Approximations. Cupido et al.~\cite{spatialpatUS,localmodelgeo} apply a spatial filtering approach to infer the latent spatial dependence in U.S.~county-level mortality. Boing et al.~\cite{boing2020quantifying} build a hierarchical model that teases out the relative importance of states versus counties versus census tracts, showing that the latter is the primary driver of mortality variability.  The consistent take-away is that taking into account spatial relationships offers better predictive power and supports the intuition that ``individuals living closer together likely have more similar lifestyles than individuals living hundreds of miles apart’’ \cite{geo-rur-mort}. In turn, similar lifestyle habits and environmental factors drive mortality.

\bigskip

Within this landscape, our contributions can be traced along two dimensions. Methodologically, we propose a new technique for studying USMDB data, namely through creating custom, targeted groupings of a handful of states at a time. Our approach builds on~\cite{gp_model1,multioutput_gp,SOA_models,huynh2021joint} and allows fusion of mortality data from similar states, while avoiding the need to directly model all 51 states jointly, a statistically daunting task. Instead, we advocate grouping states based on their geographical neighbors and a collection of socio-economic covariates. Such targeted groupings simultaneously improve computational efficiency and model efficacy.  To this end, we compute a weighted Euclidean distance between state-wise Principal Component Analysis scores on the collected state covariates. This technique to identify similar states can also be used for further settings, such as grouping other sub-national jurisdictions (counties, provinces, federal states), or for grouping countries, e.g.~within the E.U. Empirically, we provide a novel exploratory analysis about the relative experience of smoothed Age-specific mortality across U.S.~states. We augment existing literature that focuses on either life expectancy or aggregate mortality (both metrics effectively averaging across many ages) with an explicit consideration of mortality as a function of Age. Moreover, we investigate the recent dynamics of mortality through inferred mortality improvement (MI) factors. We not only rank and compare states against each other, but moreover correlate our projections with the collected external covariates. In sum, we confirm several previous aggregate analyses  (such as strong correlation between mortality and income/obesity/geographic region), and also document several new insights, such as a heterogeneity of improvement factors as a function of Age, and the strong disparities between Male and Female MIs.

\begin{remark}
Our goal in this project is to explore and compare mortality across U.S.~states. As such, we focus on exploratory analysis and do not claim to provide the \emph{best} model for this purpose. Neither do we do any significant future forecasting, rather concentrating on studying the mortality experience during the 2010s. The proposed model is designed and developed with these aims in mind, namely to enable meaningful smoothing and nowcasting of Age-specific mortality state-by-state, with the emphasis on pooling states for purposes of their \emph{relative} comparisons. Conversely, extensive experiments during the writing of this paper indicate that the presented results are largely invariant to the specific mortality modeling framework. 
\end{remark}

The rest of the paper is organized as follows. In Section \ref{sec:model}  we summarize the raw data provided by the USMDB and introduce the stochastic model  of Multi-Output Gaussian Processes and methodology used to create smoothed mortality surfaces. Section~\ref{sec:groups} describes the state grouping algorithm. Section~\ref{sec:analysis} presents results regarding state-level mortality relative ranks. Sections~\ref{sec:IR} and~\ref{SUBSEC:explorcov} in turn analyze the respective improvement factors and correlation with state-level covariates. Section~\ref{sec:conclude} concludes. Several Appendices present further plots, table, and covariate definitions. To promote analysis of USMDB, the visualizations below are augmented with the publicly available RShiny tool \cite{Shiny}. The dashboard replicates some of the shown figures and offers a starting point for other researchers to  directly explore the outputs of our models across states, genders and years.

\section{Data and Statistical Model}\label{sec:model}
\subsection{Dataset}

Built by the HMD team, the United States Mortality Database (USMDB)~\cite{usmdb} contains a complete historical set of state-level life tables for every calendar year during 1959--2019 for all 50 U.S.~states and the District of Columbia (D.C.). The USMDB covers ages 0--110+ and includes separate datasets for the male and female populations. The raw data contains birth and death counts from the U.S.~vital statistics system, and incorporates the census counts and population estimates from the U.S.~Census Bureau. 

Our objective is to estimate and smooth historical mortality rates; then forecast short-term calendar trends through analyzing and estimating mortality improvement factors. Throughout this paper we thus focus on the following subset of the USMDB: (a) Calendar Years: 1990--2018; (b) Population: males and females, considered separately; and (c) Ages: 60--84. This subset of older ages and recent years is the most relevant for actuarial applications. Omitting older data from the 20th century is in line with the data-driven machine-learning framework we employ, so that  long-past mortality experience is not only less relevant but potentially misleading for our model construction. We omit very old Ages since the population data underlying USMDB is only available up to an open age interval at 85+ years \cite{Barbieri2020}.

Due to working with raw data, nearly all of the works cited above consider aggregated mortality across many Ages, e.g.~bins of 15 years (50--64, 65--79, etc.), 10 years or 5 years, or look at life expectancy at a given age ($e_0,  e_{50}, e_{65}$, etc.). This fundamentally obscures the non-constant \emph{relative} impact of Age, which is in fact well documented. Indeed, multiple studies conclude that spatial and socioeconomic inequalities decline with age: ``health disparities narrow with age''  \cite{vierboom2019rising} and ``the gap [in probabilities of dying by SIS deciles] declines progressively after age 55 years and becomes small (less than 10 percent) at ages 85 and above'' \cite{Barbieri2020}. At the same time, most references disaggregate mortality by other, say socioeconomic or cause-of-death, factors, or into smaller spatial units. In contrast, we follow the USMDB to fully disaggregate into 1-year bins by Age, but otherwise consider aggregated mortality across all individuals in the state.

Let $\mathcal{S} = (s_i)_{1 \le i \le 51}$ represent the 51 U.S.~states (throughout we count D.C.~as the 51st ``state''). The information regarding each state $s \in \mathcal{S}$ provided by the USMDB is organized as follows: 
\begin{enumerate}[(i)]
    \item \textsl{Independent variables:} calendar year $x_{t}$ and age $x_{a}$. That is, for each year $t \in \{1990,\hdots, 2018\}$ we have $a$ associated ages with $a \in \{60,\hdots,84\}$. The pair $(x_{a},x_t)$ refers to the set of persons from state $s$ aged $a$ during year $t$. 
    \item \textsl{Dependent variables:} The death counts $D$ along with the total number of persons lived (exposed to risk) $E$ at a given age-time interval $(x_{a},x_t)$. We record the log-mortality rate 
    \begin{equation}
        y^{(a,t)} \coloneqq \log \left[ \frac{\text{$\#$ of Deaths during $(x_{a},x_t)$ age-time Interval}}{\text{$\#$ of Exposed to Risk during $(x_{a},x_t)$ age-time Interval}}\right] \equiv \log \left[ \frac{D^{(a,t)}}{E^{(a,t)}} \right].
    \end{equation}
    To simplify notation, when the age and calendar year are clear from the context, we drop the superscript $(a,t)$. While deaths counts are generally highly accurate, exposed- to-risk are based on census figures and systematically undercount undocumented immigrants.
\end{enumerate}
In summary, the USMDB provides 5 inputs, $(s,x_a,x_t,D,E) \equiv$ (state, age, year, deaths, exposed-to-risk) and the associated output $y \equiv$ log-mortality, see Table \ref{tbl:usmdb}. The complete dataset is denoted as $\mathcal{D}$, with $\mathcal{D}_s  \subset \mathcal{D}$ representing the subset of rows associated with state $s \in \mathcal{S}$.

\begin{table}[H]
\centering
\begin{tabular}{ccc|cc}
\toprule
\multicolumn{5}{c}{USMDB $\mathcal{D}$ for Males} \\
\midrule
State $(s)$ & Age $(x_{a})$   & Year $(x_{t})$  & Mortality Rate    & Log-Mortality (y)    \\
CA& 60   &  1990 & 0.014  & -4.248  \\
CA& 60   &  1991 & 0.014   & -4.261  \\
\vdots   &\vdots   & \vdots  &  \vdots   & \vdots  \\
CA& 84 & 2017 & 0.070 & -2.664 \\
CA& 84 & 2018 & 0.072 & -2.631 \\
AR& 60 & 1990 & 0.014 & -4.262 \\
AR& 60 & 1991 & 0.013 & -4.327 \\
\bottomrule
\end{tabular}
\caption{Sample row entries from the USMDB dataset $\mathcal{D}$. \label{tbl:usmdb}}
\end{table}

\subsection{Multi-Population Gaussian Process Models}

Multi-population modeling aims to identify and capture mortality dependence patterns among several populations in order to fuse data and achieve coherent forecasts. We follow Huynh et al.~\cite{SOA_models} and \cite{multioutput_gp} in implementing a Multi-Output Gaussian Process (MOGP) to model USMDB longevity data. The MOGP model quantifies mortality uncertainty by probabilistically smoothing raw data, and simultaneously  generates stochastic out-of-sample forecasts by projecting mortality surfaces across the Age- and Year-dimensions. For multi-population analysis, MOGP imposes a transparent correlation structure which disentangles it from the Age-Period pattern in each population. The data fusion employed in a MOGP allows us to improve the overall model fit, reduce model risk, mitigate hyperparameter uncertainty, and provide insights into the discrepancies among mortality trends across populations.

\subsubsection{Gaussian Process Regression} \label{SUBSUBSEC:assumptionsdata}

First, we  describe the mechanics of the Single-Output Gaussian Process (SOGP) model; see \cite{gp_model1} for a more detailed description. Fix an arbitrary state $l \in \mathcal{S}$. We are given a sample of $i = 1, \ldots, n = 29 \times 25 = 725$ observed log-mortality rates. Mortality is described as a function of age and time: 
\begin{enumerate}[$\bullet$]
    \item $\mathbf{x} \coloneqq (x^1, \hdots, x^n)$ where $x^i \coloneqq  (x_a^i,x_t^i)$ for $a \in \{60, \hdots, 84\}$ and $t \in \{ 1990,\hdots,2018\}$. 
    \item  $\mathbf{y}_l(\mathbf{x}) \equiv \mathbf{y}_l \coloneqq (y_l^1, \hdots, y_l^n)$ where $y^i_l$ denotes the observed log-mortality rate provided by the USMDB for a state $l \in \mathcal{S}$ at age $x_a^i$ during year $x_t^i$.
\end{enumerate}

\textbf{Observation Likelihood.} We assume that the relationship between $y_l^i$ and $x^i$ can be described with a latent black-box function $f_l(\cdot)$ and white noise term,
\begin{equation}
    y_l^i = f_l(x^i) + \epsilon_l^i; \qquad i =1, \ldots, n.
\end{equation}
The observation noise is Gaussian with $\bm{\epsilon}_l = (\epsilon_l^1,\hdots, \epsilon_l^n) \sim \mathcal{N}(\bm{0}, \bm{\Sigma}_l \coloneqq \diag(\sigma_l^2))$. Here, we assume an \textit{observation likelihood} $\sigma_l = \text{StDev}(\epsilon^i_l) \; \forall i$ that is population-dependent but constant in age and year. The underlying function $f_l(x^i)$ intuitively represents the true mortality rate which would materialize in the absence of random shocks. 

\textbf{Distribution of the Latent Process.} A priori, for any sample $\mathbf{x}$ of $m \ge 1$ observations, the finite dimensional distributions (fdds) of $f_l(\mathbf{x}) = (f_l(x^1), \hdots, f_l(x^m))$ are postulated to follow a multivariate Gaussian law $f_l \sim \gp\big(m_l, C_l\big)$ with prior (parametric) mean function, $m_l(\cdot)$, and covariance matrix $C_l(\cdot,\cdot)$,
\[
m_l(\mathbf{x}) \coloneqq \mathbb{E}[f_l(\mathbf{x})] =\big( \mu_l(x^1), \hdots, \mu_l(x^m) \big) \quad \text{and} \quad C_l({x}, {x}') \coloneqq \mathbb{E}\Big[\big(f_l({x}) - m_l({x})\big) \big(f_l({x}') - m_l({x}')\big) \Big].
\]
 Assuming that $\epsilon(x^i)$'s are independent across $x^i$'s and from $f$, it follows that
\[
\mathbf{y}_l \sim \mathcal{N}\left(m_l, C_l + \bm{\Sigma}_l \right), \qquad \bm{\Sigma}_l = \diag( \sigma_l^2),
\]
since Cov$(y_l^i,y_l^j) = \text{Cov}\big(f_l(x^i), f_l(x^j)\big) + \sigma_l^2 \delta(x^i,x^j)$, where $\delta(x^i,x^j)$ is the Kronecker delta.

\textbf{Mean and Covariance Structure.}  We assume functional representations for the mean and covariance functions $m_l(\cdot)$, $C_l(\cdot,\cdot)$ which represent the prior  beliefs about the dataset.  Recall that all the observed properties of a stochastic process with Gaussian fdds are characterized by $m_l(\cdot)$ and $C_l(\cdot,\cdot)$. 
 \begin{enumerate}[i)]
    \item The GP \textit{Mean function} describes the prior trend in log-mortality rates.  We use a parametric prior mean function, $m_l(x^i) = \beta_{0,l} + \sum_{j=1}^p \beta_{j,l}h_j({x})$,
    where $h_j({x})$'s are given basis functions and the $\beta_{j,l}$’s are unknown coefficients to be estimated. Letting $\bm{\beta}_l = \big( \beta_{0,l}, \hdots, \beta_{p,l})^T$, $\bm{h}(x) = \big( h_1(x), \hdots, h_p(x) \big)$, we use the shorthand $m_l(x) = \bm{h}(x) \bm{\beta}_\ell$. Below, we postulate a linear trend in the Age dimension:
    \begin{equation} \label{EQU:meanfunction}
        m_l(x^i) = \beta_{0,l} +\beta_{1,l}^a \cdot x^i_a. 
    \end{equation}
    The choice \eqref{EQU:meanfunction} is used to de-trend the data according to an exponential increase in mortality as a function of Age (the so-called Gompertz Law of Mortality) in our segment of interest $x_a \in \{60, \ldots, 84\}$.
    
    \item The GP \textit{Covariance kernel} captures the dependence of the response surface $f_l$ on the varying Age and Year dimensions $x_a, x_t$. The GP kernel characterizes the smoothing process by quantifying the influence of inputs on the likelihood of the output. Our kernels are {distance-based}, capturing the logic that  the mortality experience should be similar at neighboring data points, and \emph{separable} across the Age and Period coordinates.
    
    We concentrate on a common family of covariance functions known as the Mat\'ern class, equipped with automatic relevance determination. The Mat\'ern-5/2 kernel defines the covariance between arbitrary inputs $x,x_*$ as:
    \begin{equation} \label{EQU:kernel}
        C^{(M52)}(x,x_*; \theta) \coloneqq  \left( 1 + \frac{\sqrt{5}}{\theta}|x- x_{*}| + \frac{5}{3 \theta^2}|x-x_{*}|^2\right) \cdot \exp\left\{-\frac{\sqrt{5}}{\theta}|x-x_{*}| \right\}.
    \end{equation}

    This kernel is parameterized by the lengthscale (hyper)parameter $\theta$, to be estimated. To construct the overall dependence structure we use a multiplicative Age-Period-Cohort (APC) structure, so that the covariance  between two mortality table entries $x^i \equiv (x_a^i,x_t^i)$, $x^{j} \equiv (x_{a}^j,x_{t}^j)$ is
    \begin{align}\label{eq:apc}
     C_l(x^i,x^j) \coloneqq \eta^2 \cdot  C^{(M52)}(x^i_a,x^j_{a}; \theta_{l,a}) \cdot C^{(M52)}(x^i_t,x^j_{t}; \theta_{l,t}) \cdot C^{(M52)}(x^i_c,x^j_{c}; \theta_{l,c}),
    \end{align}
 where $x_c \coloneqq x_t - x_a$ is the Birth Cohort (i.e.~year of birth of an individual who is $x_a$-old in year $x_t$), and $\eta^2$ is the process variance hyperparameter, scaling covariances to capture the typical amplitude of the response. The last cohort term in \eqref{eq:apc} is essential to capture well-known generational effects, such as the special 1918 and 1939 cohorts.
The product structure is analogous to the Age-times-Year terms in the classical Lee-Carter framework. See \cite{ludkovski2023expressive} for a further discussion of appropriate GP kernels for mortality. 
 
\end{enumerate}

\subsubsection{GP Posterior} \label{SUBSUBSECTION:posteriorres}

The GP paradigm models input-output relationships by algebraically conditioning its prior distribution on the training data. The resulting posterior yields a probabilistic projection regarding  the latent log-mortality surface at desired inputs $\mathbf{x}_*$, given the information in the USMDB. Note that $\mathbf{x}_*$ can refer to in-sample cells (historical smoothing) or out-of-sample cells (future forecasts), both obtained from exactly the same formulas below. 

Given a prior distribution $f_l \sim \gp(m_l, C_l)$ and a training set $\mathcal{T} = (\mathbf{x}, \mathbf{y}_l)$, we calculate the posterior distribution $\mathbf{y}_{l,*}|\mathcal{T} \equiv \mathbf{y}_l(\mathbf{x}_*)|\mathcal{T}$ at predictive cells $\mathbf{x}_*$. Observe that $(\mathbf{y}_l,\mathbf{y}_{l,*})$ follows the Multivariate Normal distribution (MVN) 
\begin{equation}
    \begin{bmatrix}
    \mathbf{y}_l \\
    \mathbf{y}_{l,*}
    \end{bmatrix}
    \sim \mathcal{N}\left( \begin{bmatrix}
    m_l(\mathbf{x}) \\
    m_{l}(\mathbf{x}_*)
    \end{bmatrix}, \begin{bmatrix}
    C_l(\mathbf{x},\mathbf{x}) + \bm{\Sigma}_l & C_l(\mathbf{x},\mathbf{x}_*) \\
    C_l(\mathbf{x}_*, \mathbf{x}) & C_l(\mathbf{x}_*, \mathbf{x}_*) + \bm{\Sigma}_l
    \end{bmatrix}\right).
\end{equation}

Applying MVN conditioning expressions, the Universal Kriging equations below provide both the estimated mean-function coefficients $\bm{\beta}_l = \big( \beta_{0,l}, \beta_{1,l})^T$  in \eqref{EQU:meanfunction} and the {posterior distribution} of $\mathbf{y}_{l,*}$ $ p\big(\mathbf{y}_{l,*} | \mathbf{y}_l \big) \sim {\cal N}\big(m_{l, *}(\mathbf{x}_*), C_{*}(\mathbf{x}_*,\mathbf{x}_*) \big)$ with the posterior mean-variance
\begin{align} \label{EQU:krigingmean}
        m_{l,*}(\mathbf{x}_*)  \coloneqq & \mathbb{E}[
        \mathbf{y}_{l,*}| \mathcal{T}] = \bm{h}(\mathbf{x}_*)^T\bm{\hat{\beta}}_l + C_l(\mathbf{x}_*, \mathbf{x})(C_l(\mathbf{x}, \mathbf{x}) + \bm{\Sigma}_l)^{-1}(\mathbf{y}_l-\bm{H}\bm{\hat{\beta}}_l); \\ \label{EQU:krigingcoef}
                \bm{\hat{\beta}}_l  \coloneqq & \big(\bm{H}^T\bm{D} \big)^{-1} \bm{H}^T \left(C_l(\mathbf{x}, \mathbf{x}) + \bm{\Sigma}_l \right)^{-1}\mathbf{y}_l;\\ \label{EQU:krigingvar}
    C_{*}(\mathbf{x}_*, \mathbf{x}_*)  \coloneqq & \text{Cov}(\mathbf{y}_{l,*} | \mathcal{T})  = C_l(\mathbf{x}^*, \mathbf{x}^*) + \bm{\Sigma}_l + \\ \notag %
       &  \quad +(\bm{h}(\mathbf{x}_*)^T - C_l(\mathbf{x}_*,\mathbf{x})\bm{D})^T (\bm{H}^T\bm{D})^{-1}\big( \bm{h}(\mathbf{x}_*)^T - C_l(\mathbf{x}_*, \mathbf{x})\bm{D} \big)
\end{align}
where the matrix ${C}_l (\mathbf{x}, \mathbf{x}_*)_{i,j} = C_l(x_i,x_{j,*})$ represents the covariance between inputs in the training set and predictive locations $\mathbf{x}_*$, $\bm{H} = \big( \bm{h}(x^1), \hdots, \bm{h}(x^n) \big)$ and $\bm{D} \coloneqq \big( C_l(\mathbf{x}, \mathbf{x}) + \bm{\Sigma}_l)^{-1}\bm{H}$. 

The predictive distribution of mortality rates at different age and time coordinates represented by $\mathbf{x}_*$ yield the smoothed mortality surfaces. In parallel, the GP model also outputs confidence intervals around future stochastic trajectories based on the posterior covariance $C_*(\cdot, \cdot)$. 

\subsubsection{Shared Covariance Structure}

Assume that we have selected a collection of $L \subset \mathcal{S}$ states and that each state-specific mortality surface  $f_l$, $1 \le l \le L$, follows a GP 
$f_l \sim \gp(m_l,C_l)$. We proceed to create a joint model for the vector $\bm{f}=(f_1,\ldots, f_L)$ through correlating its components. 
The motivation is that similar states should share alike mortality rates. Therefore, we impose a shared covariance structure which captures the dependencies between mortality rates in multiple states. 

To jointly model $L$ outputs, we need to specify the mean and covariance kernel of the joint GP $\bm{f}$. More specifically, let ${\vec{x}}^i \equiv  (x_a^i,x_t^i, x^i_{1}, \hdots, x^{i}_L)$ where $x^i_l = \mathbb{I}_{\{\text{population = }l\}}$; then we take
\begin{equation}
    \bm{f} \sim \gp(\bm{m}, \bm{\mathcal{C}}) \quad \text{ where } \bm{f}(\mathbf{x}) = 
    \big(f_1(\vec{\mathbf{x}}), \hdots, f_L(\vec{\mathbf{x}}) \big),
\end{equation}
$\bm{m} \in \mathbb{R}^{Ln \times 1}$ is the mean vector whose elements represent the mean functions of each population $l \in L$, $\{ m_l(\vec{x}) \}_{l=1}^L$, and $\bm{\mathcal{C}} \in \mathbb{R}^{Ln \times Ln}$ denotes the covariance matrix across the entire system. 

\textbf{Intrinsic Coregionalization Model (ICM).} Directly specifying the cross-covariances of each output pair $f_l,f_{l'}$ becomes unwieldy for $L>3$, so instead we rely on coregionalized kernels \cite{multioutput_gp} which assume that each output $f_l$, $1 \le l \le L$ is a linear combination of $Q$ independent latent Gaussian Processes $\bm{u} = \big(u_1(\mathbf{x}), \hdots, u_Q(\mathbf{x}) \big)$ with shared covariance kernel $C^{(u)}({x},{x}')$. In our case, we use the $C^{(u)}= C^{(M52)}$ APC kernel from \eqref{eq:apc}. 

Let $\bm{a}^*_q = (a_{1,q}, \hdots, a_{L,q})^T$, $1 \le q \le Q$, be the vector containing the $q$-th factor loadings across all populations $L$. Then  $\bm{f}(\mathbf{x}) = \sum_{q=1}^Q  \bm{a}_q^* u_q(\mathbf{x})$, or for the $l$-th population,
\begin{equation}
    f_l(\mathbf{x}) = a_{l,1}u_1(\mathbf{x}) + \hdots + a_{l,Q}u_{Q}(\mathbf{x}).
\end{equation}
The switch from $L$ separate GPs to $Q$ GPs ($u_1,\ldots, u_Q$) is similar to a PCA or singular value decomposition (SVD) approach and allows to reduce the number of hyperparameters in the cross-population covariance matrix from $\frac{L(L-1)}{2}$ to $Q \times L$: 
\begin{align}
    & \hspace{-2.4cm}\bm{\mathcal{C}}({x}, {x}') = \text{Cov}\big( \bm{f}({x}), \bm{f}({x}') \big) = \text{Cov} \left( \sum_{q=1}^Q \bm{a}_q^* u_q({x}), \sum_{q=1}^Q \bm{a}_q^*u_q({x}') \right) \nonumber \\
    & \hspace{2cm} = \left( \sum_{q=1}^Q \bm{a}_q^* \bm{a}_q^{*T} \right) \otimes \text{Cov}\big(u_q({x}), u_q({x}') \big) \nonumber 
     \equiv B \otimes C^{(u)}({x},{x}').
\end{align}
The $L \times L$ \textit{coregionalization matrix} $B \coloneqq A A^T$ with entries $B_{l,k}= \sum_{q=1}^Q a_{l,q}a_{k,q}$ has rank $Q$.

\textbf{Hyperparameters.} The MOGP computations and inference ultimately reduce to several linear-algebraic formulas, see \eqref{EQU:krigingmean}-\eqref{EQU:krigingvar}. The modeling task is to learn the spatial covariance structure, i.e.~the mean and kernel functions based on the training data. The overall set of the ICM MOGP hyperparameters is $\Theta= ( (\theta_{j})_{j \in \{a,t,c\} }, (a_{l,q})_{l=1,\ldots, L, q=1,\ldots, Q}, (\sigma_l)^2_{l=1,\ldots,L}, \beta_0, \beta_{1})$.
In our results below, the \texttt{R} package \texttt{kergp} \cite{deville2015package} is used to carry out the respective Maximum Likelihood estimation through Kronecker decompositions, see \cite{multioutput_gp,huynh2021joint} for more details.

\section{Model Grouping}\label{sec:groups}

The MOGP model from the previous section works on a group of states. The goal of grouping is to  maximize data fusion and maintain computational tractability. On the one hand, joint models lead to more accurate longevity modeling than individual-state models. This is especially noticeable for the smaller states where state-level data is very noisy and trends are hard to decipher. Moreover, a joint model facilitates making comparisons about the \emph{relative} experience of states. On the other hand, directly modeling all 51 states is computationally intractable within the MOGP framework and is unlikely to perform well anyway. The respective model would have hundreds of hyperparameters and is likely to suffer from unstable inference and identifiability issues. Furthermore, as discussed in \cite{gp_model1}, fusing  information from different populations through a MOGP can be expected to improve predictions only when these populations are similar. Thus, it makes little sense to group, say, Massachusetts (East Coast, urbanized, wealthy state) with Alabama (South, rural, poor). Indeed, 
to account for spatial relationships described in the Introduction, we would like whenever possible to group \emph{neighboring} states  following the maxim ``Everything is related to everything else, but near things are more related than distant things'' \cite{Tobler1970}. This operationalizes the hypothesis that neighboring states often share similar economic and demographic characteristics and thus should have cognate mortality trends and experiences (see also the aforementioned spatial-based analysis of U.S.~counties in \cite{geo-rur-mort,countyUS}).

With the above in mind, we seek to create groups of 3--6 \emph{similar} states, with a preference for geographic contiguity. The grouping algorithm determines which states are alike, so as to incorporate the ``right information'' to provide more accurate predictions and reduce predictive uncertainty, excluding irrelevant information. Our groupings $\mathscr{O}_s$ are state-specific, i.e., are not mutually exclusive across $s$'s, and are based on identifying similar state characteristics.
A secondary concern is making sure that groupings provide enough ``critical mass'', namely a large enough aggregate population to distinguish signal from noise.

\subsection{Motivation} \label{SUBSEC:motiv}

To motivate the issue of how to group states, we briefly discuss two GP-based alternatives. First,  we recall the base case of constructing a Single Output GP (SOGP) model state-by-state. This can be done using the methodology in Ludkovski et al.~\cite{gp_model1} and yields independently-fitted GP models. Specifically, we fit 51 SOGP models utilizing the kernel \eqref{eq:apc} with hyperparameters $\theta_a, \theta_t, \theta_c, \sigma^2, \eta^2, \beta_0, \beta_1$. Second, we construct MOGP models based on a \emph{geographic grouping}, namely the U.S.~Census regions, see Appendix~\ref{APPEND:geogroup}. These regions are 3--9 states in size and are purely geographically aligned. From a statistical perspective, the widely varying group size and the large size of some groups (e.g.~the South Atlantic region includes 9 states with a total population of over 50 million) are challenging and slow down MOGP performance.

\begin{figure}[!ht]
\centering
\begin{subfigure}{.475\textwidth}   
  \centering
  \includegraphics[width=1\linewidth]{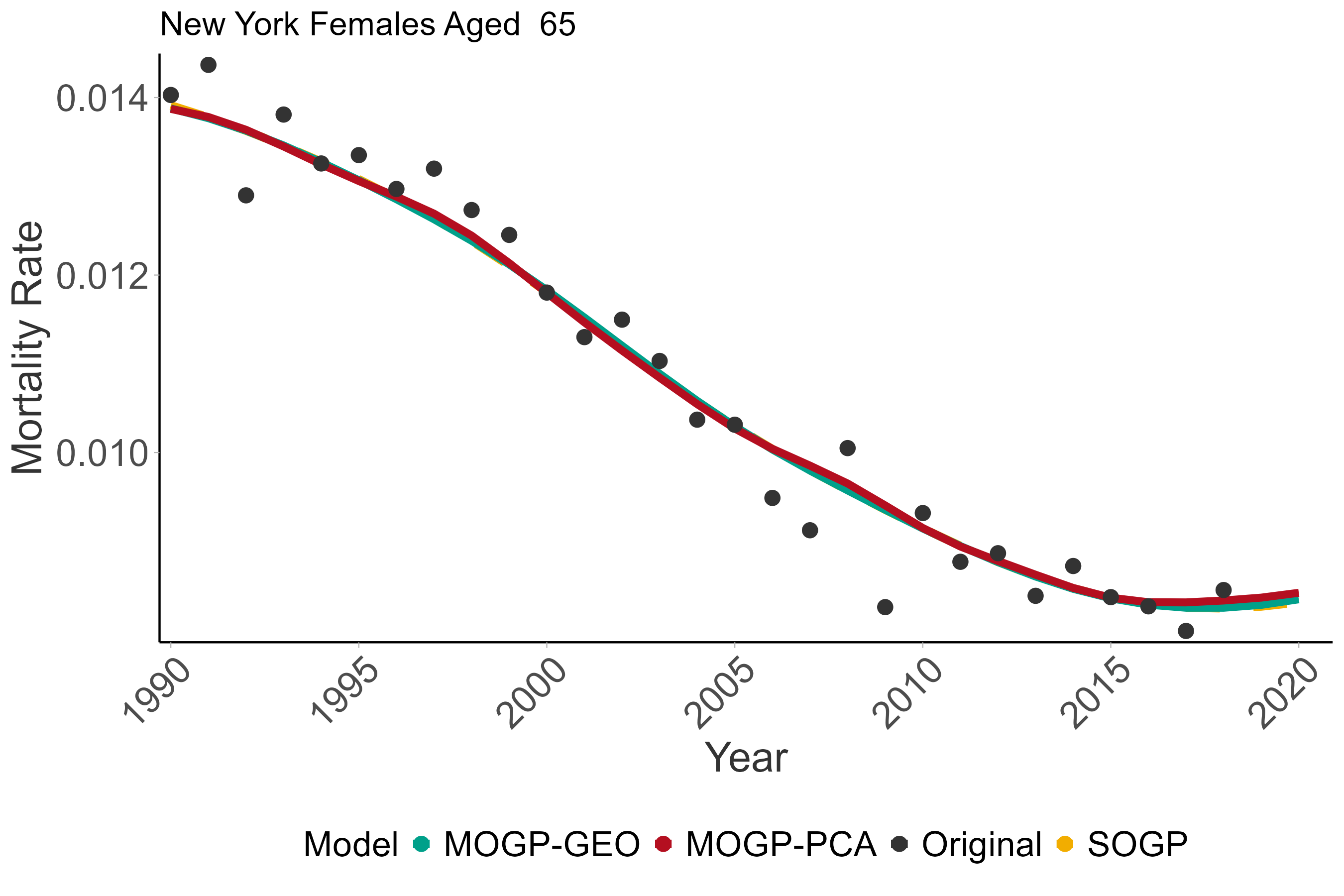}
  \caption{New York Females Aged 65}
\end{subfigure}
\begin{subfigure}{.475\textwidth}
  \centering
  \includegraphics[width=1\linewidth]{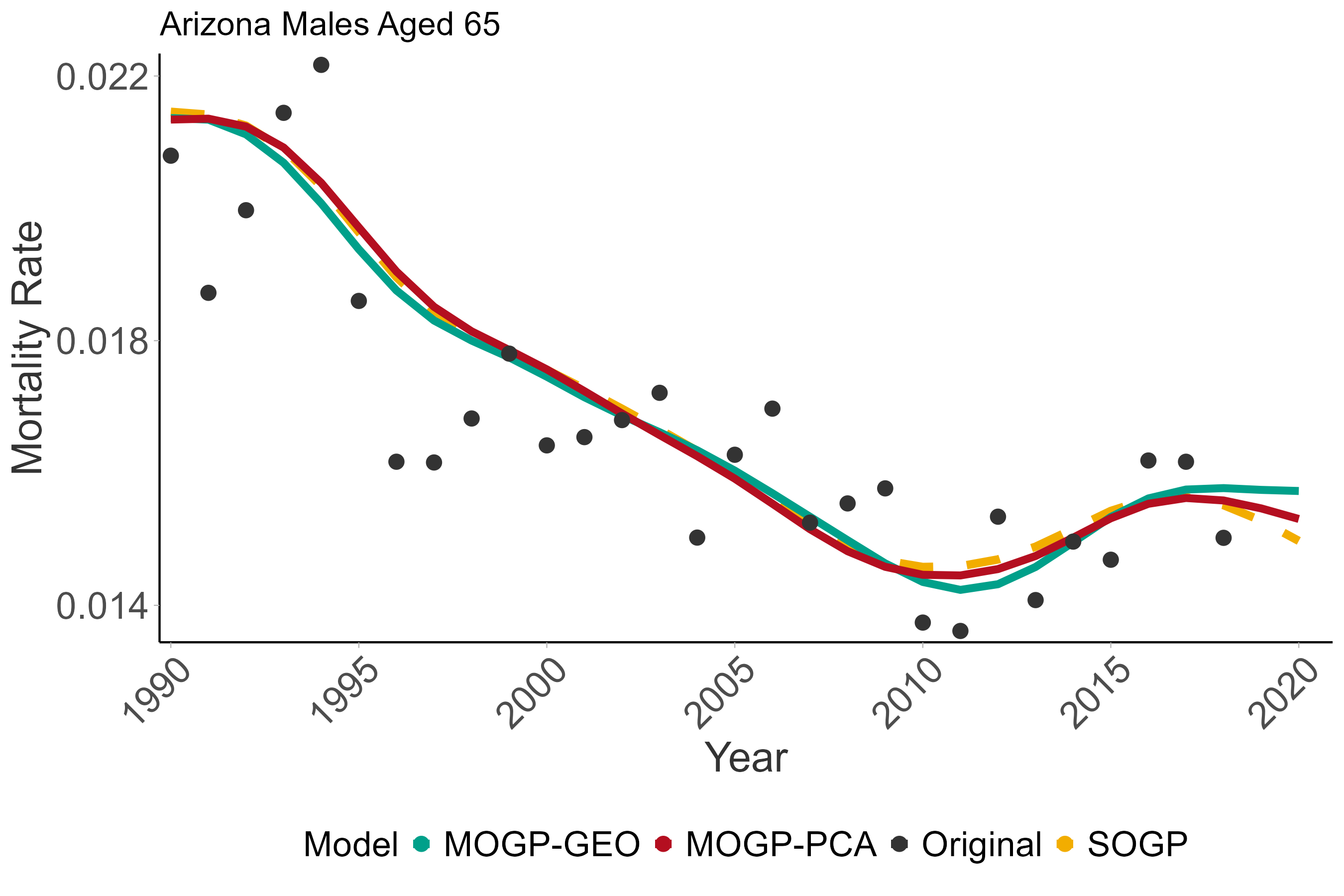}
  \caption{Arizona Males Aged 65}
\end{subfigure}
   \caption{Raw data (black circles, covering years 1990--2018) and smoothed/projected mortality curves (for years 1990--2020) for two representative states based on three different groupings:  (i) a single-state SOGP model; (ii) MOGP-GEO with geographic groupings based on U.S.~Census regions in App.~\ref{APPEND:geogroup}; (iii) proposed MOGP-PCA.} \label{FIG:groupingmatters}   
\end{figure}

Figure~\ref{FIG:groupingmatters} displays mortality experience at Age 65 in two states as a function of year for the above two choices, as well as the proposed 
state-characteristic grouping from Section \ref{sec:groups}. These smoothed mortality predictions are contrasted with the  raw mortality rates shown as black circles.  For the latter,
we observe that the observation noise is much higher in the right panel of Figure~\ref{FIG:groupingmatters} compared to the left panel. This matches the intuition that the noise in mortality data is roughly proportional to the underlying exposed population. Since Arizona's population is 7 million compared to about 20 million in New York, we expect triple the respective observation variance. The estimated magnitude of this noise is the GP hyperparameter $\sigma_l = 3.42\%$ for N.Y.~Females, and $\sigma_l = 5.09\%$ for Ariz.~Males.

All three GP forecasts are statistically unbiased and carry out viable non-parametric penalized curve fitting to make mortality predictions. In particular, their in-sample estimated mortality rates (for years 1990--2018) closely match each other. At the same time, there are non-trivial differences among the models when it comes to out-of-sample prediction, best understood as differences in mortality improvement factors. 
Figure~\ref{FIG:comp-more} in the Appendix highlights some of the discrepancies. In general, we find that single-state models tend to produce more extreme MI
and tend to over-smooth the data, 
while the MOGP-GEO models occasionally overfit. 
For example, in Figure~\ref{FIG:comp-more} Vermont Males and Minnesota Females' geographic groupings lead to a negative MI estimate, while the PCA-based grouping projects a positive MI.  The take-away is that better MOGP groupings can help maximize stability and interpretability.

\subsection{Covariates} \label{SUBSEC:covariates}

As documented in prior studies and confirmed in our results below, there is a strong correlation between economic and demographic variables and observed state-level mortality discrepancies. Hence we use various state characteristics to compute a customized similarity metric that drives our group selection.

To group states we consider a diverse set ${\cal C}$ of 18 non-mortality state-level covariates, chosen to be a representative collection of (1) economic, (2) demographic, and (3) geographic characteristics. The respective state-level data are obtained from several sources, including \cite{uscensus}, \cite{fred}, and \cite{bea}. These sources do not  distinguish between the Male and Female subpopulations, hence all the covariates are shared among the genders.
 Table~\ref{tab:statecov} lists the covariates we work with; Appendix~\ref{APPEND:datasources} provides a complete description. Our covariates overlap with and are broadly similar to the 20 used in Chetty et al.~\cite[Table 8]{chetty2016association} (who grouped them into  Health Behaviors, Healthcare, Environmental, Labor Market, Social Cohesion and Other Factors), the six used in Couillard et al.~\cite{risinggeodis} and the 11 used in Barbieri~\cite{Barbieri2020}.

\begin{table}[ht]
    \centering
{\small \begin{tabular}{ p{1.9in}||p{1.85in}||p{1.85in}}
 \hline
 Economic Covariates & Demographic Covariates & Geographic Covariates \\
 \hline
 Educational Attainment \textcolor{economic-var}{(EA)}  & Non-minority Pop'n \textcolor{demographic-var}{(NMP)}    & Average Temperature \textcolor{geographic-var}{(TP)} \\
 Percent Change in GDP \textcolor{economic-var}{(GDP)} &   Percentage Elderly \textcolor{demographic-var}{(ED)} & Average Rel. Humidity \textcolor{geographic-var}{(RH)}   \\
 Median Income \textcolor{economic-var}{(MI)} & Without Health Insurance \textcolor{demographic-var}{(HI)} & Average Dew Point \textcolor{geographic-var}{(DP)} \\
 Regional Price Parities \textcolor{economic-var}{(RPP)}  & Obesity Rate \textcolor{demographic-var}{(OR)} & Population Density \textcolor{geographic-var}{(TPD)}\\
 Poverty Rate \textcolor{economic-var}{(PR)} &   Political Preference \textcolor{demographic-var}{(PP)} & Land in Farms \textcolor{geographic-var}{(LF)} \\
 Urbanization Percentage \textcolor{economic-var}{(UP)} & Percent Religious \textcolor{demographic-var}{(R)} & Share of Immigrant Pop'n \textcolor{geographic-var}{(IP)} \\
 \bottomrule 
\end{tabular}}
  \caption{Summary of the 18 selected state covariates in $\mathcal{C}$ , grouped by the categories of Economic, Demographic and Geographic variables. See Appendix~\ref{APPEND:datasources} for definitions of each covariate.}
    \label{tab:statecov}
\end{table}

\begin{figure}[ht] \centering
\minipage{0.33\textwidth}
  \includegraphics[width=\linewidth,trim=0.65in 0.2in 0.05in 0.15in]{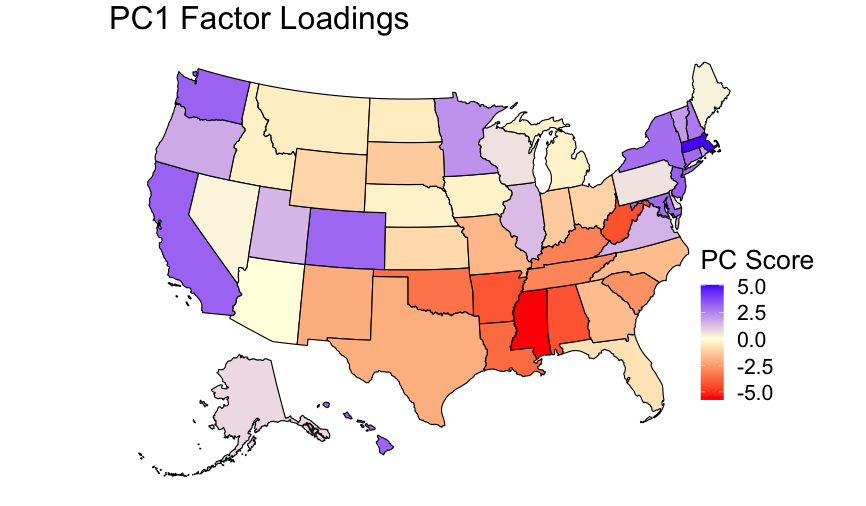}
\endminipage
\minipage{0.33\textwidth}
  \includegraphics[width=\linewidth,trim=0.65in 0.2in 0.05in 0.15in]{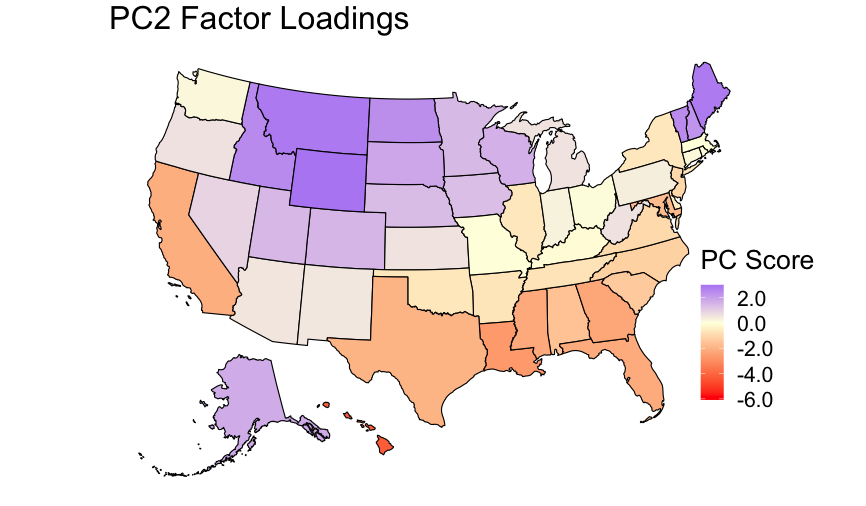}
\endminipage
\minipage{0.33\textwidth}
  \includegraphics[width=\linewidth,trim=0.6in 0.2in 0.2in 0.15in]{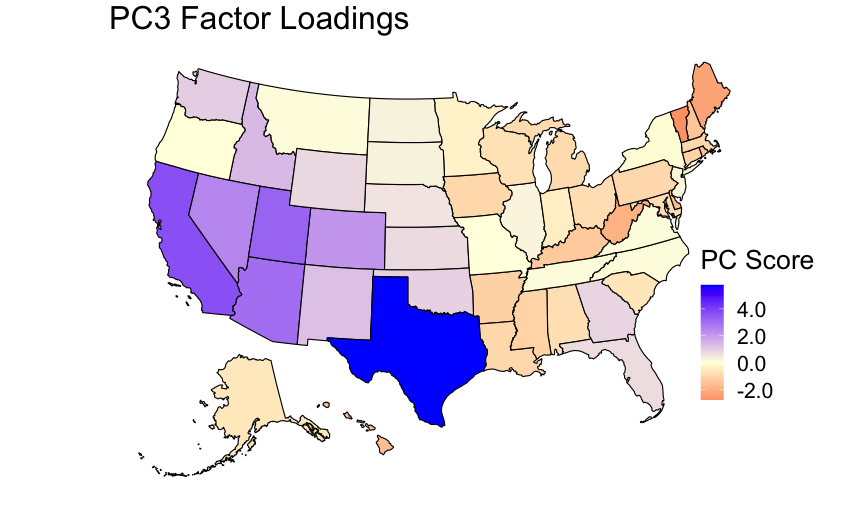}
\endminipage
 \caption{State-wise PCA factor loadings $\mathcal{P}_k(s)$, $k=1,2,3$. } \label{FIG:pcscores}
\end{figure}

In order to minimize collinearity and identify the main sources of differences across states, we apply Principal Component Analysis (PCA) to the above $51 \times 18$ matrix ${\cal C}$ (states times covariates). Use of PCA to summarize covariates is also advocated in Barbieri~\cite{Barbieri2020}. 
Figure \ref{FIG:pcscores} and Table \ref{table:factorload} summarize the PCA results, where we focus on the first three PCA components $PC1, PC2, PC3$ that together explain over 66\% of the variance in ${\cal C}$ and allow us to succinctly summarize the drivers of heterogeneity.

\begin{table}[H]
\centering
\begin{tabular}{r|ccc}
& PC1 & PC2 & PC3\\\hline
Standard Deviation& 2.53 & 1.70 &1.59\\
Proportion of Variance & 0.36 & 0.16 & 0.14\\ 
Eigenvalue $\lambda_i$ &  6.486 &  3.424 &  2.548 \\
\bottomrule
\end{tabular}
\caption{Summary of Principal Component Analysis for state covariates. \label{tbl:pca}}
\end{table}

The PC1 component can be interpreted as an economic/wealth factor: the most expensive states (N.Y., Cali.) display the highest PC1 loadings, while states like Miss.~and La.~have lower PC1 loadings. As confirmation, (Table \ref{table:factorload} in Appendix \ref{APPEND:le-pca}) most of the \textsl{economic covariates} from $\mathcal{C}$ are positively correlated with the first PC component. The PC2 component can be interpreted as a climate-related factor, showing a North-South trend, with the Southern-most, warmest states having highest PC2 loadings. Lastly, PC3 component loosely corresponds to the Sun Belt states: the Southwest plus Texas, Georgia and Florida. The population in this region is rapidly growing due to internal migration and immigration. 

The presented covariates and PCA factors are expected to be linked to mortality experience.  Table \ref{table:factorload} in the Appendix shows the correlation between each covariate and LE at birth, $e_0$, as constructed by the CDC \cite{cdc}. We observe that economic variables are the ones most correlated to LE, confirming the intuition that the PC1 factor loadings are helpful in grouping for mortality analysis. Similar statistical association between income, education and age-specific mortality is documented in \cite{li2021assessing}.

\subsection{Grouping by Covariate Similarity} \label{SUBSECT:groupingcovsim}

The PCA factor loadings obtained above are used to construct groupings of states for the MOGP. We first define a distance metric $\mathfrak{D}(s_1,s_2)$ between any given two states $s_1,s_2 \in \mathcal{S}$. We then aim to group $s$ with the most similar states according to $\mathfrak{D}(s,\cdot)$ while respecting  geographical contiguity of the resulting group. Geographical contiguity is preferred since neighboring states naturally share similar mortality experiences. The groups are constructed in a stepwise agglomerative manner, aiming to have 3--5 states in each group. Note that since the covariates are the same across genders, all the groupings pertain both for Males and for Females.

\textbf{Distance between States.} We use PCA components from previous section to define a distance between states. To this end, we compare the respective factor loadings ${\cal P}_{k}(s)$, generating a Euclidean metric weighted by the eigenvalues associated with each PCA component:
\begin{equation} \label{EQU:distance}
    \mathfrak{D}(s_1,s_2)^2 \coloneqq  \lambda_1\big(\mathcal{P}_{1}(s_1) - \mathcal{P}_{1}(s_2)\big)^2+ \lambda_2\big(\mathcal{P}_{2}(s_1) - \mathcal{P}_{2}(s_2)\big)^2+ \lambda_3\big(\mathcal{P}_{3}(s_1) - \mathcal{P}_{3}(s_2)\big)^2,
\end{equation}
where  $\lambda_k$ is the eigenvalue associated with PC component $k= 1,2,3$. The intuition is to prioritize the first factor that  explains the majority of the observed variability in our covariates. 

\textbf{Identifying Similar States.} Our next goal is to create an algorithm that simultaneously identifies which states are similar and are close geographically. 
Fix a state $s \in \mathcal{S}$. We construct the set $\mathcal{N}_{s}$ of  \textsl{nearest neighbors} of $s$ as follows:
\begin{enumerate}[(i)]
    \item Compute the geographical neighbors of $s$, denoted \\ $O_{1}(s) \equiv O_1 = \{ s_* :  s \text{ and }s_* \text{ are geographically contiguous}\}$.   That is, $O_{1}$ is the collection of states which share a physical boundary with $s$. 
    \item Add  to $\mathcal{N}_{s}$ the state $s_{1}$ that minimizes the PCA-based distance to $s$ in $O_1$, $s_{1} = \arg\min_{s_* \in O_{1}} \mathfrak{D}(s,s_*)$.
    \item Update the neighborhood definition to include $s_1$: $$O_2 \equiv O_{2}(s\cup s_{1}) = \{ s_* : s_* \text{ is geographically contiguous with either } s \text { or } s_1 \} .$$ 
    \item Repeat steps (ii) and (iii) with $s_n = \arg \min_{s_* \in O_{n}}\mathfrak{D}(s,s_*)$ for $n=2,\ldots, 10$.
\end{enumerate}

\begin{figure}[H]
\minipage{0.32\textwidth}
  \includegraphics[width=\linewidth]{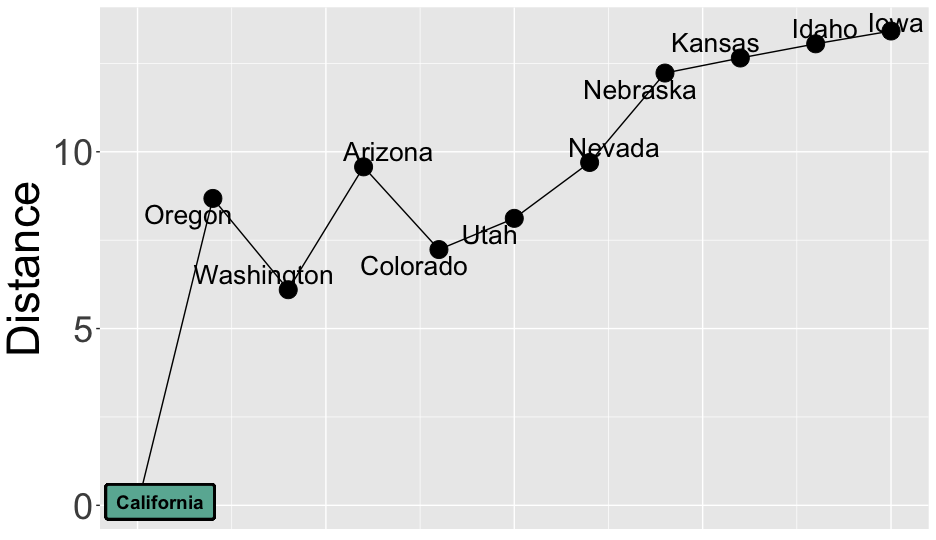} \\ \hspace*{1in}  CA
\endminipage\hfill
\minipage{0.32\textwidth}
  \includegraphics[width=\linewidth]{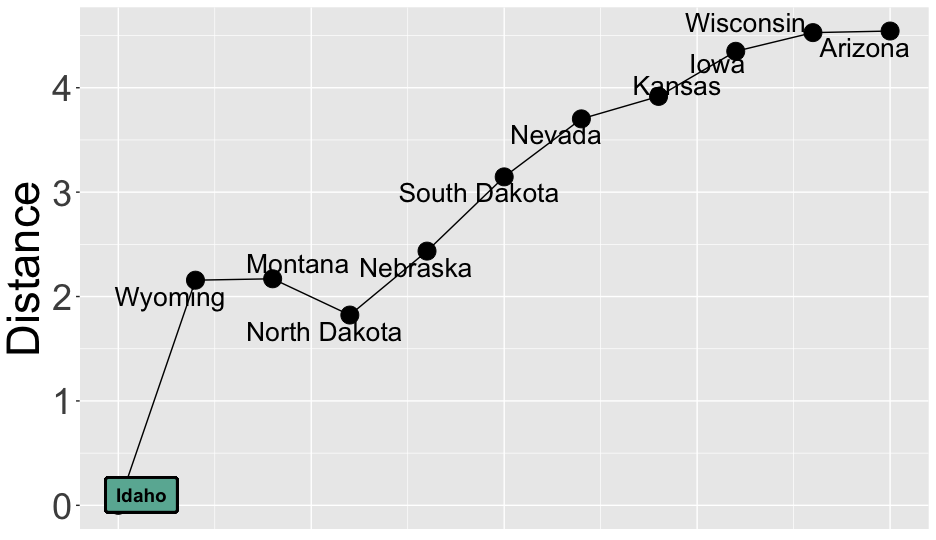} \\ \hspace*{1in} ID
\endminipage\hfill
\minipage{0.32\textwidth}%
  \includegraphics[width=\linewidth]{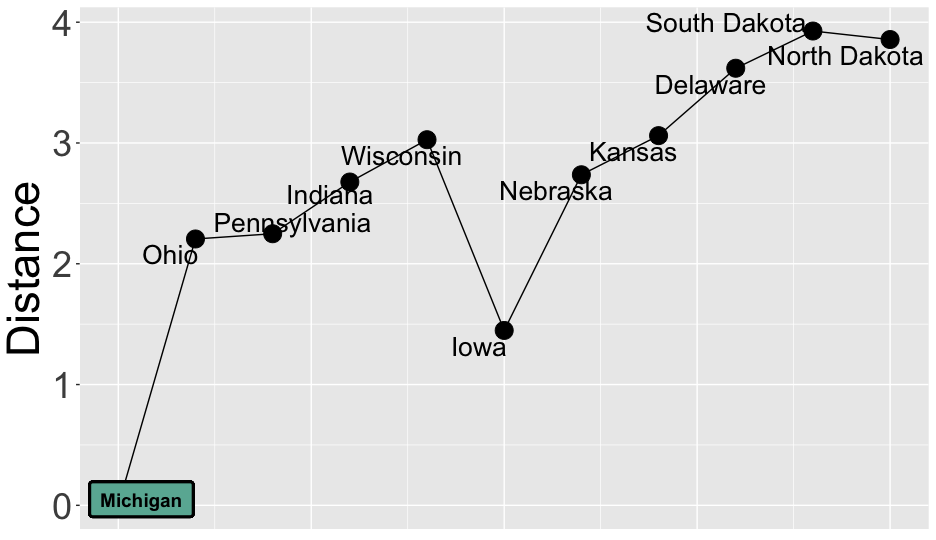} \\ \hspace*{1in} MI
\endminipage
\caption{The ten most similar states $\mathscr{N}_s$ for $s=$Cali.~(\emph{left}), Idaho (\emph{middle}), and Mich.~(\emph{right}).}\label{FIG:nn10-lines}
\end{figure}

Figure \ref{FIG:nn10-lines} visualizes the results from running the above algorithm on a few representative states and until $|{\cal N}_s| = 10$. The y-axis shows the distance $\mathfrak{D}(s,s_n)$ between the target state  $s \in \mathcal{S}$ and the ``neighbor'' added in round $n$. We observe the common structure of some `zig-zags', indicating that geographical contiguity does not capture the whole story. That is, there exist states which are further apart in physical distance, yet closer with respect to $\mathfrak{D}$; see, for example, Michigan and Iowa, or Idaho and North Dakota. Indeed, N.Dak.~is the closest to Idaho in terms of $\mathfrak{D}(s,\cdot)$ even though they do not share a geographical boundary. Such observations give us the inspiration to eschew strict geographical contiguity in the proposed group selection procedure below.

\begin{figure}[H]
\minipage{0.32\textwidth}
  \includegraphics[width=.8\linewidth]{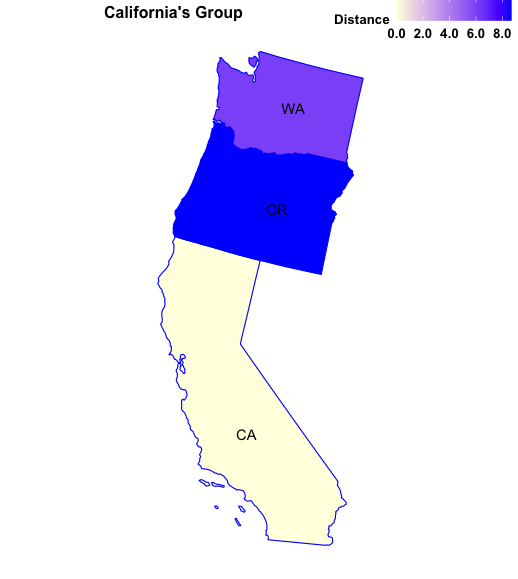}
\endminipage\hfill
\minipage{0.32\textwidth}
  \includegraphics[width=\linewidth]{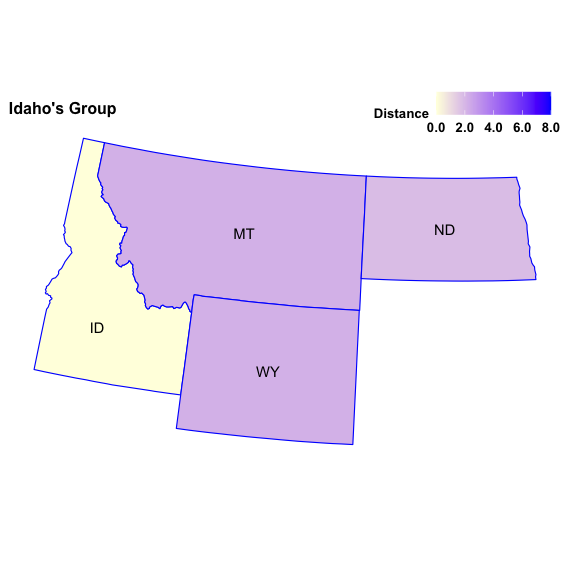}
\endminipage\hfill
\minipage{0.32\textwidth}%
  \includegraphics[width=\linewidth]{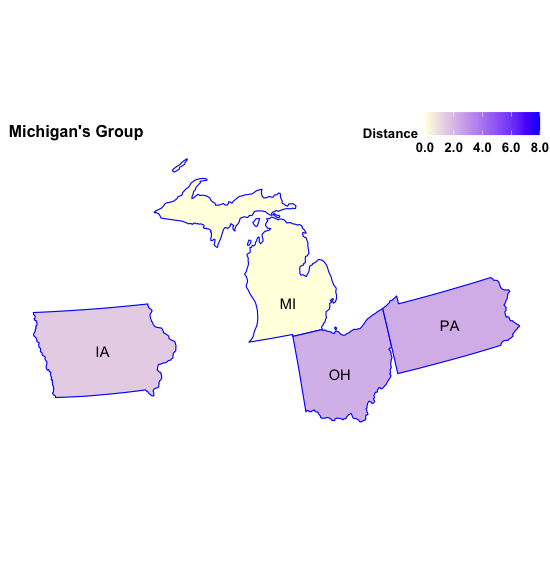}
\endminipage
 \caption{Selected state groupings $\mathscr{O}_s, \, s = $ Cali.~(left panel), Idaho (middle), and Mich.~(right).} \label{FIG:optimalgroups}  
\end{figure}

\textbf{Proposed Grouping $\mathscr{O}_s$  for $s \in \mathcal{S}$}. We start with ${\cal N}_ s = \{ s, s_1, s_2, \ldots \}$ as above and select  $s, s_1, s_2 \in \mathscr{O}_s$; that is, the first three states from $\mathcal{N}_s$ are in the group. Next, let $s_3' = \arg\min_{s^* \in {\cal N}_s \backslash (s_1,s_2)} \mathfrak{D}(s^*,s)$ be the next closest state to $s$ in terms of $\mathfrak{D}(\cdot, s)$. We include  $s_3' \in \mathscr{O}_s$, if and only if $\mathfrak{D}(s_3',s) <  \max_{i=1,2} \mathfrak{D}(s_i, s)$; so that we create a group of 4 states if $s_3'$ is a better (closer) fit in distance to $s$ than either of the two neighbors $s_1, s_2$ already in the group.  The above procedure enforces a high degree of geographic contiguity (at least two states are guaranteed to be contiguous with $s$) but also allows to add one more state that is not geographically close but is very similar to $s$ in terms of the PCA loadings. As an example, Iowa is added to Mich.'s group (Figure~\ref{FIG:optimalgroups}, right), and N.~Dak.~is added to Idaho's group (middle panel),  while Cali.~stays in a group of 3 (no further state is closer than $s_3=$Wash.~in that case, left panel). 

As a final step, we ensure that all groups have sufficient underlying population to yield credible estimation. To this end, we augment additional states (in order of their distance $\mathfrak{D}$) until the group $\mathscr{O}_s$ has a total population of at least 5 million. This is particularly relevant for the Mountain West region, where we have Idaho, Montana, Wyoming, North and South Dakota (all with populations under 1.5M) tending to group together.

\begin{remark}
Alaska and Hawaii lack natural geographical neighbors. See Appendix \ref{APPEND:sg} for the methodology (and results) used to calculate AK and HI groupings.
\end{remark}

A sample of the resulting groupings $\mathscr{O}_s$ are shown in Figure \ref{FIG:optimalgroups}; the full list of 51 $\mathscr{O}_s$'s  is in App.~\ref{APPEND:sg}. The colors correspond to $\mathfrak{D}(s_*,s)$, i.e.~how `close' a given state is to its selected neighbors. We note that closeness in terms of $\mathfrak{D}$ varies; there are many very similar (in terms of PCA factor loadings) states in the Midwest, while California's neighbors are all much less similar to it.

Figure \ref{GIF:G3VSG4} in Appendix~\ref{APPEND:sg} visualizes the size of the constructed groups. Recall that by default $|\mathscr{O}_s| = 3$; groups of four arise when one of the two most similar  (in terms of $\mathfrak{D}$) states is not contiguous with $s$, but farther away. Most of these cases arise in the Midwest and Mid-Atlantic regions. In addition, in Mountain West we have groups of 5 or 6 due to low state population counts. 

The groups $\mathscr{O}_s$ are constructed separately for each state. To get a sense of how similar are the proposed groupings for different target states, we define the concept of reciprocity.
 States $A$ and $B$ are in positive reciprocity (PR) if $B \in \mathscr{O}_A$ and $A \in \mathscr{O}_B$, i.e.~they are mutual members in the respective groups. The states that do not experience any PR are La., Miss., R.I., S.Dak., Texas, and W.Va.. This generally means that these states are quite different from all their geographical neighbors and are not getting selected for their neighbors' groups. Conversely, the following states end up with exactly the same groups of three:
\begin{equation*}
    \big(\text{Arizona, Nevada, Utah}\big).
\end{equation*}
This triple is both contiguous geographically and is very similar across covariates, forming a ``mini-cluster". Similarly, the following pairs of states result in the same constructed groups $\mathscr{O}_s$: 
\begin{equation*}
\big(\text{Montana, Wyoming}\big);  \quad
    \big(\text{Connecticut, New York}\big). 
\end{equation*}

\section{State Mortality Predictions}\label{sec:analysis}

Following the grouping method in the previous section, we construct 51 MOGP models for the 51 states. Throughout we use the ICM covariance structure with rank $Q=3$. We then make predictions for years 1990--2020, focusing on retrospective analysis of U.S.~state mortality. Recall that the training USMDB data is up to 2018, so the models are based on pre-COVID19 data and are not aware of the pandemic. Consequently the 2020 predictions can be viewed as a statistical baseline for how 2020 would have looked like without COVID19.
 
 The mortality surfaces obtained from the MOGP models describe the mortality trends for each individual state, across age groups and time horizons.  In this Section we analyze the relative mortality projections across calendar Years (section \ref{SUBSEC:MORTPREDYR}) and then individual Ages (section \ref{SUBSEC:AGESTURCMI}). In  Section \ref{sec:IR} we then investigate the \textit{mortality improvement factors} (MI), obtained as the time-gradient of these mortality surfaces.

\subsection{Time Structure} \label{SUBSEC:MORTPREDYR}

\begin{figure}[ht] \centering
\minipage{0.46\textwidth}
   \includegraphics[width=\linewidth]{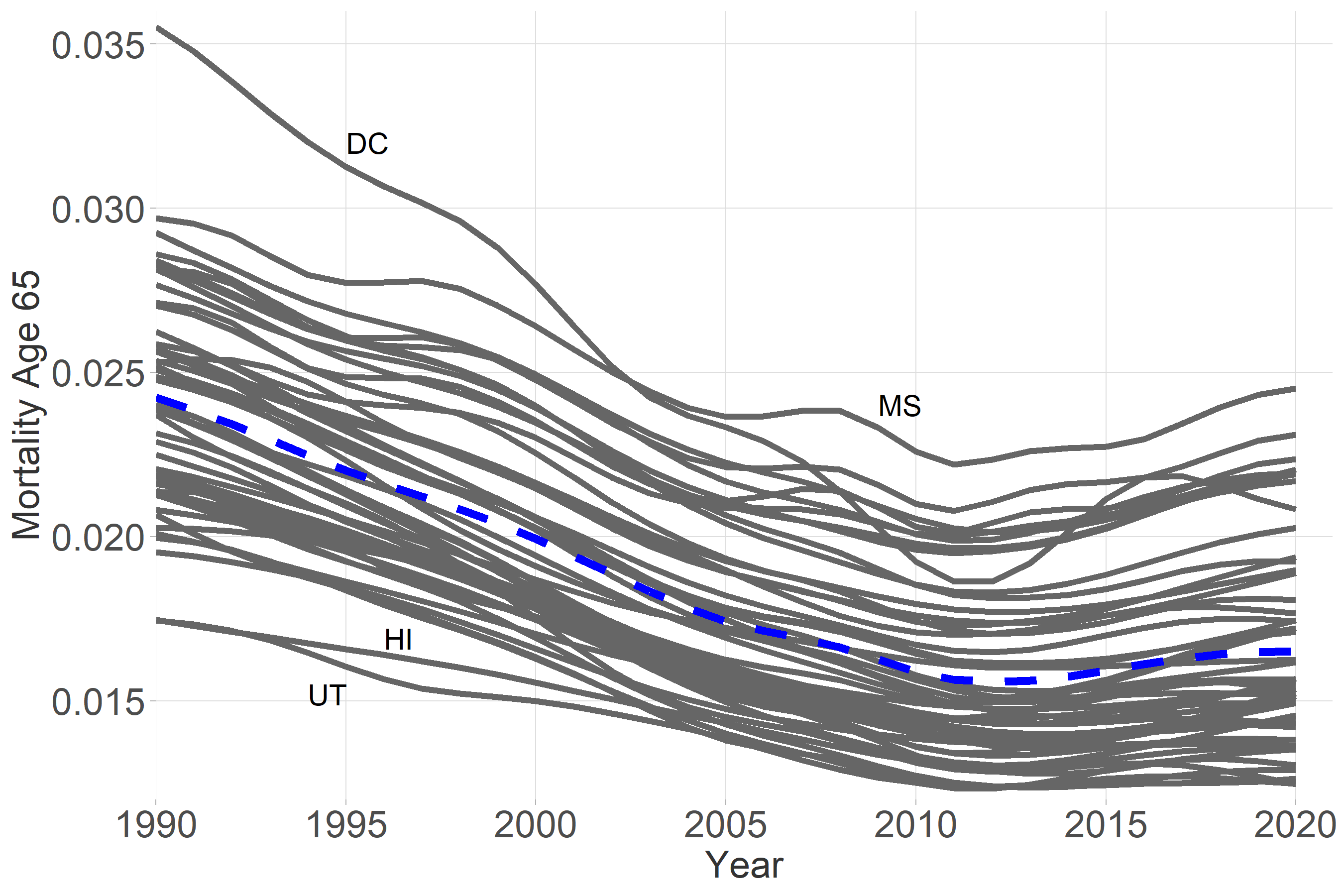}\\ 
  \hspace*{1in} Males Age 65
\endminipage\hfill
\minipage{0.46\textwidth}
 \includegraphics[width=\linewidth]{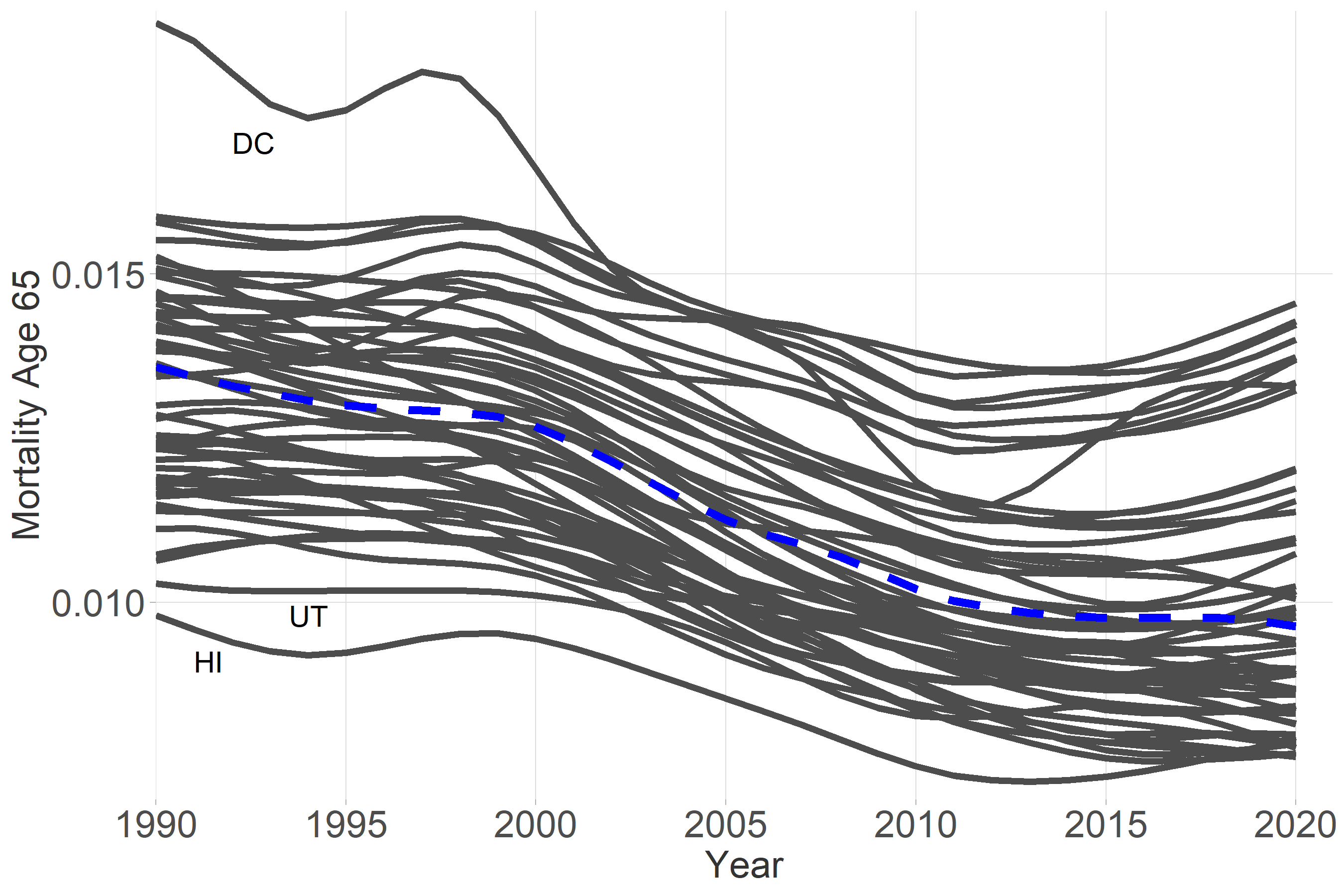} \\
  \hspace*{1in} Females Age 65
\endminipage
\caption{MOGP-PCA mortality rates for Age 65 Males (\emph{left} panel) and Females  (\emph{right}) and Years 1990--2020. U.S.~national smoothed rate is shown as dashed blue.} \label{MR-years1990-2020-65}
\end{figure}

Figure \ref{MR-years1990-2020-65} shows the bulk behavior of state mortality rates across years 1990--2020 fixing the Age, namely at age 65. As a baseline, we also compute and show the national-level U.S.~mortality rate. This curve (dark blue dashed line in the plot) is generated by fitting a SOGP model to the aggregated U.S.~mortality experience. As expected, it lies in the core of the state curves and can be interpreted as  the population-weighted average of the state-level mortality projections. See also Figure \ref{MR-years1990-2020-70} in Appendix~\ref{app:mortality-75} for Age 75 counterpart, and the interactive Shiny widget \cite{Shiny} where users can select any other desired Age.   
Notice that the mortality trend before 2010 is universally positive, while in the past decade mortality has been either stagnating or deteriorating. 
We also observe that although there is a strong common trend, mortality evolution has been rather variable across states. There are several outlying curves in Figure \ref{MR-years1990-2020-65}, notably D.C.~and Hawaii, as well as many ``cross-overs'' where states change relative ranks over time. This heterogeneity  increased  during the 2010s compared to 2000s, and there are also more cross-overs in the Male populations compared to Females. Moreover, the spread among states is very substantial: e.g.~from just over 0.75\% mortality in 2020 for 65-year old Females in N.Dak.~and Conn., to 1.45\% in Miss.~and Okla.---a ratio of 1.88 at the right edge of the right panel. 

\begin{figure}[ht]
\begin{center}
\minipage{0.46\textwidth}
   \includegraphics[width=\linewidth,trim=0.35in 0.2in 0.35in 0in]{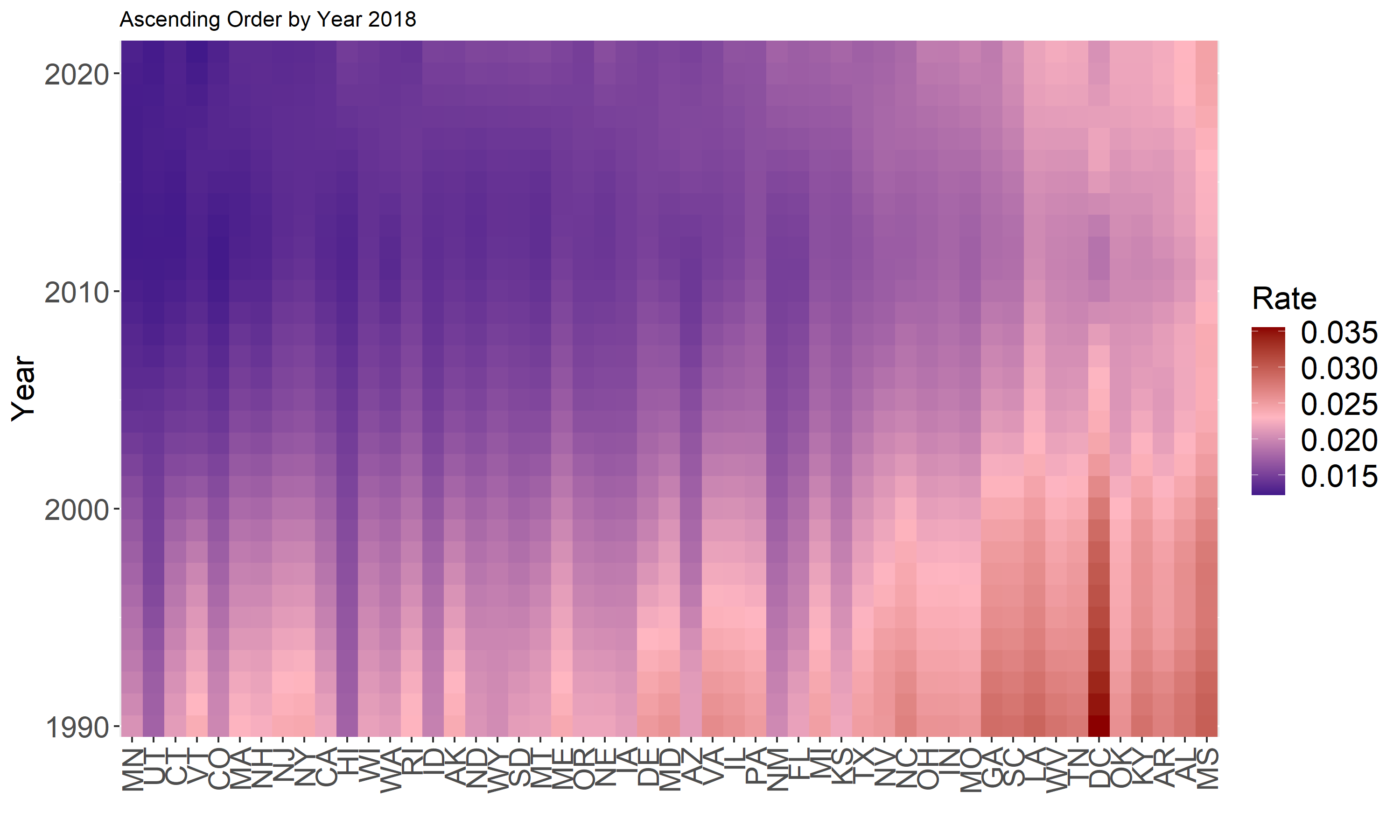} \\
   \hspace*{1in} Males Age 65
\endminipage\hspace*{14pt}
\minipage{0.46\textwidth}
   \includegraphics[width=\linewidth,trim=0.35in 0.2in 0.35in 0in]{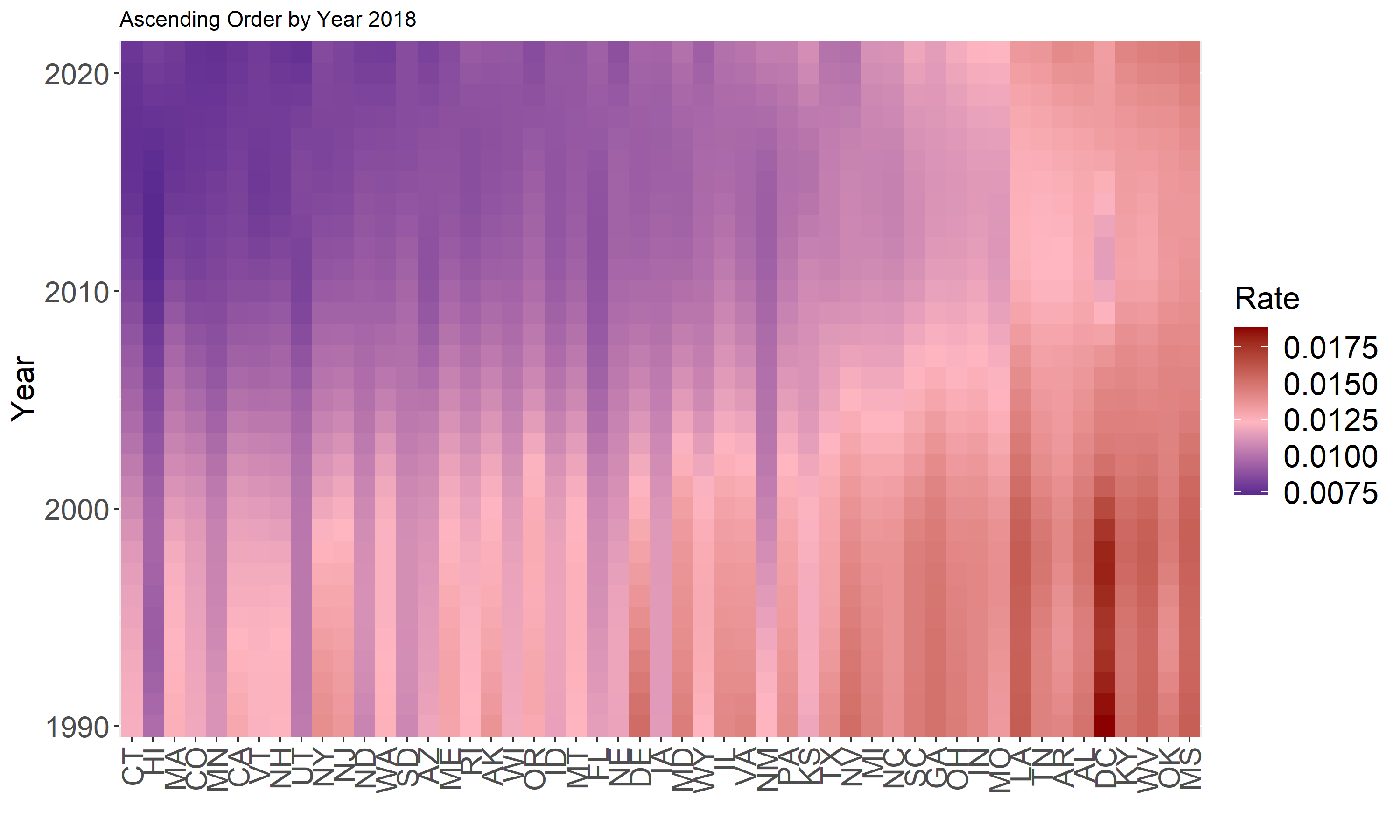} \\
  \hspace*{1in}  Females Age 65
\endminipage
\end{center} \vspace*{-10pt}
\caption{MOGP-PCA smoothed Mortality Rates for Age 65 Males (\emph{left} panel) and Females (\emph{right}) and Years 1990--2021. States are ordered from left to right by their mortality in 2018.}
\label{MR-2010HM-65}
\end{figure}

A more in-depth visualization is provided by the heatmaps in Figure \ref{MR-2010HM-65} where columns denote states and rows denote years; darker gradient corresponds to lower mortality. The states are sorted in increasing order of their mortality as of 2018.
The expected behavior as we move up across a column is a smooth transition from red to blue/purple corresponding to improving mortality throughout 1990--2020. While this occurs for some states (notably District of Columbia is displaying this pattern, having gone from one of the biggest laggards to being middle-ranked by 2020), for many columns the pattern is much more checkered. Note the colored stripes that indicate different historical mortality paths for states that nowadays have similar rates. Only a few states, such as Utah, Colo.~and Hawaii show a {consistent} positive mortality trend throughout the past three decades.  

Relative ranks of states are of great interest.
By looking at the first few/last columns of the heatmaps in Figure \ref{MR-2010HM-65}, we can read off the states with the worst/best mortality. Table~\ref{TABLEbestworst202065-pred} summarizes which states are projected to have the highest (worst) and lowest (best) mortality in 2020. Northeast and Pacific states have the overall best mortality, and the Southern states are at the bottom. There is a lot of consistency between the genders as well, with Miss., Ala., and Ky.~being among the bottom-5 in all four columns on  the left of Table~\ref{TABLEbestworst202065-pred}. We caution against reading too much into the precise rankings: our fitted models yield predictive standard deviations of latent log-mortality on the order of 0.04-0.1 (depending on size of the state), which corresponds to about 0.002-0.003 on the original scale of mortality rates. For example, Colo.~'s Male mortality in 2020 is forecast to be 1.304\% (ranked fifth) but with a 95\% predictive interval of $[1.141\%,1.489\%]$ which would be anywhere within the top-20. There is a complex correlation between the projections of different states which makes relative ranks less volatile, but the upshot is that it is not statistically possible to decide whether a given state is in top-5 or top-10. Nevertheless, these finds closely echo the socio-economic analysis in Chetty et al..~\cite{chetty2016association} on life expectancy for individuals in the bottom income quartile, who have singled out Tenn., Ark., Okla.~as the worst performers and Cali.~and Vt.~as being the top-ranked states.
 
\begin{table}[ht]
    \centering
\begin{tabular}{llll}
\multicolumn{4}{c}{Best States} \\
\cline{1-4}
\multicolumn{2}{c}{Males}    & \multicolumn{2}{c}{Females}  \\
Age 65 & Age 75 & Age 65 & Age 75 \\
\hline   
Vt.   & Colo.   & Minn.   & Hawaii    \\
Utah  & Hawaii      & Colo. & Cali.       \\
Minn.  & Vt.     & Conn.  & Ariz.       \\
Conn.   & Conn.   & Utah  & Fla.     \\
Colo. & Cali. & Cali.    & Conn.    \\
\hline 
\end{tabular}
\hspace{0.4in}
\begin{tabular}{llll}
\multicolumn{4}{c}{Worst States} \\
\cline{1-4}
\multicolumn{2}{c}{Males}    & \multicolumn{2}{c}{Females} \\
Age 65 & Age 75 & Age 65 & Age 75 \\
\hline  
Miss.  & Miss.  & Miss.  &  Ky.   \\
Ala.  & Ala.     & Okla.  & W.Va.        \\
Ark. & Ky.     & W.Va. & Miss.       \\
W.Va. & Okla. & Ky.    & Okla.      \\
Ky. & Tenn. & Ala.    & Ala.      \\
\hline  
\end{tabular}  
    \caption{Top-5 and Bottom-5 states ranked by MOGP-PCA projected mortality rate in 2020 at Ages 65 and 75. The best and worst states are in the first rows.}
    \label{TABLEbestworst202065-pred}
\end{table}

\begin{figure}[ht]  \hfill
\minipage{0.47\textwidth}
  \includegraphics[width=\textwidth,trim=0.05in 0.05in 0.15in 0.5in, clip=true]{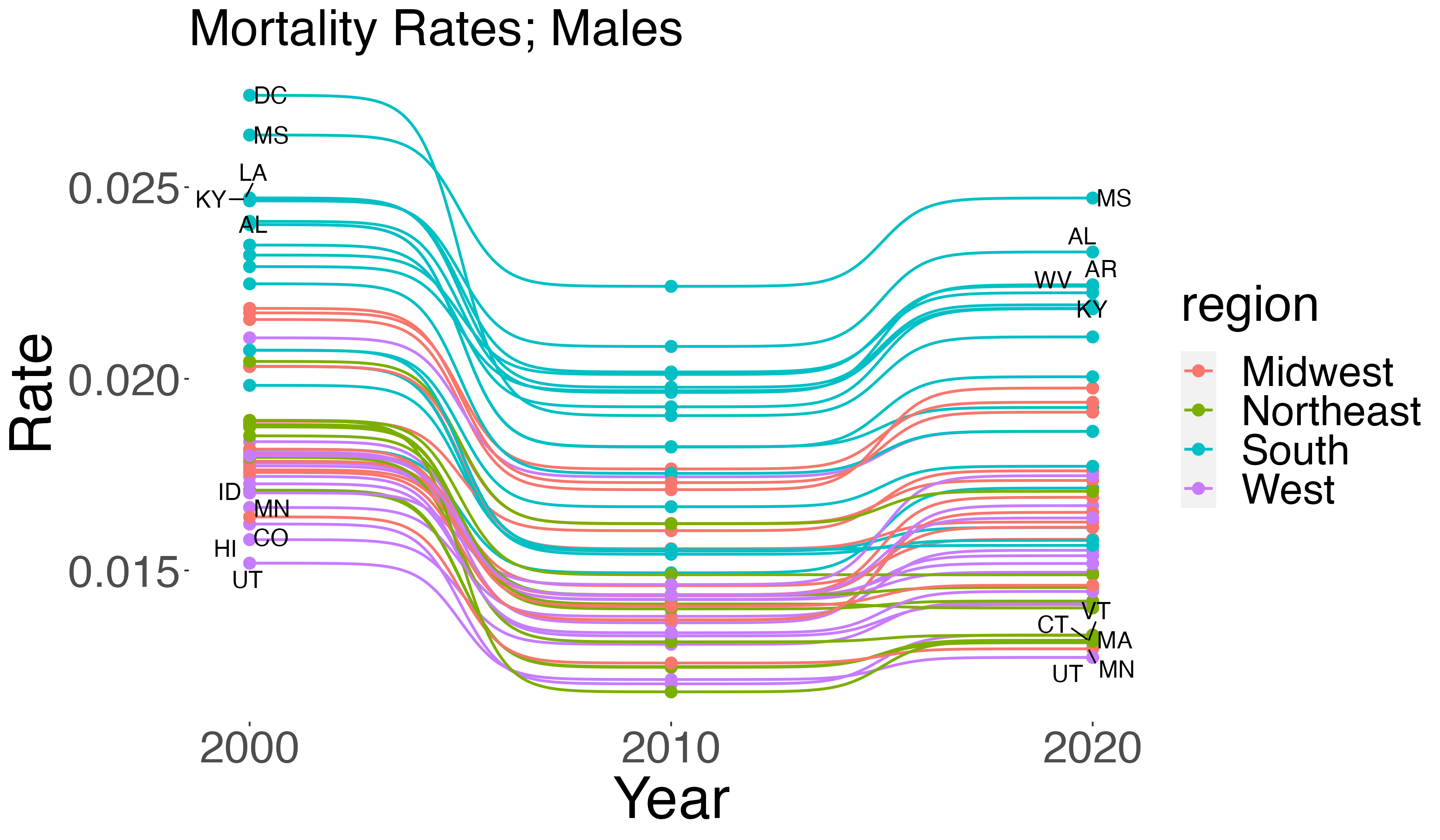}  \\
  \hspace*{1in} Males Aged 65
\endminipage \hspace*{0.2in} 
\minipage{0.47\textwidth}
  \includegraphics[width=\textwidth,trim=0.05in 0.05in 0.15in 0.5in, clip=true]{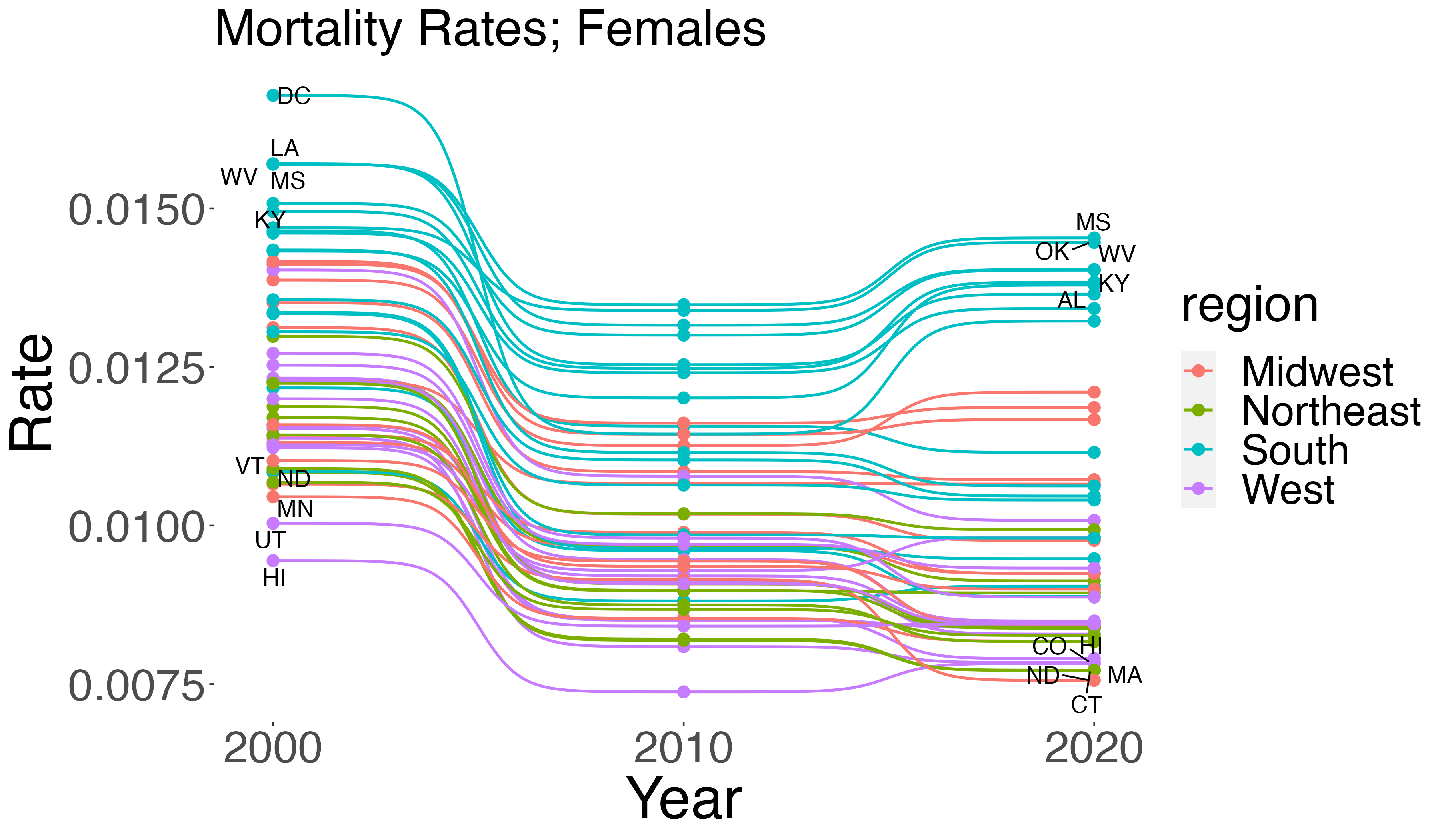} \\
  \hspace*{1in} Females Aged 65
\endminipage
\caption{State rankings in terms of Age 65 mortality rate across years 2000, 2010 and 2020.} \label{fig:MR-BC-65}
\end{figure}

To better visualize the relative ranks of states across time, Figure \ref{fig:MR-BC-65} shows a bump chart ranking states by their mortality rates in 2000, 2010, and 2020 at Age 65. We observe that all states experienced a decrease in their mortality rates between the years 2000 and 2010. However, for the 2010s the picture is largely reversed.
 In fact for Males, only Maine and New York are projected to improve between 2010 and 2020. For Females, the picture is mixed; 16 states (AL, AR, DC, FL, HI, IN, KY, LA, MO, MS, NM, OH, OK, TN, UT, WV) are projected to have a worse Female Age 65 mortality in 2020 compared to 2010 and the rest improve.  Our analysis corroborates the county-level findings in \cite{countyUS} who report an improvement in mortality between 2000 and (approximately) 2014, followed by a recent deterioration in county-level rates, attributed to  \emph{deaths of despair}.

Moreover, Figure \ref{fig:MR-BC-65} indicates a certain split among Female mortality: laggard states, primarily in the South, have deteriorating Female mortality in 2020 compared to 2010, while the best-performing states in the Northeast and West continue to experience improving mortality. In other words, there is a \emph{divergent} pattern where states with low (Female) mortality are doing relatively better compared to states with higher mortality, with this pattern aligning with regional partitions. Similar age-aggregated conclusions appear in \cite{Fenelon2013,chetty2016association}.

\subsection{Age Structure} \label{SUBSEC:AGESTURCMI}

\begin{figure}[ht]
\minipage{0.46\textwidth}
  \includegraphics[width=\linewidth]{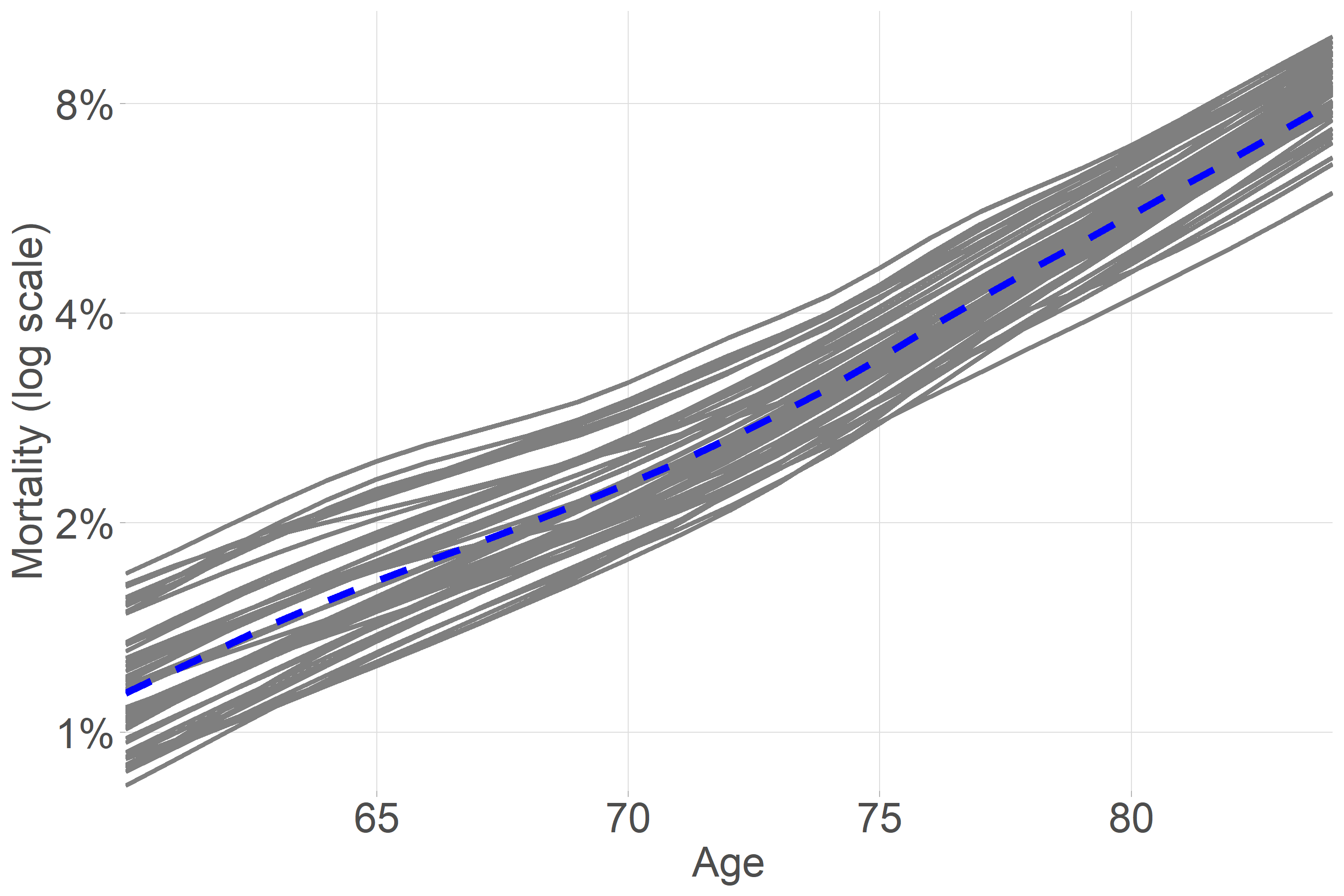} \\
  \hspace*{1.25in} Males 
\endminipage\hspace*{0.3in}
\minipage{0.46\textwidth}
  \includegraphics[width=\linewidth]{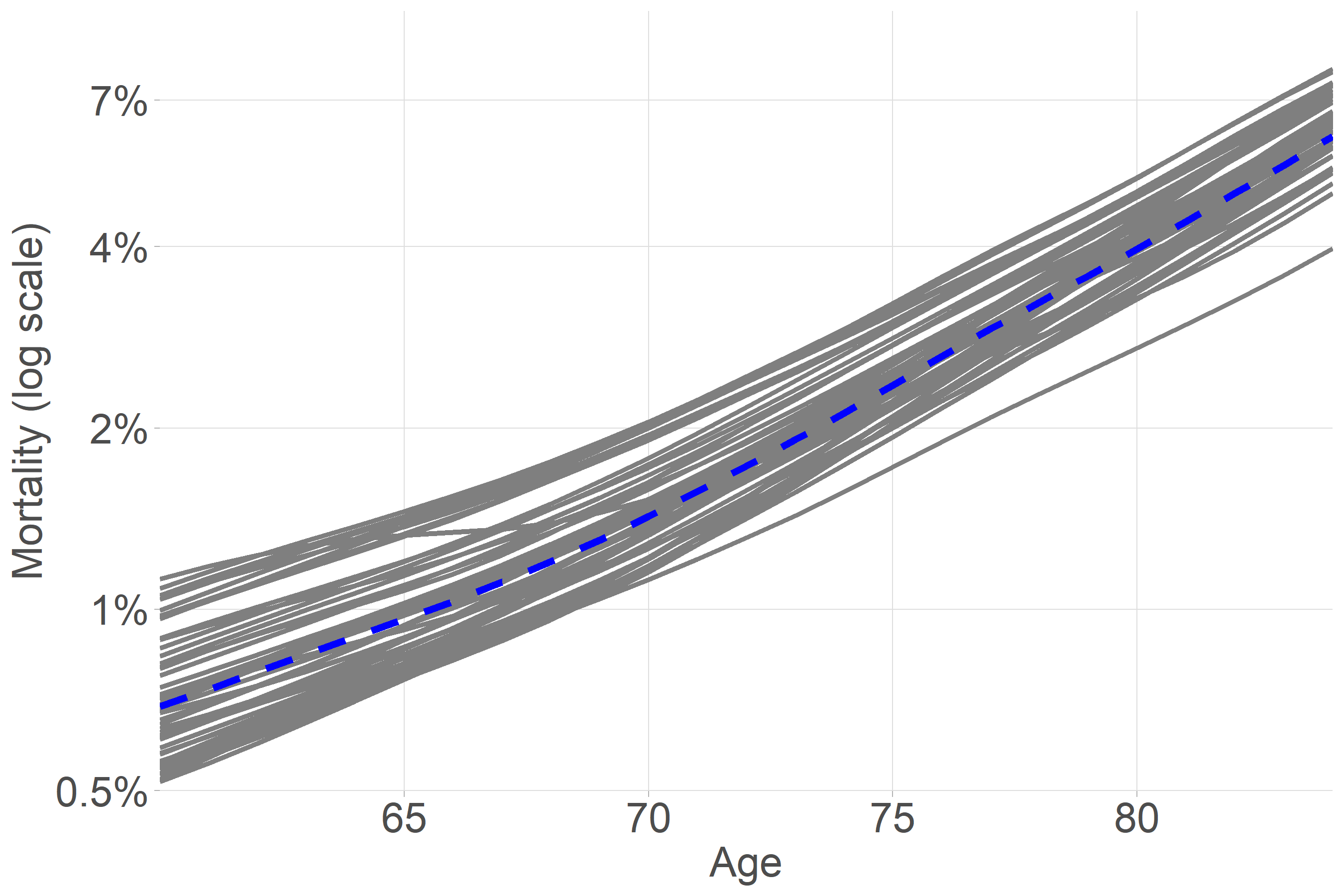} \\
  \hspace*{1.25in} Females
\endminipage
\caption{MOGP-PCA projected mortality rates (on the log scale) across Ages 60--85 for Year 2020. The dashed blue line shows the fitted SOGP model for the U.S.~nationwide dataset. \label{M-AGE-MALES-FEM-append}}
\end{figure}

To complement our discussion of the temporal pattern of mortality, we next discuss its structure in Age.  Figure \ref{M-AGE-MALES-FEM-append} presents the MOGP-smoothed age structure of mortality for Males and Females. The online RShiny app \cite{Shiny} shows an interactive version of this Figure for any user-specified Year. Given that mortality increases exponentially in Age, we plot log-mortality rates, which are roughly linear matching the prior mean function in \eqref{EQU:meanfunction}. We observe that in bulk the Age structure is very consistent across states, meaning that there is a high correlation between relative mortality experience at different Ages. Consequently, state-level mortality rankings are largely age invariant. For example, Connecticut Male mortality is \#1 lowest in the nation at Age 60, and is 4th, 5th and 4th lowest at Ages 65, 70, 75 (Females similarly rank 3rd, 3rd, 4th and 5th at those ages).  At the other end of the spectrum Mississippi has the highest Male mortality across \emph{all} ages and is in the worst-five for all ages for Females as well.  

\begin{figure}[ht] \centering
\minipage{0.45\textwidth}
  \includegraphics[width=\linewidth]{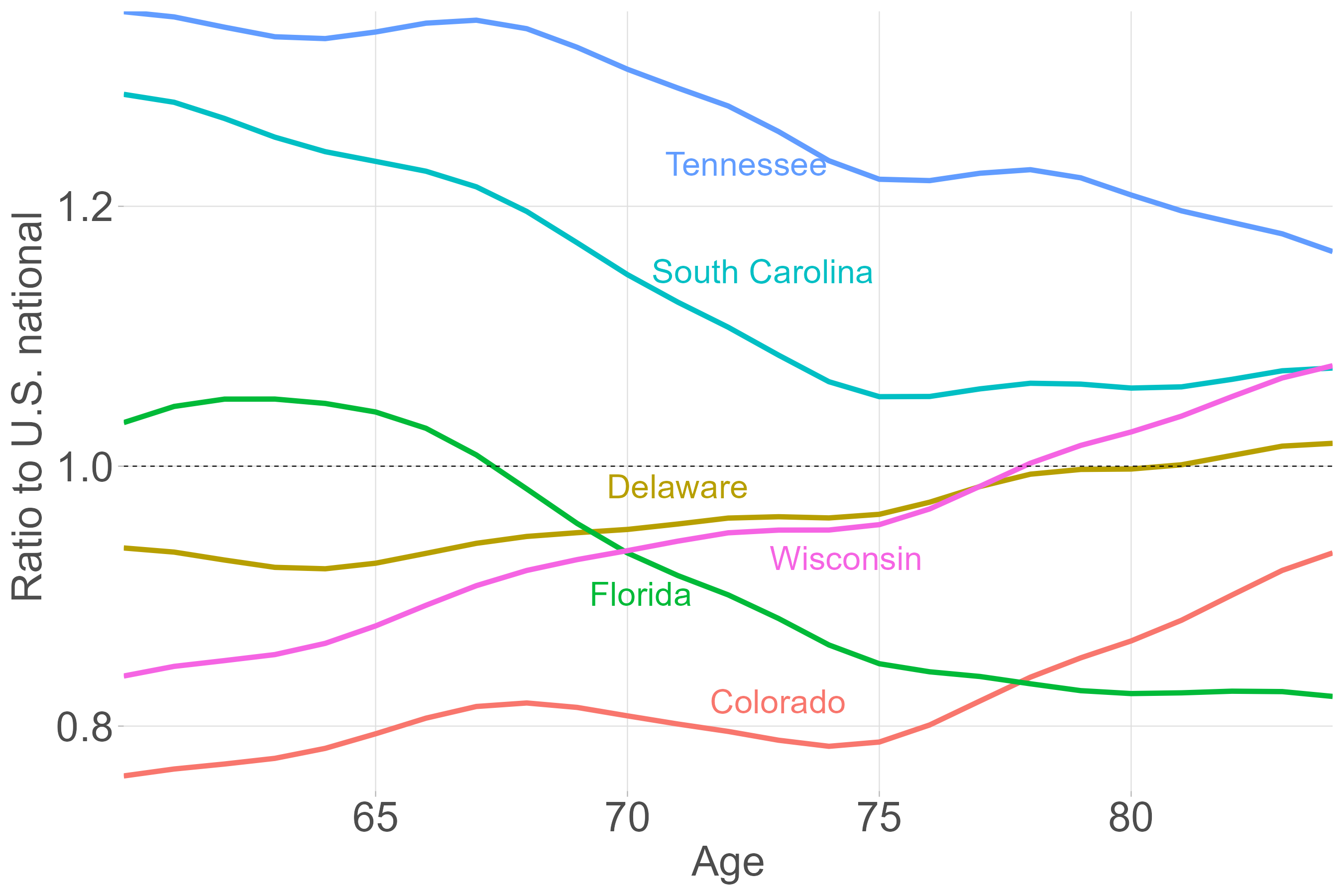} 
\endminipage\hspace*{0.3in}
\minipage{0.45\textwidth}
  \includegraphics[width=\linewidth]{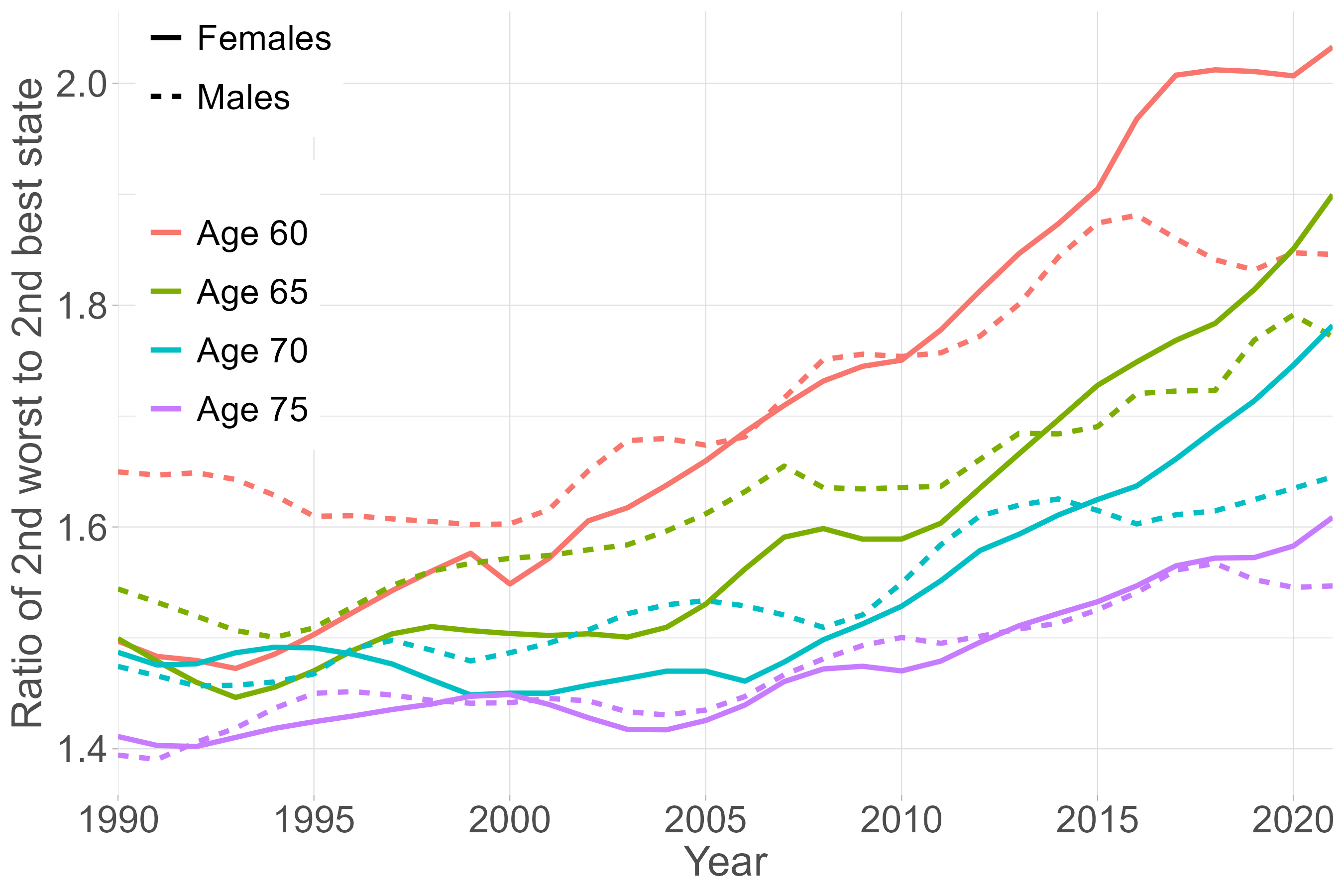} 
\endminipage
\caption{\emph{Left:} Mortality rates for Males in 2020 expressed as a ratio of U.S.~national average, as a function of Age for 6 representative states. \emph{Right:} Ratio of second-worst to second-best state mortality at four different Ages, as a function of Year. \label{fig:ratios}}
\end{figure}

Although the overall shape in Figure \ref{M-AGE-MALES-FEM-append} is cylindrical (implying a fixed spread between log-mortality of the best and worst state, as a function of Age), the individual behavior of states relative to the U.S.~average exhibits highly heterogeneous patterns. The right panel of Figure \ref{fig:ratios} (also available interactively in RShiny) shows those ratios for Males in 2018, for a selection of representative states. For some states, younger Ages are doing better than average, while older Ages are worse than average, see e.g.~Wisconsin where Male mortality at Age 60 is 84\% of national average, while mortality at Age 80 is 103\% of national average. 
For other states, younger Ages do relatively worse than old Ages; this is the pattern for Tenn.~and S.C.~in Figure~\ref{fig:ratios}. Male Tennesseans aged 60 have mortality that is 37\% above U.S.~average, while those aged 80 are only 21\% above average. Yet other states have no discernible trend: Delaware Males are within 8\% of U.S.~average rates across all ages.

In all, we observe distinct \emph{clusters} of state mortality age structure: the increasing pattern of Wisc.~is repeated for AK, ID, IA, ME, NE, NH, RI, UT, VT, VA; the decreasing pattern of S.C.~and Tenn.~is repeated in  AR, DC, KY, LA, MS, NV, OK, and WV. CA, CO, CT, MA, MN, MT, NJ, NY, ND, OR, SD, WA and WY have a largely Age-invariant negative gap to national average (doing better at all ages), while IN, KS, MO, NC, OH, PA and TX have a largely Age-invariant positive gap to national average (doing a bit worse at all Ages). Finally, there are a couple of idiosyncratic outliers like Florida, where younger ages are worse than average but older ages are much better than average. The above reveals a novel similarity metric among states. 

Moreover, we document a widening gap between mortality rates of best/worst states across time. Thus, the spread between states in Figure \ref{M-AGE-MALES-FEM-append} has been increasing over the past two decades, something already observed in Figure \ref{MR-years1990-2020-65} at Age 65. The right panel of Figure~\ref{fig:ratios} shows the ratio of mortality of the second best state vs.~the second worst state (we remove the best and the worst to stabilize this metric, akin to winsorizing) at four representative Ages and across years. As expected, the relative spread in mortality is lower for older ages because the underlying rates themselves increase in Age (so the absolute spread is in fact growing in Age). However, the main and perhaps unexpected pattern is that the dispersion dramatically increased since 2005 or so. In 2020, the Female mortality in the worst states at Age 60 is more than double compared to the best states, see the right edge of the right panel in Figure~\ref{fig:ratios}, while it used to be only 50\% thirty years ago. At Age 70 it is more than 75\% higher in 2020 compared to a 45\% gap in 2000. 

\section{Mortality Improvement}  \label{sec:IR}
 
In this section we study  \textsl{mortality improvement} (MI) factors, i.e.~the relative change in mortality rates across years. The most common metric is a year-over-year change  expressed in annualized percentage terms (such that lower mortality corresponds to a positive improvement), 
\begin{align}
    & MI_s \big( x_a, x_t\big) = 1- \exp \bigl( m_{*,s}(x_a, x_t)-m_*(x_a, x_{t-1}) \bigr)  \equiv \frac{e^{m_{*,s}(x_a,x_{t}-1)}-e^{m_{*,s}(x_a,x_t)}}{e^{m_{*,s}(x_a,x_{t}-1)}},
\end{align}
where the posterior mean $m_{*,s}(\cdot)$ is defined in \eqref{EQU:krigingmean}. More broadly, we denote by $MI$ the (either instantaneous or discrete) gradient of the mortality surface in the Year coordinate.

We find that the Mat\'ern-5/2 kernel in \eqref{EQU:kernel} is not ideal for MI analysis. As can be seen in Appendix~\ref{APPEND:matern-MI}, the respective surface tends to have high-level fluctuations (see spurious ``bands'' around age 70--75, and many ``speckles'' in the heatmap), which are not noticeable when looking at $m_{*,s}(\cdot)$ but become significant when considering MI. This feature can be understood by recalling that the smoothness of a GP as determined by the behavior of its kernel $C(x,x')$ as $x' \to x$. Mat\'ern-$\nu$ kernels yield  fits that are $\nu-1/2$ times differentiable, so  the $M52$ kernel in~\eqref{EQU:kernel} leads to a predictive surface that is exactly twice differentiable.  

To remedy this issue, we refit our MOGP models using the following Squared-Exponential-type kernel \eqref{eq:sqexp} in all coordinates (Age, Year, Cohort):
\begin{align}\label{eq:se-mogp}
   C_l(x^i,x^i_*) \coloneqq \eta^2 C^{(SqExp)}(x^i_a,x^i_{a,*}; \theta_{l,a}) \cdot C^{(SqExp)}(x^i_t,x^i_{t,*};  \theta_{l,t}) \cdot C^{(SqExp)}(x^i_c,x^i_{c,*};  \theta_{l,c})
\end{align}
where
\begin{align}\label{eq:sqexp}
    C^{(SqExp)}(x,x'; \theta) \coloneqq \exp\Big\{-\frac{(x-x')^2}{2\theta^2}\Big\}.
\end{align}
The SqExp kernel yields infinitely differentiable fitted mortality surfaces, which are thus smoother temporally and more interpretable.

\begin{remark}
The conceptual aim of MI is to understand the Year trend in mortality. By construction, the GP applies smoothing by removing  noise that is \emph{idiosyncratic} for each observation. In practice, mortality experience is subject to significant ``common shocks'', such as heat waves or epidemics that might cause high mortality in one year, but not the next. Consequently, observations at different Ages and same Year are correlated, affecting the GP projection $m_{*}$. Our ``over-smoothing'' through the SqExp kernel matches the actuarial task of detecting the time structure of mortality after removing short-term temporal fluctuations. 
\end{remark}

\begin{figure}[H]
\minipage{0.48\textwidth}
  \includegraphics[width=\linewidth,trim=0.3in 0.15in 0.5in 0.25in]{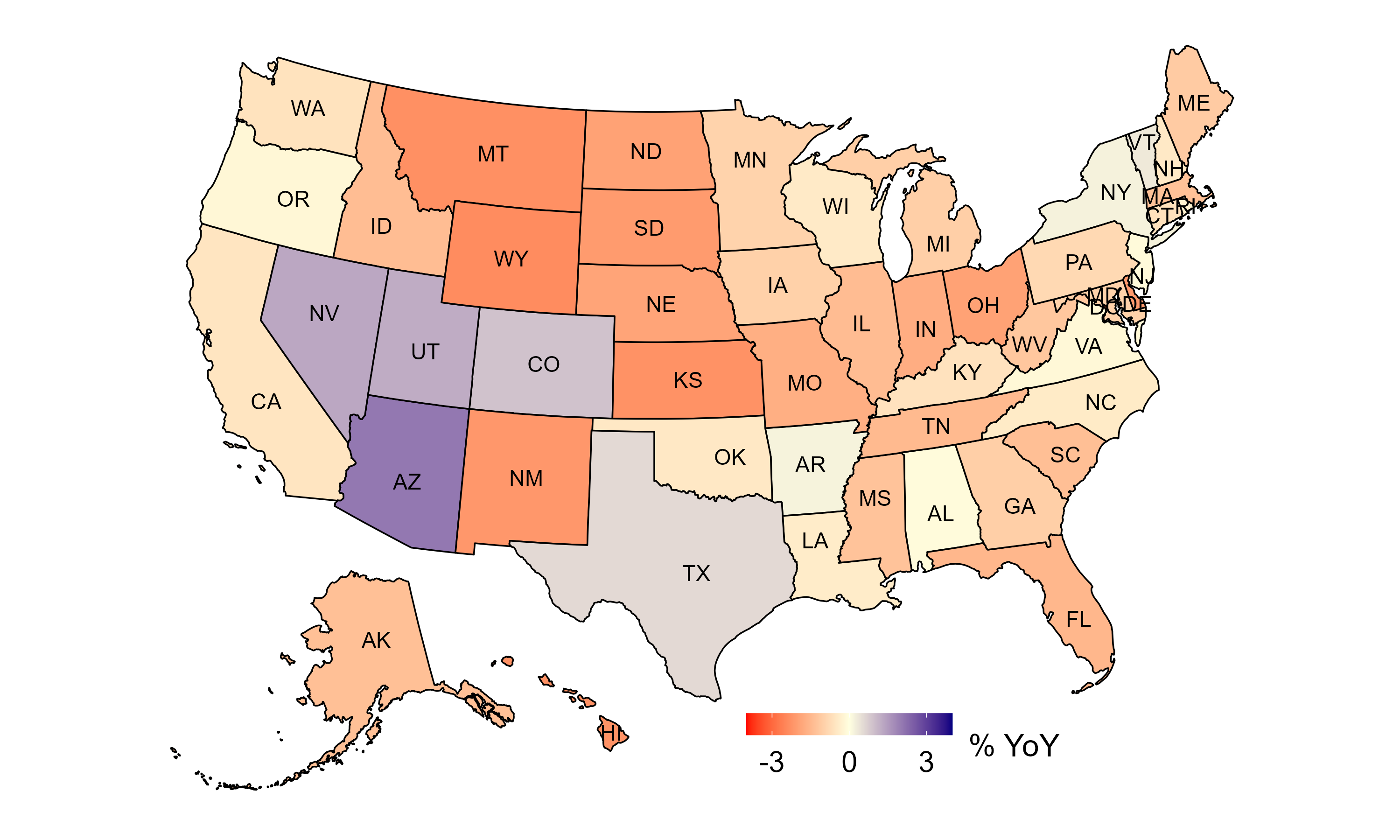} \\
  \hspace*{1in} Males Age 65
\endminipage
\minipage{0.48\textwidth}
  \includegraphics[width=\linewidth,trim=0.3in 0.15in 0.5in 0.25in]{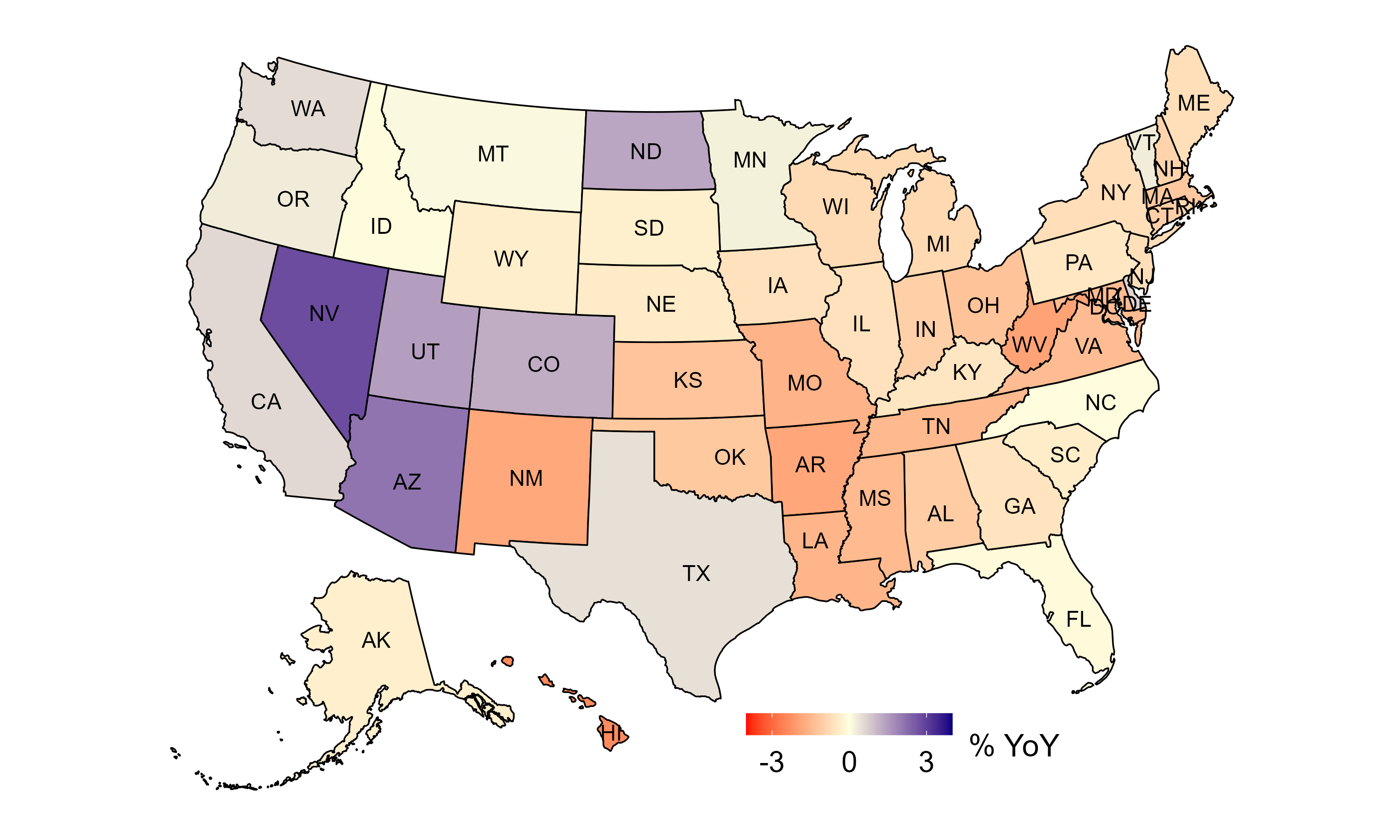} \\
  \hspace*{1in} Females Age 65
\endminipage
\caption{MOGP-based annualized mortality improvement factors in 2020 and  Age 65.}  \label{FIG:IR-65}
\end{figure}

Figure \ref{FIG:IR-65} visualizes the model-based MI's at Age 65 and Year 2020 across the 51 states. Recall that the training USMDB data is up to 2018, so that we show the MOGP-PCA forecast for the MI trend two years into the future, on the eve of the pandemic. The results display a lot of heterogeneity in state-level MI's, with some states experiencing improvements  (blue shades) and others stagnation (gray) or deterioration (negative improvement rate, orange). In general, MI for Females is higher compared to Males who in most states are experiencing a deterioration in their mortality. At Age 65 (resp.~75) Females have higher MI than Males in 32 (resp.~37) out of 51 states. This implies a widening gap in Male-Female mortality.
Notably, the statistical significance of estimated MIs is not high; the MOGP provides standard errors of about $\pm 1\%$, so for many states and Ages it is impossible to conclusively decide whether its MI is positive or negative, see right panel of Figure~\ref{fig:mi-scatter} in Appendix~\ref{APPEND:matern-MI}.  At the 95\% posterior significance level, only AZ, NV and UT  definitely have positive MI for Females (and only AZ for Males) at Age 65, while Females in 9 states (and Males in 14 states) have conclusively negative MI at Age 65.


The  regional patterns in Figure \ref{FIG:IR-65} support the hypothesis that  populations which share similar economic, geographic, and demographic characteristics have also similar mortality \emph{trends}. For example, the Sun Belt states tend to experience a positive MI; most of the South experiences deteriorating mortality. There are several  ``outlier'' states in the Southwest: Nevada and Arizona have exceptionally positive mortality improvement, while New Mexico has negative MI across both genders. Table~\ref{TABLE:best&worstIR} summarizes the top- and bottom-5 states in terms of MI at $x_*=(65, 2020)$ across genders. 
Comparing with the previous section, the MI pattern implies that the Sun Belt is catching up to the lower-mortality Northeast states; at the same time, Midwest and Southern States are falling behind, amplifying the national discrepancies. This can be linked to the causal analysis in  Chetty et al.~\cite{chetty2016association} who 
report that (based on raw data and aggregating all working ages) ``Hawaii, Maine, and Massachusetts had the largest gains in LE [between 2001 and 2014] (gaining $>0.19$ years annually) when men and women in the bottom income quartile were averaged. The states in which low-income individuals experienced the largest losses in LE (losing $>0.09$ years annually) were Alaska, Iowa, and Wyoming,'' and an earlier similar conclusion for changes in LE between 1983 and 1999 in \cite{ezzati2008reversal}.

\begin{table}[h]
    \centering
\begin{tabular}{llll}
\multicolumn{4}{c}{Best MI} \\
\cline{1-4}
\multicolumn{2}{c}{Males}    & \multicolumn{2}{c}{Females}  \\
Age 65 & Age 75 & Age 65 & Age 75 \\
\hline
Ariz. & Vt.      & Nev.   &   Colo. \\
Nev. & Utah       & Ariz.    &  Nev.      \\
Utah & Colo.      & Utah   &  Alaska    \\
Colo. & Ky.    &  N.Dak.   &   Or. \\
Texas & Ariz.  & Colo.   &   Ariz.   \\
\hline
\end{tabular}
\hspace{1cm}
\begin{tabular}{llll}
\multicolumn{4}{c}{Worst MI} \\
\cline{1-4}    

\multicolumn{2}{c}{Males}    & \multicolumn{2}{c}{Females}  \\
Age 65 & Age 75 & Age 65 & Age 75 \\ \hline
D.C.    & D.C.  &  D.C  & D.C.\\
Wyo. & N.Mex.       & Hawaii & S.Dak.          \\
Mont. & Hawaii     & W.Va.    & Hawaii    \\
Kan.  & Md.    &  Ark.    & N.Mex.  \\
Hawaii &  Va. & N.Mex.     & Kan.     \\
\hline
\end{tabular}
    \caption{Top-5 and bottom-5 states in terms of MOGP PCA-SqExp based projected annual mortality improvement at Age 65 in Year 2020. The best and worst states are in the first row.}
    \label{TABLE:best&worstIR}
\end{table}

\begin{figure}[htb] \hspace*{0.22in}
\minipage{0.42\textwidth}
  \includegraphics[trim=0.75in 0.2in 0.75in 0in,width=\linewidth]{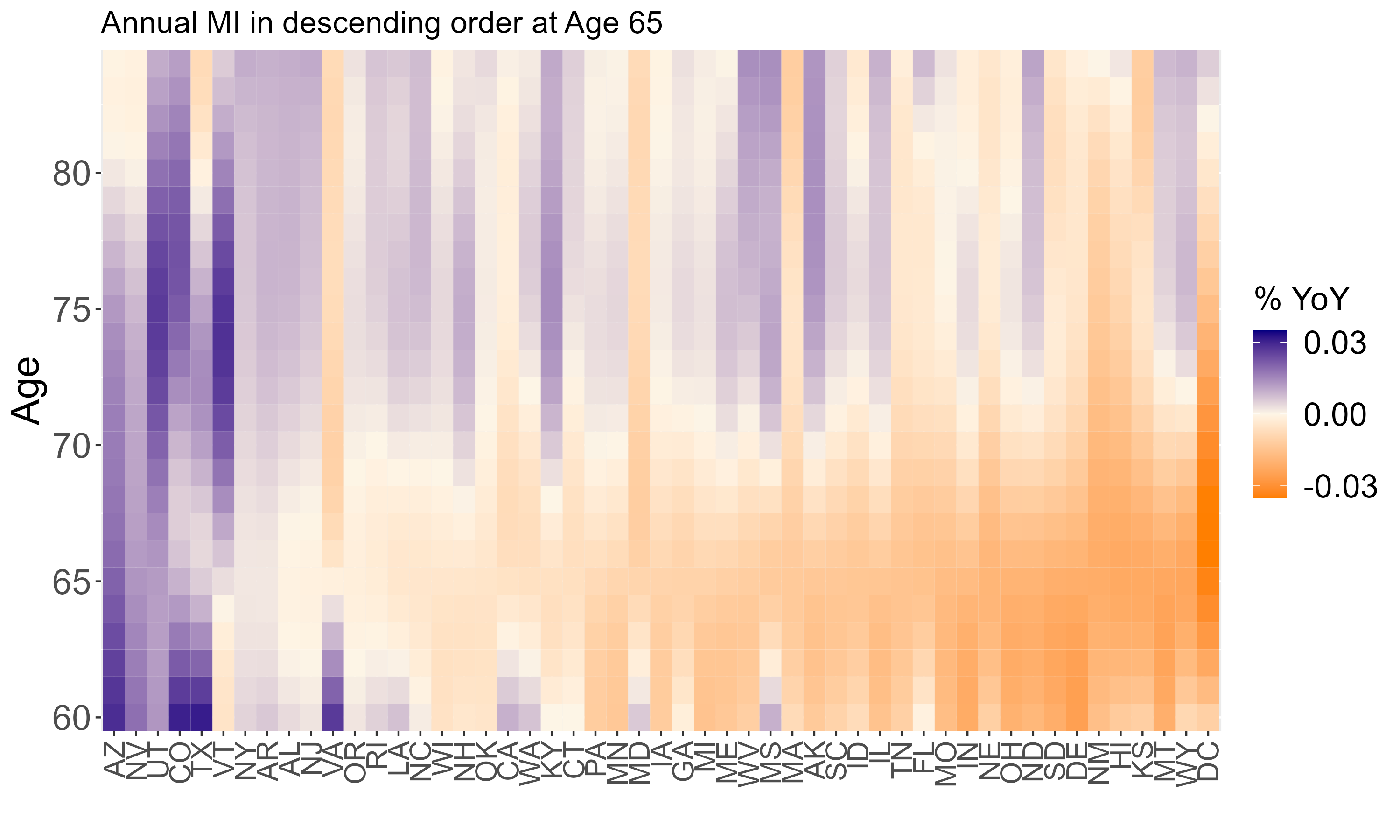} \\
  \hspace*{1in} Males
\endminipage\hspace*{0.5in}
\minipage{0.42\textwidth}
  \includegraphics[trim=0.75in 0.2in 0.75in 0in,width=\linewidth]{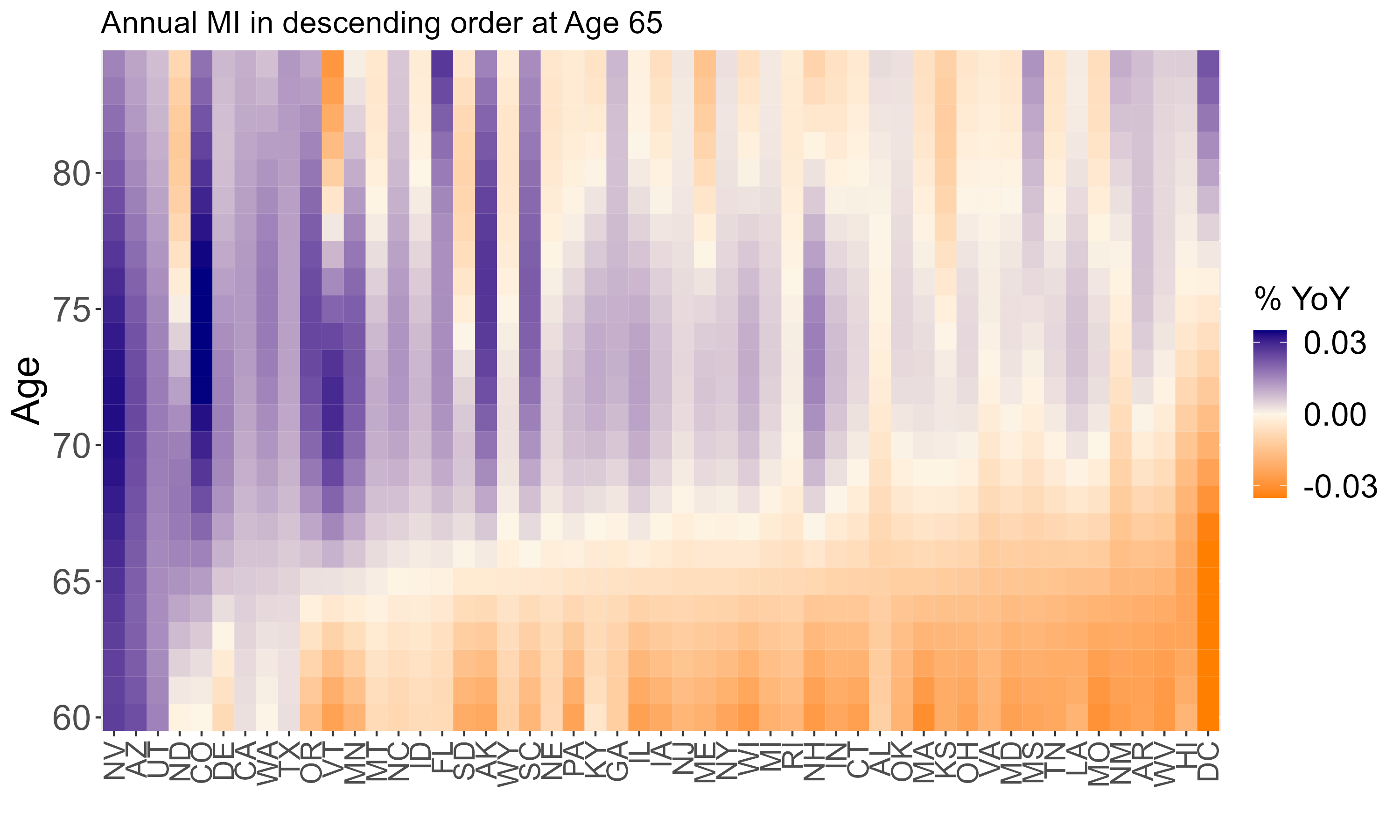} \\
  \hspace*{1in} Females
\endminipage
\caption{MOGP-generated Mortality Improvement rates in 2020. States are sorted by MI at Age 65.} \label{FIG:65-ir-heatmap}
\end{figure}

Next we investigate the \textbf{age structure} of annualized mortality improvement rates. Figure~\ref{FIG:65-ir-heatmap} displays MI
for Males and Females between ages 60--84, sorted according to MI at Age 65.  
Looking across Ages, we see a mix of improvement/deterioration for nearly all columns, exceptions being Females in the Southwest and Pacific (UT, NV, AZ, CA, WA, TX) which display improvement for all ages, as well as a few West North Central states (KS, NE, SD, also TN and MA)  where Males have negative MI for all ages. 

The most common pattern in Figure~\ref{FIG:65-ir-heatmap} is deterioration of mortality (orange gradient) for the younger ages and positive MI (blue gradient) for the elderly. This means that the mortality curves are flattening in Age, and is likely explained by the recent medical advancements that tend to target older individuals, continuing to lower their mortality. A second explanation could be generational, reflecting better cumulative health of the older baby boomers (who are 65-75 in 2020) compared to their younger counterparts. For Females, we also often observe a convex shape, with Ages $\le 67$ and $\ge 80$ deteriorating, and Ages around 70--75 improving. 

Some states exhibit idiosyncratic age structure of MI. For example,  Males in Ariz., Nev.~, Va.~and Texas are experiencing strong mortality improvement at younger ages, and mortality deterioration at older ages, the opposite pattern to above. That could reflect inter-state migration patterns between the working-age and retiree populations. In a few states (OR, OK, WI, RI, CT Males, NE Females), the MI factors are all very close to zero, indicating static mortality experience. 

As mentioned, Females tend to display a higher mortality improvement (across most states) than their Male counterparts,
though the precise MI patterns often differ significantly between genders. For instance, at Age 65, only in 8 states is Male MI positive, while $MI > 0$ for Females in 13 states. This difference is even more stark at Age 70 where 39 states have positive Female MI (of which 9 are positive with more than 95\% posterior probability), but only 20 states (and just 2 with 95\%+ credibility) have positive Male MI. The left panel of Figure~\ref{fig:mi-scatter} in the Appendix (see also the RShiny dashboard) visualizes Male vs Female MIs. The respective right panel compares Male and Female MIs at Age 65, furthermore showing the respective posterior 90\% credible intervals.

\section{Explanatory Covariates} \label{SUBSEC:explorcov}

Next, we document the relationship between our predicted mortality rates/improvement factors and state covariates from Section \ref{SUBSEC:covariates}. The objective is to gain further insights into the drivers of mortality levels and trends via correlation to state characteristics. To this end, we compare 22 variables (18 covariates, 3 PC factors, and life expectancy at birth from \cite{cdc}) against three different MOGP-PCA model outputs: (a) 2020 mortality predictions, (b) 2020 MI factors, (c) 2020 vs 2010 improvement ratio. We use the SqExp kernel \eqref{eq:se-mogp} for computing improvement factors and the M52 kernel \eqref{EQU:kernel} for mortality rates.

\begin{figure}[H] 
  \centering
  \begin{tabular}{lccr}
      \includegraphics[width=0.24\textwidth,trim=0.3in 0in 0in 0in,height=1.17in]{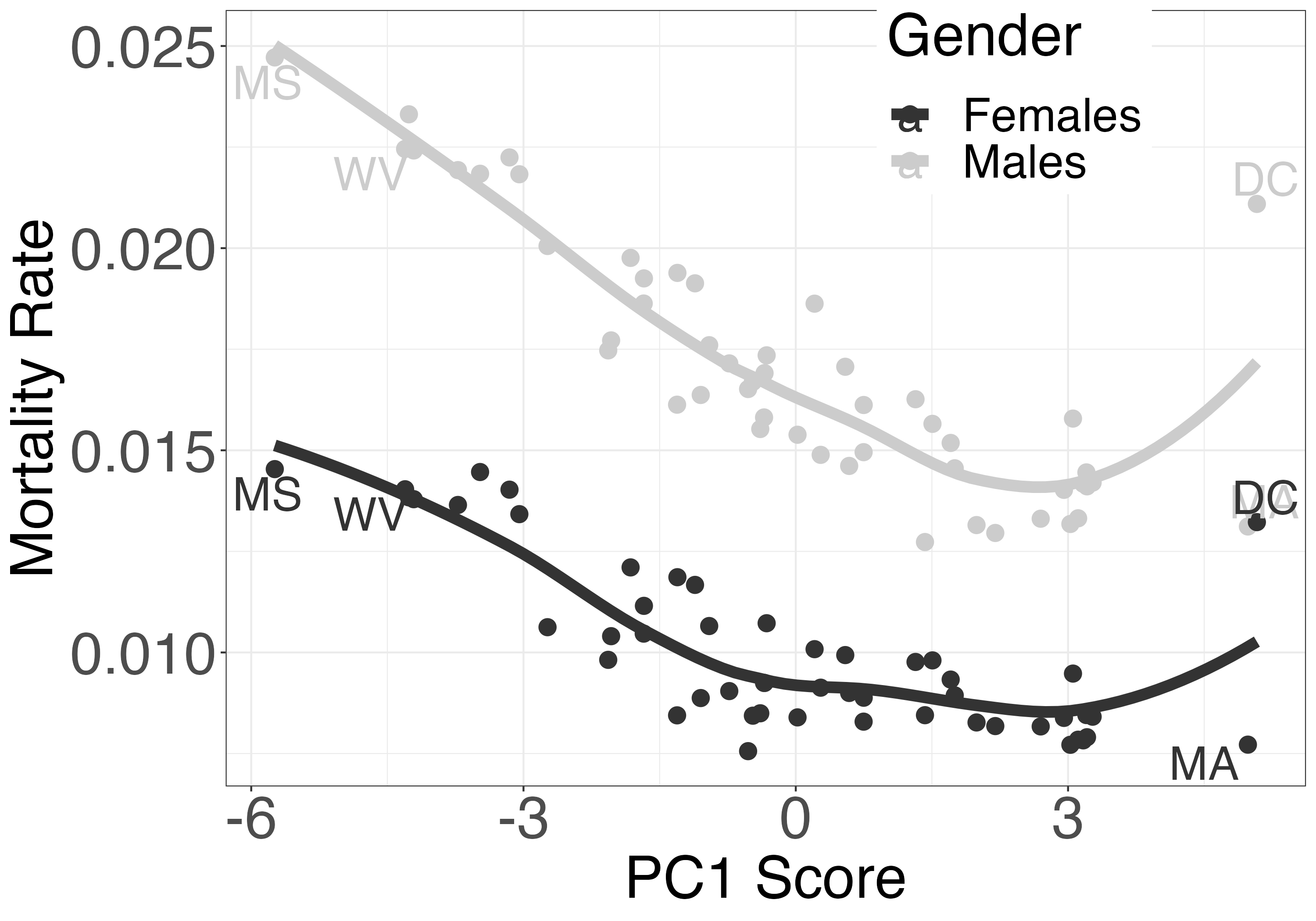} & 
      \includegraphics[width=0.225\textwidth,trim=1.64in 0in 0in 0in,clip]{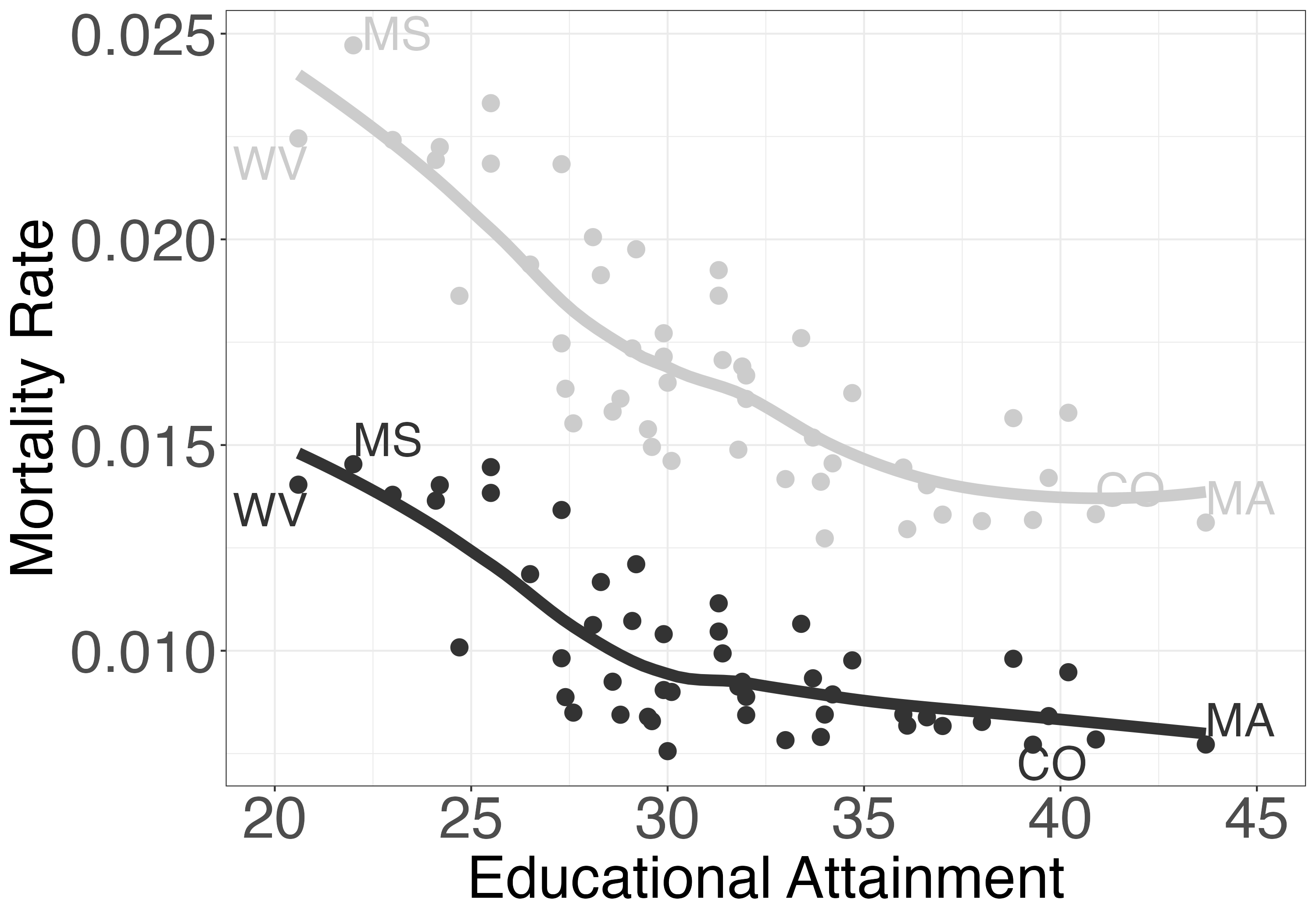} & 
      \includegraphics[width=0.225\textwidth,trim=1.64in 0in 0in 0in,clip]{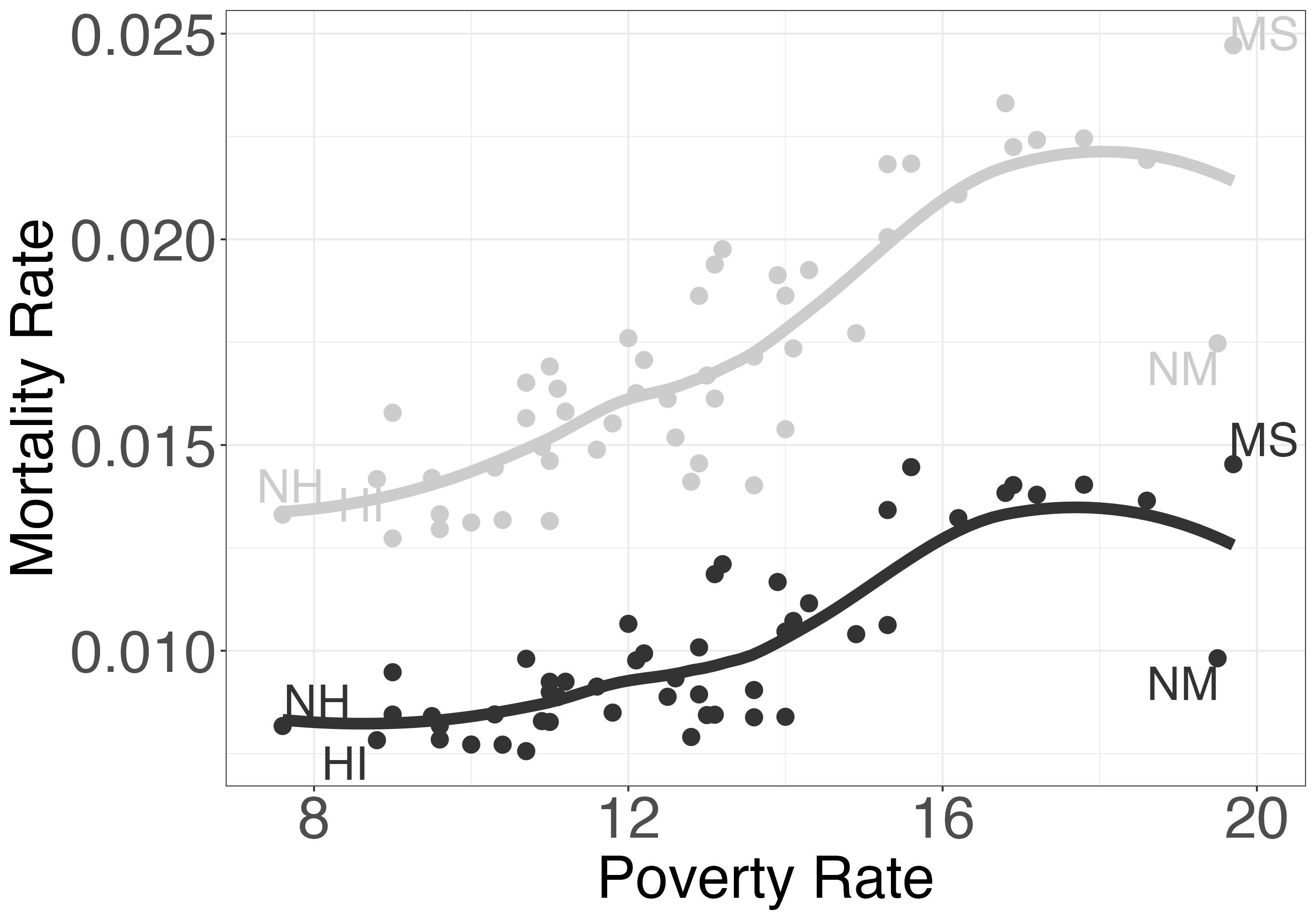} & 
      \includegraphics[width=0.225\textwidth,trim=1.64in 0in 0in 0in, clip]{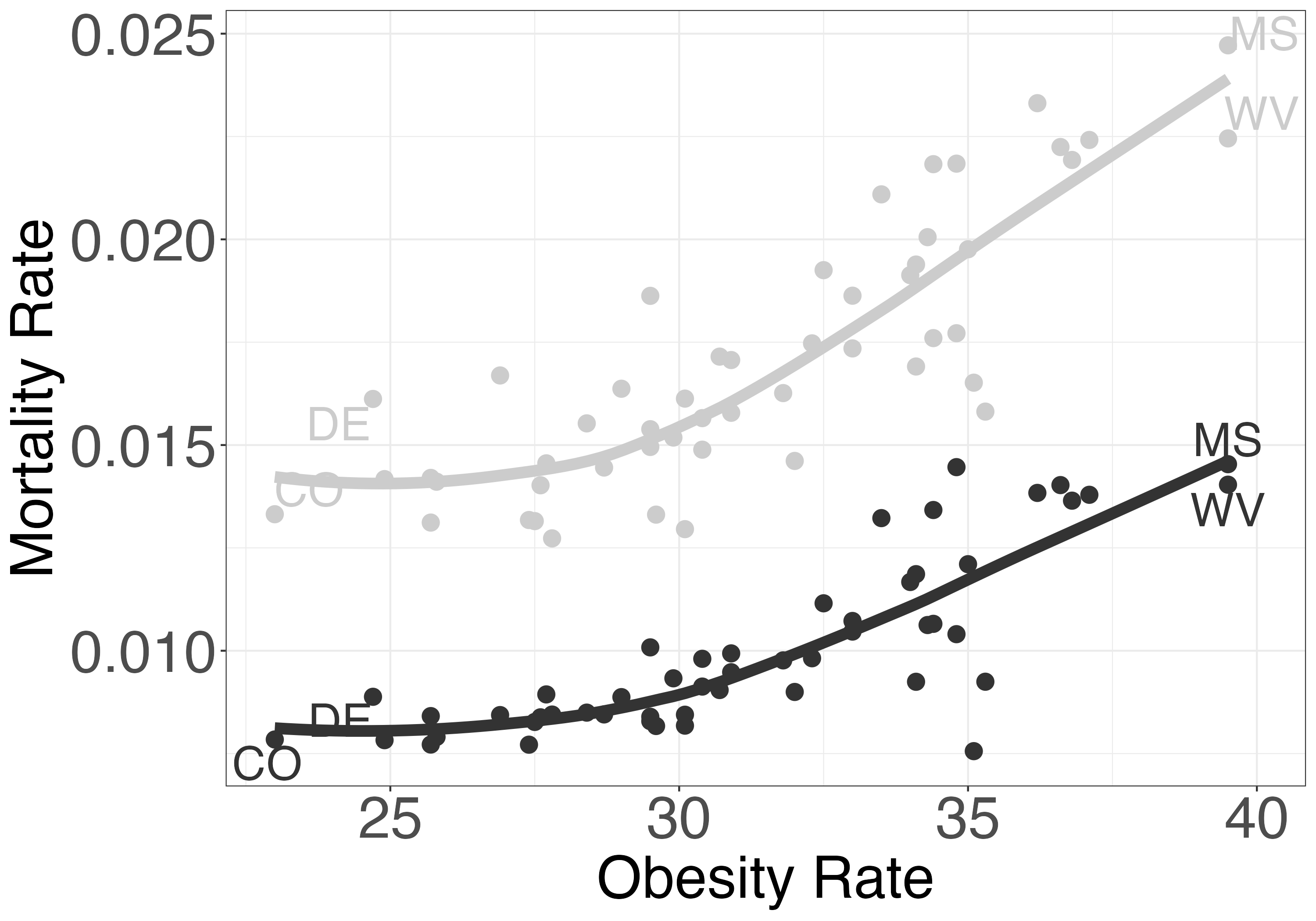} \\
      $\qquad\qquad$ (a) PC1 & (b) EA & (c) PR & (d) OR $\qquad\qquad$ 
      \end{tabular}
   \caption{MOGP-PCA smoothed mortality rates at Age 65 in Year 2018 against four selected state covariates.}
  \label{FIG:mortcovandpc1} 
\end{figure}

\begin{figure}[ht]
  \centering
  
  \begin{subfigure}[b]{0.32\textwidth}
    \includegraphics[width=\textwidth]{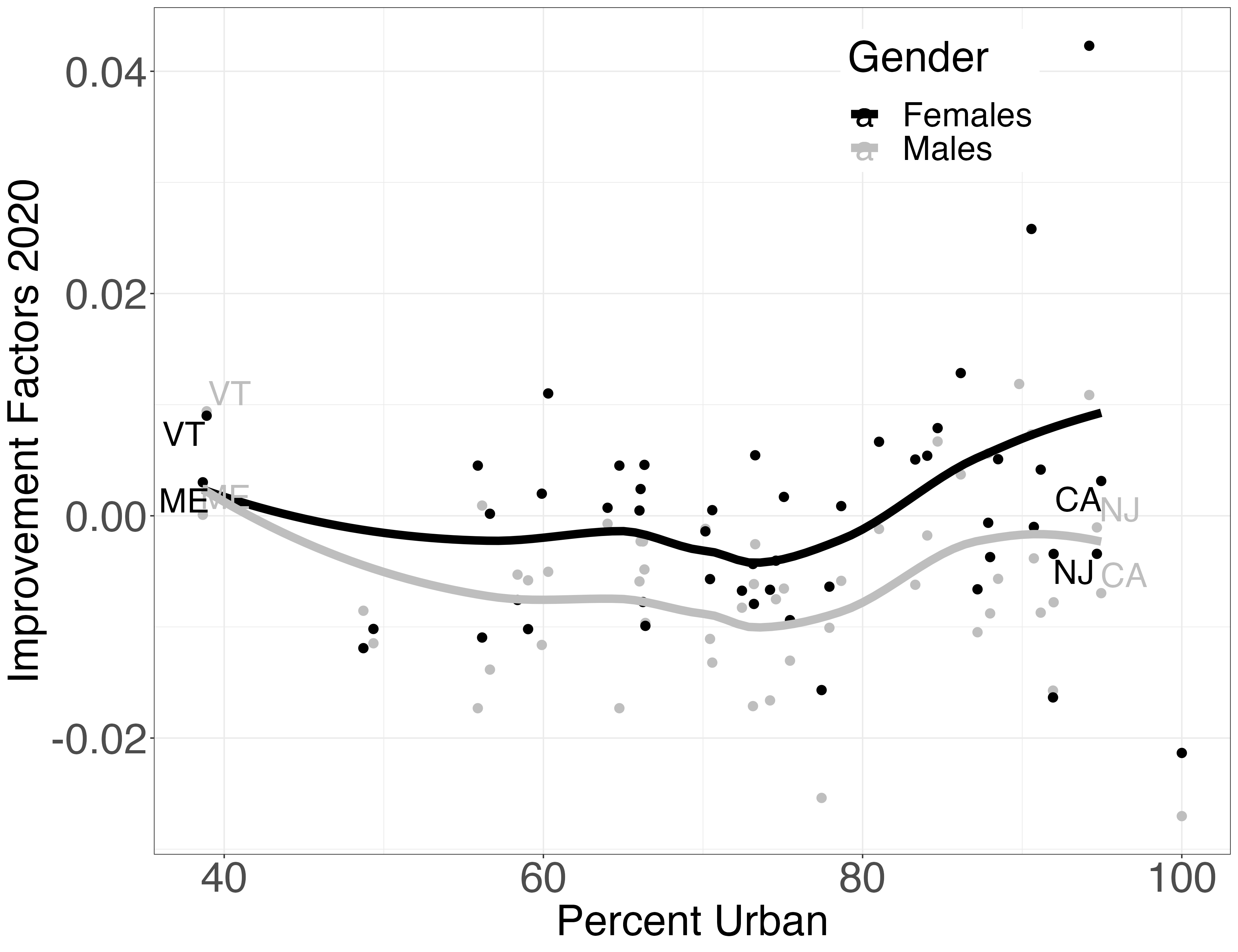}
    \caption{Urbanization Percentage}
  \end{subfigure}
  \hfill
  \begin{subfigure}[b]{0.32\textwidth}
    \includegraphics[width=\textwidth]{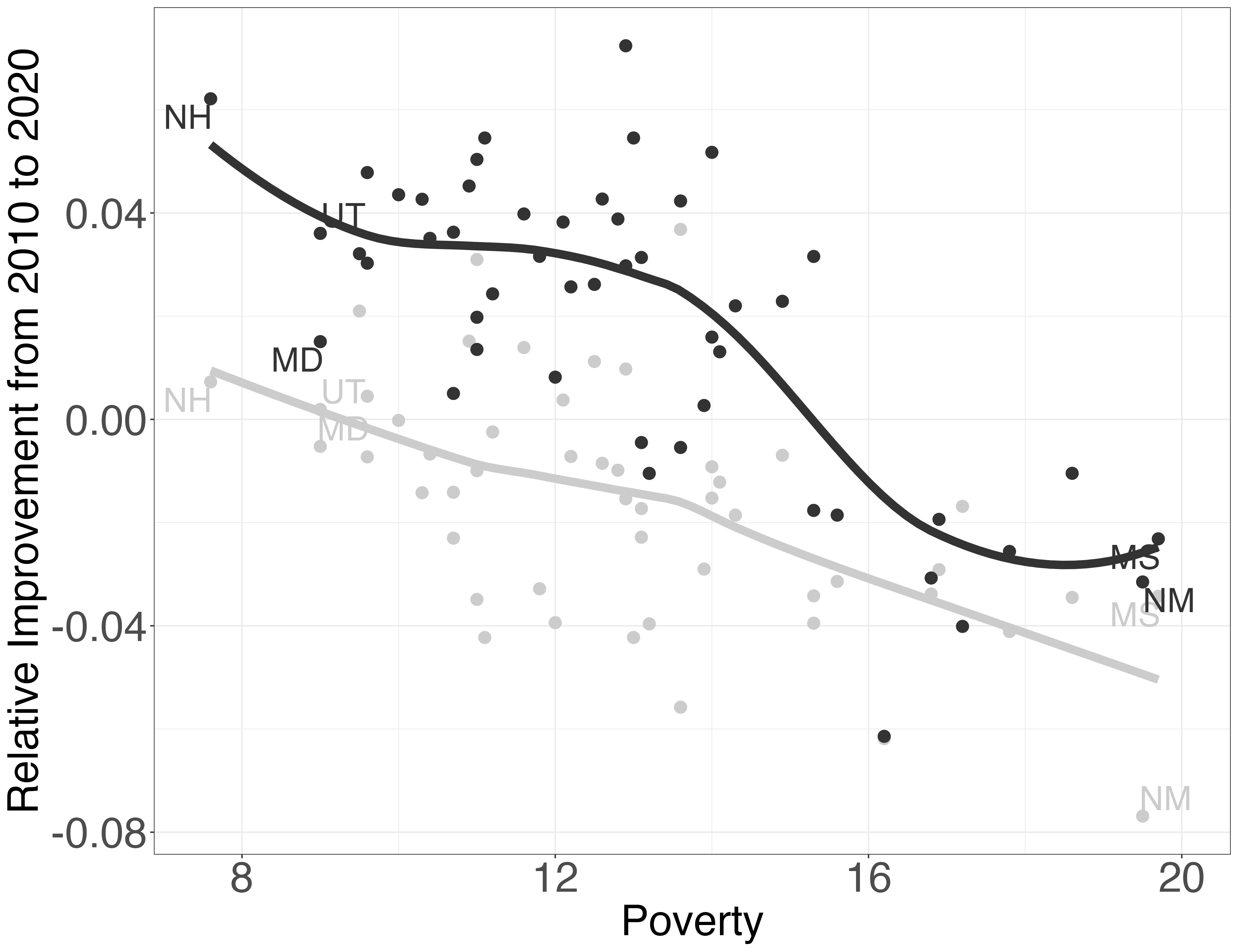}
    \caption{Poverty Rate}
    \label{fig:subfig3}
  \end{subfigure}
  \hfill
    \begin{subfigure}[b]{0.32\textwidth}
    \includegraphics[width=\textwidth]{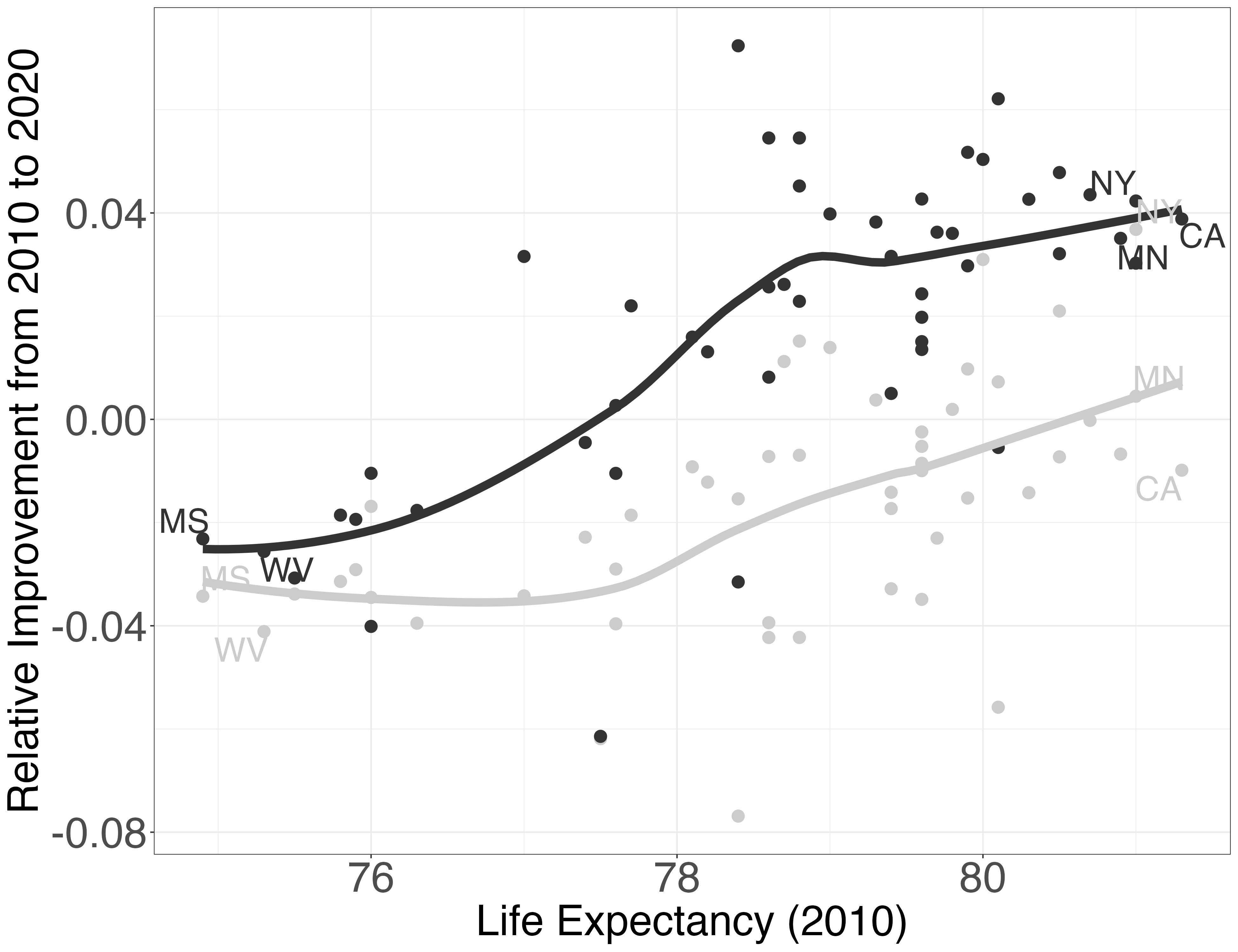}
    \caption{Life Expectancy in 2010}
  \end{subfigure}

  \caption{MOGP PCA-SqExp MI factors at Age 65 in Year 2018 against three selected state covariates.}
  \label{PIC:improvfacagnstcov}
\end{figure}

Figure \ref{FIG:mortcovandpc1} displays the relationship between predicted mortality rates in 2020 and several covariates, namely educational attainment, poverty rate, obesity rate, and PC1 scores.  To visualize the nonlinear dependence, we include a regression curve estimated using LOESS regression. The latter excludes several outliers at the extreme left/right of the plots: D.C.~often has markedly different covariates (due to it being a 100\% urban region), see e.g.~Figures \ref{FIG:mortcovandpc1}a and \ref{PIC:improvfacagnstcov}a, and so does Mississippi (Figs.~\ref{FIG:mortcovandpc1}b,\ref{FIG:mortcovandpc1}c  and \ref{PIC:improvfacagnstcov}b). 

First, we observe that the smoothed curves are essentially parallel for the Male and Female populations, implying that the impact of different state covariates (which are shared across genders) is very similar for both genders. Second, we observe a strong correlation between economic variables and mortality rates. Recall, as suggested in Section \ref{SUBSEC:covariates}, that the states' PC1 factor loadings are correlated with economic prosperity, with a Pearson correlation coefficient of $-0.75$. Therefore, the relationship observed in Figure \ref{FIG:mortcovandpc1}a implies that mortality levels are positively correlated with the state's economic well-being, wealthier states having lower mortality. This matches the finding in Chetty et al.~\cite{chetty2016association} regarding the strong pattern between economic prosperity and longevity, see also the claims made in  \cite{harris2021high,risinginequLEsocio, risinggeodis} (who all fit linear relationships to age-aggregated data).

Additional related insights are in panels (b)-(d) of Figure \ref{FIG:mortcovandpc1}. Figure \ref{FIG:mortcovandpc1}b shows that U.S.~states with higher educational attainment tend to experience lower mortality; Figures \ref{FIG:mortcovandpc1}c-d show that higher state-wide poverty rates /higher obesity rates are associated with higher mortality. Of note, the above patterns are often \emph{nonlinear}. For example, the positive association between state obesity rate and mortality is substantially weakened for states with obesity rate below 30\% (left edge of Figure~\ref{FIG:mortcovandpc1}d). Likewise, lower mortality associated with higher education no longer follows once over 35\% of the state's population has Bachelor's degrees (right edge of Figure~\ref{FIG:mortcovandpc1}b). These facts can be interpreted as one-sided risk drivers: states that have a lot of obese individuals (or few college-educated individuals) suffer higher mortality, but having ``exceptionally'' non-obese or highly educated populace is not associated with lower mortality.

Next, we analyze the relationships between our 22 covariates and the annual improvement factors in 2020 along with the relative improvement in mortality rates during the decade of 2010s. We find that for the most part, there is little significant correlation between state characteristics and their latest MI. Nevertheless, Figure~\ref{PIC:improvfacagnstcov} displays several interesting patterns. We observe a positive relationship between MI and urbanization, and a negative relationship between MI and poverty (panels $a$ and $b$). This implies that rural and poorer states tend to exhibit worse mortality trends, also observed in \cite{risinggeodis}. Furthermore, Figure \ref{PIC:improvfacagnstcov}$c$  shows that decadal MIs are positively associated with a 
 higher life expectancy. This reinforces our earlier discussion on divergence: states that do well (high LE) keep improving, and states that do poorly (low LE) are deteriorating. As a result, the gap between ``best'' and ``worst'' performing states {increased} during the final decade, see Section \ref{sec:analysis}. We note that none of the bottom-7 states by LE experience a positive MI in the 2010s.

\section{Conclusion}\label{sec:conclude}

In this paper we have developed MOGP models geared for an actuarial analysis of Age- and Year-specific mortality rates across the 51 U.S.~states. Our work complements existing Age-aggregated studies in the economics literature, but goes much further through using the developed statistical framework to study Age-specific trends. In particular, the MOGP framework allows to analyse smoothed state-wide Mortality Improvement factors and the various Age-structures.

At a basic level, we confirm the well-known and well-documented  features of U.S.~mortality in the late 2010s, such as deterioration of mortality at pre-retirement ages, and the vast gaps between states based on economic and health characteristics. We also observe the familiar geographic patterns for Appalachia, Southwest, the Deep South, etc.

Going deeper, the oft-repeated heterogeneity manifests itself through two different channels. First, it leads to a lot of relative moves, as states move up and down the rankings. These shifts are highly Age-dependent. Second, the Age structures can exhibit very different behaviors, showing the limitations of using a single summary statistic like Life Expectancy to explain state differences. In some states younger Ages are doing better, in others those in the 70s, in yet others, the oldest. 

An important insight of our exploratory analysis is the growing \emph{divergence} across states. We document that the gaps between best and worst states are getting wider (Fig.~\ref{fig:ratios}), and that many (but not all) of the ``laggard'' states are falling further behind (due to below-average MI) while states with lowest mortality are often pulling even more ahead (Fig.~\ref{fig:MR-BC-65}). One exception is the Southwest that is catching up and moving up the ranks. We also record the divergence between Male and Female mortality trends.

Our analysis also shows that the U.S.~can be broken up into clusters of states that share similar patterns both in mortality rates and in respective trends (MI). These clusters are loosely geographical, and include ``Appalachia'' (TN, KY, SC, AR), ``Deep South'' (AL, MS, WV, LA), ``Pacific'' (CA, OR, CO), ``Southwest'' (UT, AR, NV), ``Upper Plains'' (MN, SD, ND), ``Midwest'' (PA, OH, IN, MO, KS), ``New England'' (ME, VT, NH, RI), ``North East'' (MA, CT, NJ, NY). These observations validate our proposed strategy of grouping into 3--5 similar \emph{and} neighboring states.  We also identify states with truly idiosyncratic patterns, including D.C., Florida, New Mexico, Hawaii and Texas.

To conclude, let us outline two avenues for future research. First, one may consider additional spatial scales.  Looking at county-level data can help to understand further intra-state patterns, especially in large states like California or New York, where there is a lot of intra-state heterogeneity. Looking at metropolitan area data can help to isolate urban/rural effects. Analysis across these different spatial units can better tease out the impact of `place' in terms of different mortality drivers. For example, health policies (such as ACA rules and smoking regulations) are set by state, while demographics vary more in terms of urban and rural locales.

Second, more in-depth analysis is warranted about Age-linked drivers of U.S.~mortality. This includes the role of income, racial characteristics, education and especially health factors. One approach would be to merge our analysis with cause-of-death data, linking to our earlier work in \cite{huynh2021joint}
and isolating cause-specific state differences. For example, it would offer a direct window to discuss \emph{deaths of despair} effects, regional cancer patterns, or cardiovascular trends which in turn correlate to wealth and urban/rural discrepancies.  Another approach would be along the lines of Hartman et al.~\cite{countyUS}, who applied spatial GLM methodology.

\appendix

\section{Appendix}
\renewcommand\thefigure{\thesection.\arabic{figure}}
\setcounter{figure}{0} 
\renewcommand\thetable{\thesection.\arabic{table}}
\setcounter{table}{0} 

\subsection{Geographic Regions} \label{APPEND:geogroup}

We use the following U.S.~divisions \cite{census_geo} when creating geographical groups for the MOGP model. 
\begin{enumerate}
    \item New England (6 states): CT, ME, MA, NH, RI, VT;
    \item Mid-Atlantic (3): NJ, NY, PA;
    \item East North Central (5): IN, IL, MI, OH, WI;
    \item West North Central (7): IA, KS, MN, MO, NE, ND, SD;
    \item South Atlantic (9): DE, DC, FL, GA, MD, NC, SC, VA, WV;
    \item East South Central (4): AL, KY, MS, TN;
    \item West South Central (4): AR, LA, OK, TX;
    \item Mountain (8): AZ, CO, ID, MT, NM, NV, UT, WY;
    \item Pacific (5): AK, CA, HI, OR, WA
\end{enumerate}

\subsection{Impact of State Groupings}\label{APPEND:com}

\begin{figure}[H]
\centering
\begin{subfigure}{.45\textwidth}
  \centering
  \includegraphics[width=1\linewidth]{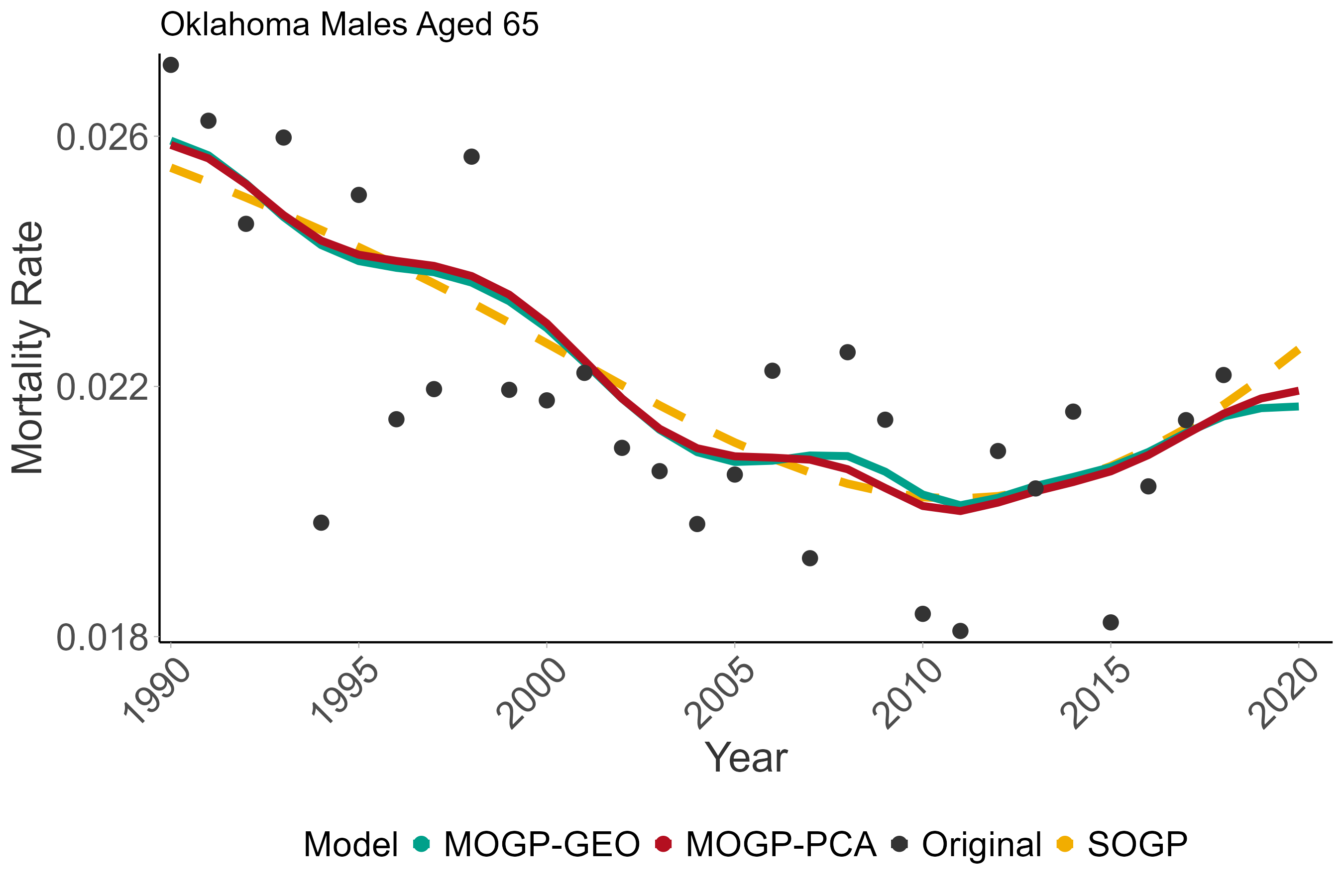}
  \caption{Oklahoma Males Aged 65}
\end{subfigure}
\begin{subfigure}{.45\textwidth}
  \centering
  \includegraphics[width=1\linewidth]{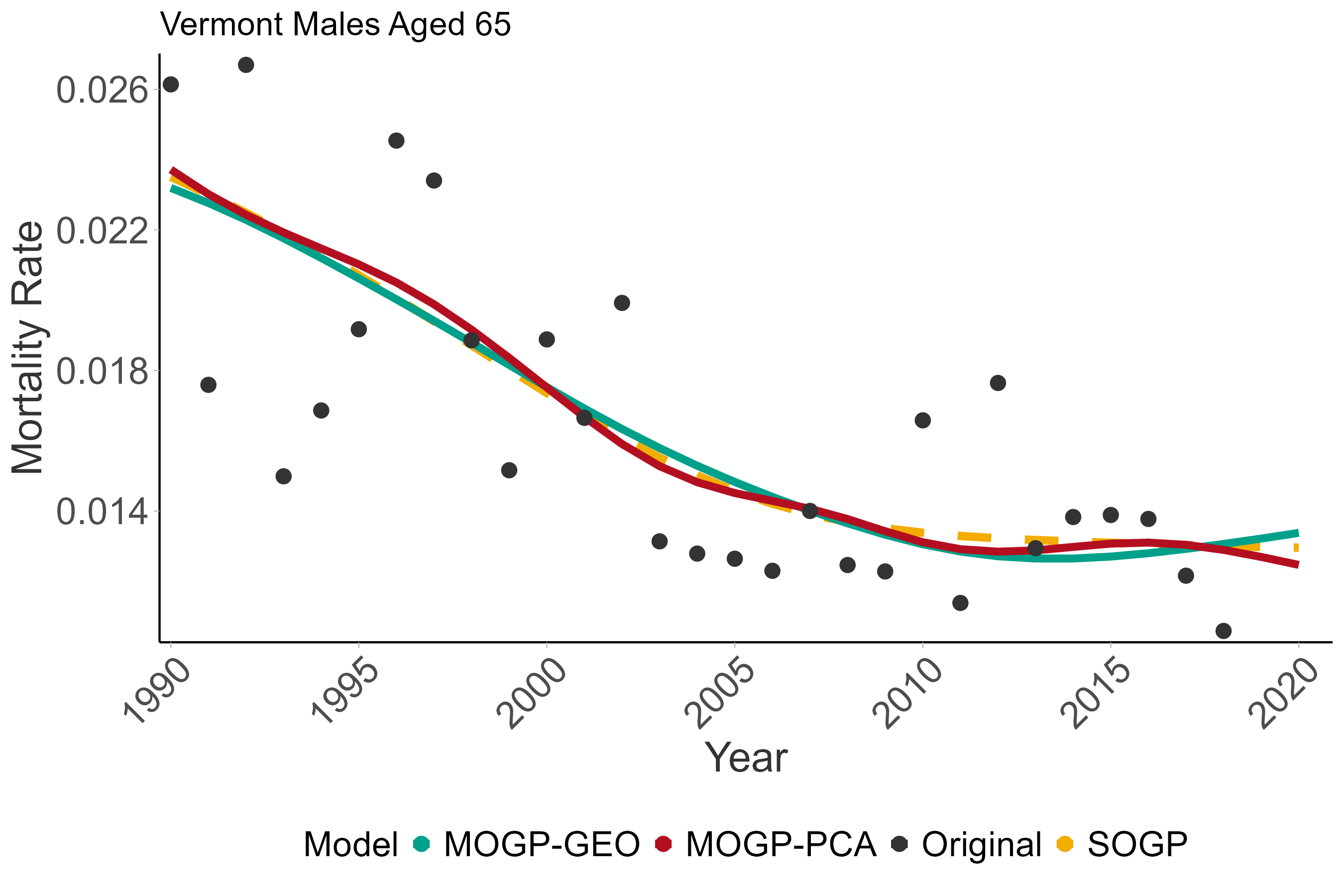}
  \caption{Vermont Males Aged 65}
\end{subfigure}
\\ \medskip 
\begin{subfigure}{.45\textwidth}
  \centering
  \includegraphics[width=1\linewidth]{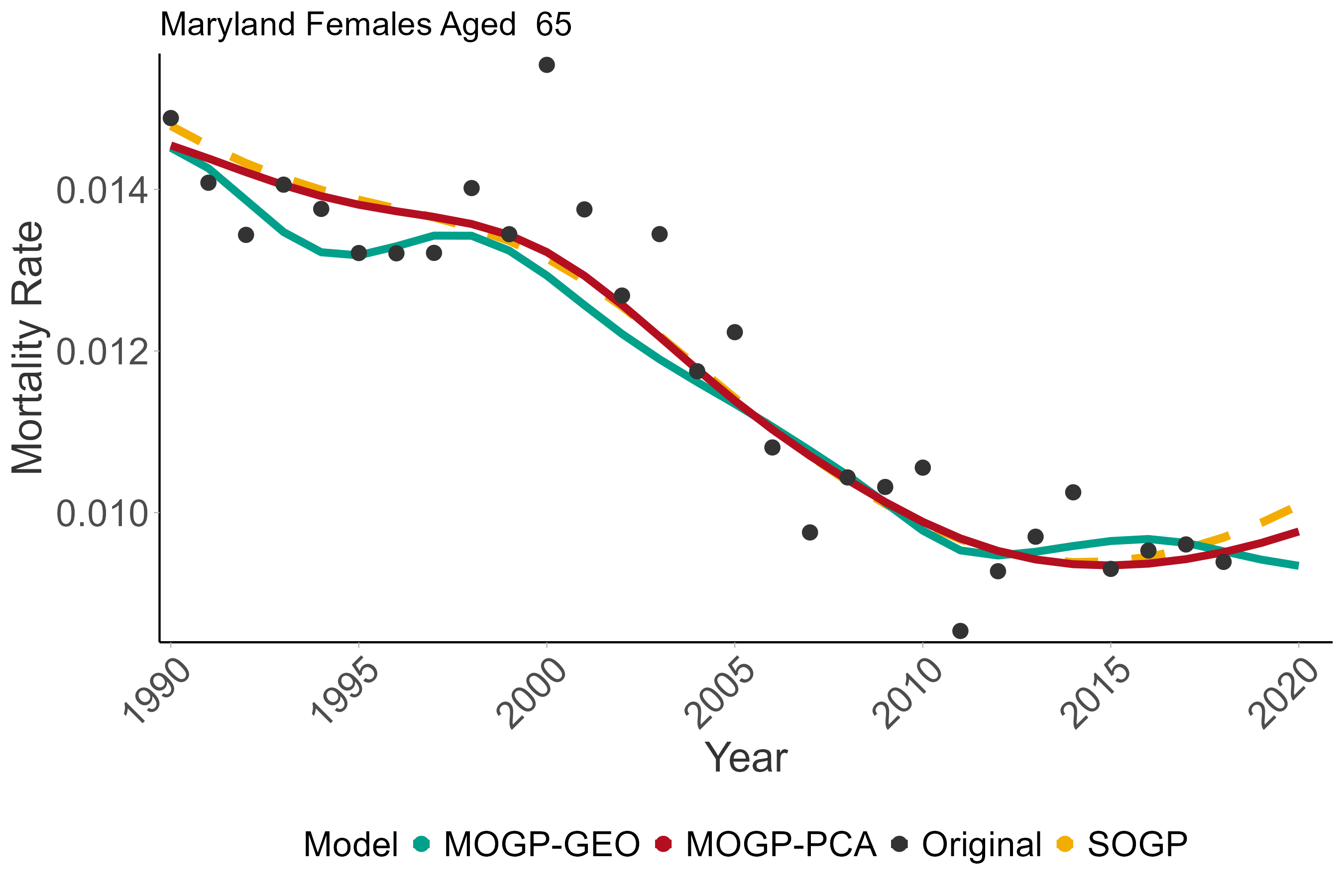}
  \caption{Maryland Females Aged 65}
\end{subfigure}
\begin{subfigure}{.45\textwidth}
  \centering
  \includegraphics[width=1\linewidth]{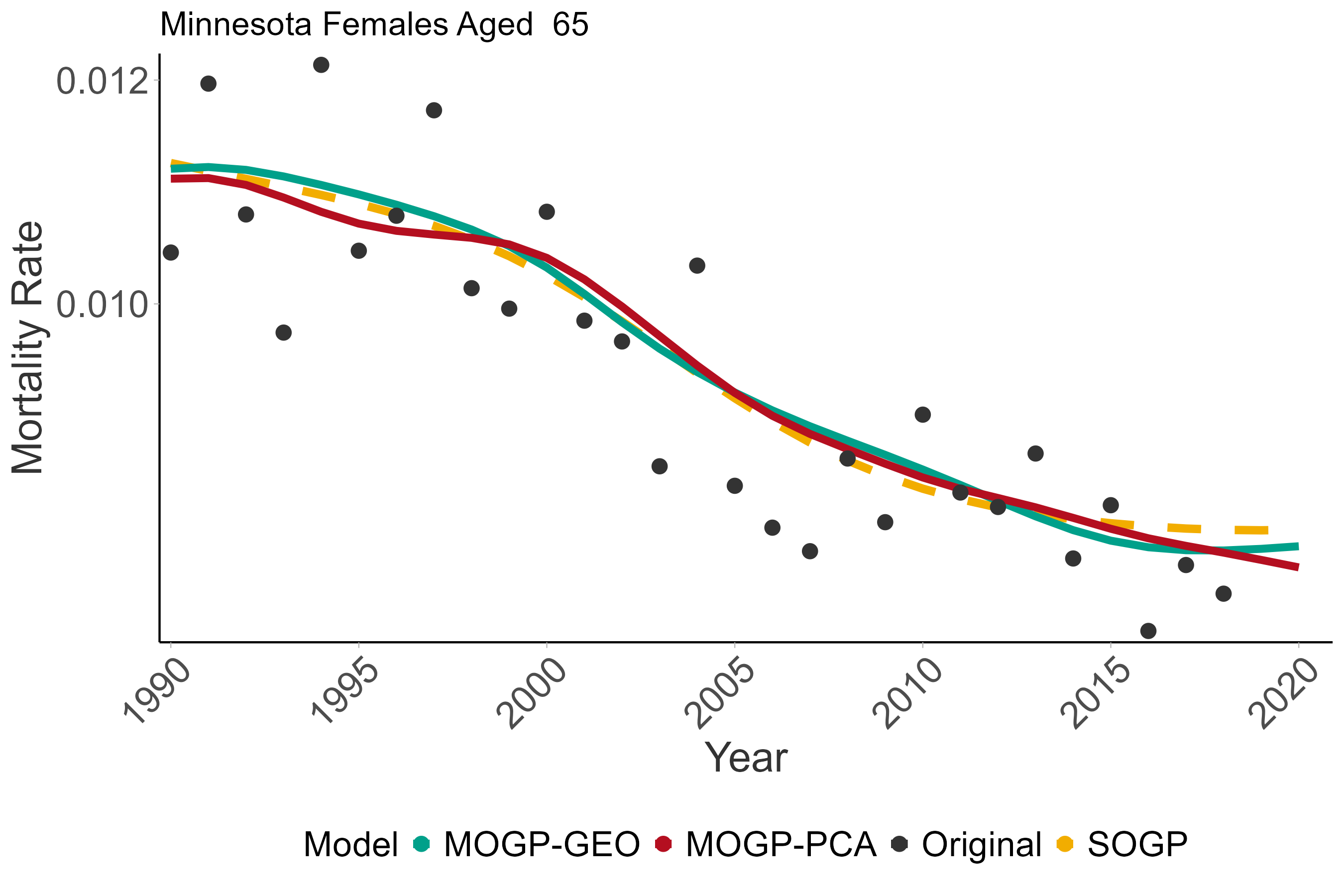}
  \caption{Minnesota Females Aged 65}
\end{subfigure}
   \caption{Raw data (black circles) and predicted mortality rate (curves) for 4 representative states based on GP models with 3 different groupings. \emph{Top row}: Males; \emph{Bottom row}: Females.} \label{FIG:comp-more}
\end{figure}

\subsection{State-Level Covariates}

\subsubsection{Data Sources} \label{APPEND:datasources}

The state-level covariates used in the PCA analysis  described in Section \ref{SUBSEC:covariates} are based on 2018 data and described as follows:

\noindent   \textbf{Economic Covariates:}
\begin{enumerate}[1.]
     \setlength\itemsep{0.0em}
    \item \underline{\textsl{Educational Attainment (EA):}}  Percentage of population aged 25+ with a bachelor's degree or higher. \textit{Source:} \href{https://www.census.gov/quickfacts/geo/chart/AL/EDU685219}{https://www.census.gov}.
    \item \underline{\textsl{Percent Change in GDP (GDP):}}  Percent change in real GDP from 2017 to 2018. Here, GDP represents the inflation-adjusted market value of goods \& services produced by the labor and property in the state. \textit{Source:} \href{https://www.bea.gov/news/2019/gross-domestic-product-state-fourth-quarter-and-annual-2018}{https://www.bea.gov}.
    \item \underline{\textsl{Median Income (MI):}} Real median household income computed by the U.S.~Census Bureau  based on data from the Current Population Survey (CPS), the American Community Survey (ACS), and other surveys. \textit{Source:} \href{https://fred.stlouisfed.org/release/tables?rid=249&eid=259515&od=2018-01-01#}{https://fred.stlouisfed.org}.
    
    \item \underline{\textsl{Regional Price Parities (RPP):}}  Price indexes that measure the geographic price level differences. For example, an RPP of 120 means the prices within the state are on average 20 percent higher than the U.S.~average. \textit{Source:} \href{https://www.bea.gov/data/prices-inflation/regional-price-parities-state-and-metro-area}{https://www.bea.gov}.
    \item \underline{\textsl{Poverty Rate (PR):}}   Poverty estimates are drawn from the Current Population Survey Annual Social and Economic Supplement (CPS ASEC), conducted three times per year with a sample of approximately 100,000 addresses. The Census Bureau determines poverty status by using an official poverty measure (OPM) that compares pre-tax cash income against a threshold. \textit{Source:} \href{https://www.epi.org/blog/poverty-continues-to-fall-in-most-states-though-progress-appears-to-be-slowing/}{https://www.epi.org}.
    \item \underline{\textsl{Urbanization Percentage (UP):}} Percentage of state population living within urban areas. The Census Bureau classifies an urban area as ``a densely settled core of census tracts and/or census blocks that meet minimum population density requirements''. \textit{Source:} \href{https://www.census.gov/programs-surveys/geography/guidance/geo-areas/urban-rural/2010-urban-rural.html}{https://www.census.gov}.
\end{enumerate}

\vspace*{-4pt}
\noindent
    \textbf{Demographic Covariates:}
\begin{enumerate}[1.]
    \setlength\itemsep{0em}\setcounter{enumi}{6}
    
    \item \underline{\textsl{Non-minority Population (NMP):}}  Percentage of state population classified as White alone.  \textit{Source:} \href{https://www.census.gov/quickfacts/geo/chart/AL/RHI125219}{https://www.census.gov}.
    \item \underline{\textsl{Percentage Elderly (ED):}}  Percentage of  state population aged 65+. \textit{Source:} \href{https://www.census.gov/quickfacts/geo/chart/AL/AGE775219}{https://www.census.gov}.
    \item \underline{\textsl{Percent Without Health Insurance (HI):}} Based on data collected for ages below 65 by the CPS ASEC and the American Community Survey (ACS). \textit{Source:} \href{https://www.census.gov/library/publications/2019/demo/p60-267.html}{https://www.census.gov}.
    \item \underline{\textsl{Obesity Rate (OR):}}  Percentage of state adult ($\ge 18$) population  with body mass index (BMI) of 30 or more, based on CDC Behavioral Risk Factor Surveillance System (BRFSS) annual telephone survey. 
        \textit{Source:} \href{https://obesity.procon.org/us-obesity-levels-by-state/}{https://obesity.procon.org}.
    \item \underline{\textsl{Political Preference (PP):}}  Percentage of state-wide eligible voters identifying as ``Democrat/Lean Democrat'' in the 2017 Gallup Daily tracking dataset. \textit{Source:} \href{ https://news.gallup.com/poll/226643/2017-party-affiliation-state.aspx}{https://news.gallup.com}.
    
    \item \underline{\textsl{Religious (R):}} Percentage of  religious population in the state according to a combined index based on four individual measures of religious observance.
 \textit{Source:} \href{https://www.pewresearch.org/fact-tank/2016/02/29/how-religious-is-your-state/?state=alabama}{https://www.pewresearch.org} which summarizes the  national \href{https://www.pewresearch.org/religion/religious-landscape-study/}{2014 Religious Landscape Study} survey. 
\end{enumerate}

\vspace*{-4pt}

\noindent\textbf{Geographic Covariates:}
\begin{enumerate}[1.]
 \setlength\itemsep{0.0em} 
 \setcounter{enumi}{12}
    \item \underline{\textsl{Average Temperature (TP):}}  
    Area-weighted state-wide averages based on climate data from the 344 \href{https://www.ncei.noaa.gov/access/monitoring/dyk/us-climate-divisions}{continental U.S.~Climate Divisions}. For each division, monthly temperatures and precipitation values are calculated from daily observations. The dataset is manually augmented for AK and HI. \textit{Source:} \href{https://www.ncdc.noaa.gov/cag/statewide/mapping/110/tavg/202101/12/value}{https://www.ncdc.noaa.gov}.
    \item \underline{\textsl{Average Relative Humidity (RH):}}  annual historical daily average of ``water vapor in the air relative to how much the air can hold'', computed based on Continental U.S.~Climate Divisional Dataset as for \textsl{Average Temperature} above. \textit{Source:} \href{https://www.ncei.noaa.gov/products/land-based-station/automated-surface-weather-observing-systems}{https://www.ncei.noaa.gov} 
    \item \underline{\textsl{Average Dew Point (DP):}}  annual historical daily average of ``the minimum temperature an airmass can achieve given the amount of moisture in the air,'' computed based on U.S.~Climate Divisional Dataset as for \textsl{Average Temperature} above. \textit{Source:} \href{https://www.ncei.noaa.gov/products/land-based-station/automated-surface-weather-observing-systems}{https://www.ncei.noaa.gov}
    
    \item \underline{\textsl{Population Density (PD):}}  Ratio of state population divided by total geographic area of the state. 
        \textit{Sources:} \href{https://www.census.gov/programs-surveys/popest/technical-documentation/research/evaluation-estimates/2020-evaluation-estimates/2010s-state-total.html}{https://www.census.gov/pop} and \href{https://en.wikipedia.org/wiki/List_of_U.S._states_and_territories_by_area}{https://en.wikipedia.org/areaofstate}.
    \item \underline{\textsl{Land in Farms (LF):}}  Includes (a) agricultural land used for crops, pasture, or grazing; (b) woodland and wasteland used in the farm operator’s total operation, and (c) land owned and operated, as well as land rented from others. 
    \textit{Source:} \href{https://www.nass.usda.gov/Publications/Todays_Reports/reports/fnlo0220.pdf}{https://www.nass.usda.gov} (Page 6).
    \item \underline{\textsl{Share of Immigrant Population (IP):}} Percentage of state population that are non-citizens. Based on the three-stage method by the Migration Policy Institute to assign legal status to noncitizen respondents in the U.S.~Census Bureau Survey Data. \textit{Source:} \href{https://www.migrationpolicy.org/programs/us-immigration-policy-program-data-hub/unauthorized-immigrant-population-profiles}{https://www.migrationpolicy.org}.
\end{enumerate}

\subsubsection{PCA Factor Loadings}\label{APPEND:le-pca}

\begin{table}[H]
\centering
\begin{tabular}{r|r|r|r|r}
  & Component 1 & Component 2 & Component 3 & LE Correlation \\ \hline
    EA & \textbf{0.86} & -0.23 & \textcolor{cor-very-weak}{-0.06} & 0.61 \\ 
   GDP & \textcolor{cor-very-weak}{0.20} & \textcolor{cor-very-weak}{-0.01} & 0.57 & 0.24 \\
 MI & \textbf{0.91} & \textcolor{cor-very-weak}{-0.04} & \textcolor{cor-very-weak}{-0.02} & \textbf{0.71}  \\
  RPP & \textbf{0.89} & -0.30 & \textcolor{cor-very-weak}{0.04} & \textbf{0.71} \\
  PR &\textbf{-0.74} & -0.38 & \textcolor{cor-very-weak}{-0.03} & \textbf{-0.79} \\
  UP & 0.59 & -0.45 & 0.43 & 0.48 \\
NMP & \textcolor{cor-very-weak}{-0.12} & \textbf{0.84} & \textcolor{cor-very-weak}{0.10} & \textcolor{cor-very-weak}{0.06}\\ 
 ED & \textcolor{cor-very-weak}{-0.14} & 0.24 & -0.53 & \textcolor{cor-very-weak}{-0.06} \\
  HI& -0.62 & \textcolor{cor-very-weak}{-0.03} & 0.57 & -0.41\\
  OR & \textbf{-0.78} & -0.19 & -0.15 & \textbf{-0.80}\\
  PP & 0.66 & -0.45 & -0.28 & 0.36\\
   R &  \textbf{-0.78} & -0.42 & \textcolor{cor-very-weak}{0.15} & \textbf{-0.75} \\
  TP & -0.30 & \textbf{-0.80} & \textcolor{cor-very-weak}{0.15} & -0.30\\
  RH & \textcolor{cor-very-weak}{-0.06} & -0.20 & -0.63 & \textcolor{cor-very-weak}{-0.10}\\
  DP & -0.31 & \textbf{-0.80} & \textcolor{cor-very-weak}{-0.17} & -0.35  \\
 PD & 0.35 & -0.51 & \textcolor{cor-very-weak}{-0.15} & \textcolor{cor-very-weak}{-0.00}\\
LF &-0.24 & \textcolor{cor-very-weak}{0.14} & 0.64 & \textcolor{cor-very-weak}{0.04}\\ 
IP & \textcolor{cor-very-weak}{0.18} & -0.32 & 0.60 & 0.28  \\
\bottomrule
\end{tabular}
\caption{Factor loadings of the 18 covariates (rows, see Table \ref{tab:statecov}) with respect the first three PCA components (columns). Values are highlighted according to the PCA scores: \textcolor{cor-very-weak}{$|\cdot|<0.2$}, \textbf{$|\cdot|>0.7$}. The last column shows the correlation between state covariates and Life Expectancy (LE) at birth in 2018 from \cite{cdc}. \label{table:factorload} }
\end{table}

\subsection{Groupings} \label{APPEND:sg}

\subsubsection{Complete List of All Groupings}
\vspace*{-6pt}
\begin{center}
 {\small  \begin{longtable}{ |l|l| }
  \hline
  State $s$ & PCA Grouping $\mathscr{O}_s$ \\
  \hline
  Alabama & Tennessee, Arkansas \\
  Arizona & Nevada, Utah \\
  Arkansas & Tennessee, Alabama \\
  California & Oregon, Washington \\
  Colorado & Utah, Nevada, Washington \\
  Connecticut & New York, New Jersey \\
  Delaware & Pennsylvania, Ohio \\
  Florida & Georgia, North Carolina \\
  Georgia & North Carolina, Florida \\
  Idaho & Wyoming, Montana, North Dakota, Nebraska \\
  Illinois & Wisconsin, Minnesota, Pennsylvania\\
  Indiana & Ohio, Michigan, Missouri\\
  Iowa & Wisconsin, Michigan \\
  Kansas & Nebraska, South Dakota, Missouri \\
  Kentucky & Tennessee, Arkansas, South Carolina \\
  Louisiana & Arkansas, Tennessee, Alabama \\
  Maine & New Hampshire, Vermont, Wisconsin \\
  Maryland & Virginia, Delaware, New Jersey  \\
  Massachusetts & Connecticut, New York, New Jersey \\
  Michigan & Ohio, Pennsylvania, Iowa \\
  Minnesota & Wisconsin, Illinois, Oregon \\
  Mississippi & Alabama, Arkansas \\
  Missouri & Kansas, Tennessee, Indiana \\
  Montana & North Dakota, Wyoming, Idaho, Nebraska \\
  Nebraska & Kansas, Iowa, North Dakota \\
  Nevada & Arizona, Utah \\
  New Hampshire & Vermont, New York, Connecticut \\
  New Jersey & New York, Connecticut, Maryland \\
  New Mexico & Oklahoma, Missouri \\
  New York & New Jersey, Connecticut \\
  North Carolina & Georgia, South Carolina, Missouri \\
  North Dakota & Montana, Idaho, Wyoming, Nebraska \\
  Ohio & Indiana, Michigan \\
  Oklahoma & Arkansas, Tennessee \\
  Oregon & Washington, Nevada, Minnesota \\
  Pennsylvania & Delaware, Ohio, Michigan \\
  Rhode Island & Connecticut, New York, Delaware \\
  South Carolina & North Carolina, Tennessee \\
  South Dakota & North Dakota, Wyoming, Nebraska, Kansas \\
  Tennessee & Kentucky, Missouri, Oklahoma \\
  Texas & New Mexico, Arizona, Georgia \\
  Utah & Nevada, Arizona \\
  Vermont & New Hampshire, Maine, New York\\
  Virginia & Maryland, Delaware, Rhode Island \\
  Washington & Oregon, California, Colorado \\
  West Virginia & Kentucky, Tennessee, Arkansas \\
  Wisconsin & Iowa, Michigan \\
  Wyoming & Montana, North Dakota, Idaho, South Dakota, Nebraska \\
  Washington D.C. & Maryland, Virginia, New Jersey \\
  \hline
\end{longtable}}
\end{center}
\vspace*{-10pt}

\textbf{Alaska and Hawaii.} To compute the optimal grouping for Alaska and Hawaii, we first identify which state out of the other 50 minimizes the distance $\mathfrak{D}$: 
\begin{align*}
    & s_1 = \arg\min_{s_* \in \mathcal{S}} \mathfrak{D}(s_*,s) = \begin{cases}
    \text{Wisconsin} & \text{if $s = $ Alaska}; \\
    \text{Maryland} & \text{if $s = $ Hawaii}.
    \end{cases}
\end{align*}
We then proceed as in Section \ref{SUBSECT:groupingcovsim}, initializing with $\mathcal{N}_s = s_1 \cup s$. The resulting groupings are (Alaska, Iowa, Wisconsin) and (Hawaii, Maryland, District of Columbia).

\subsubsection{Group Size}

\begin{figure}[H]
    \centering
    \includegraphics[width = 3in]{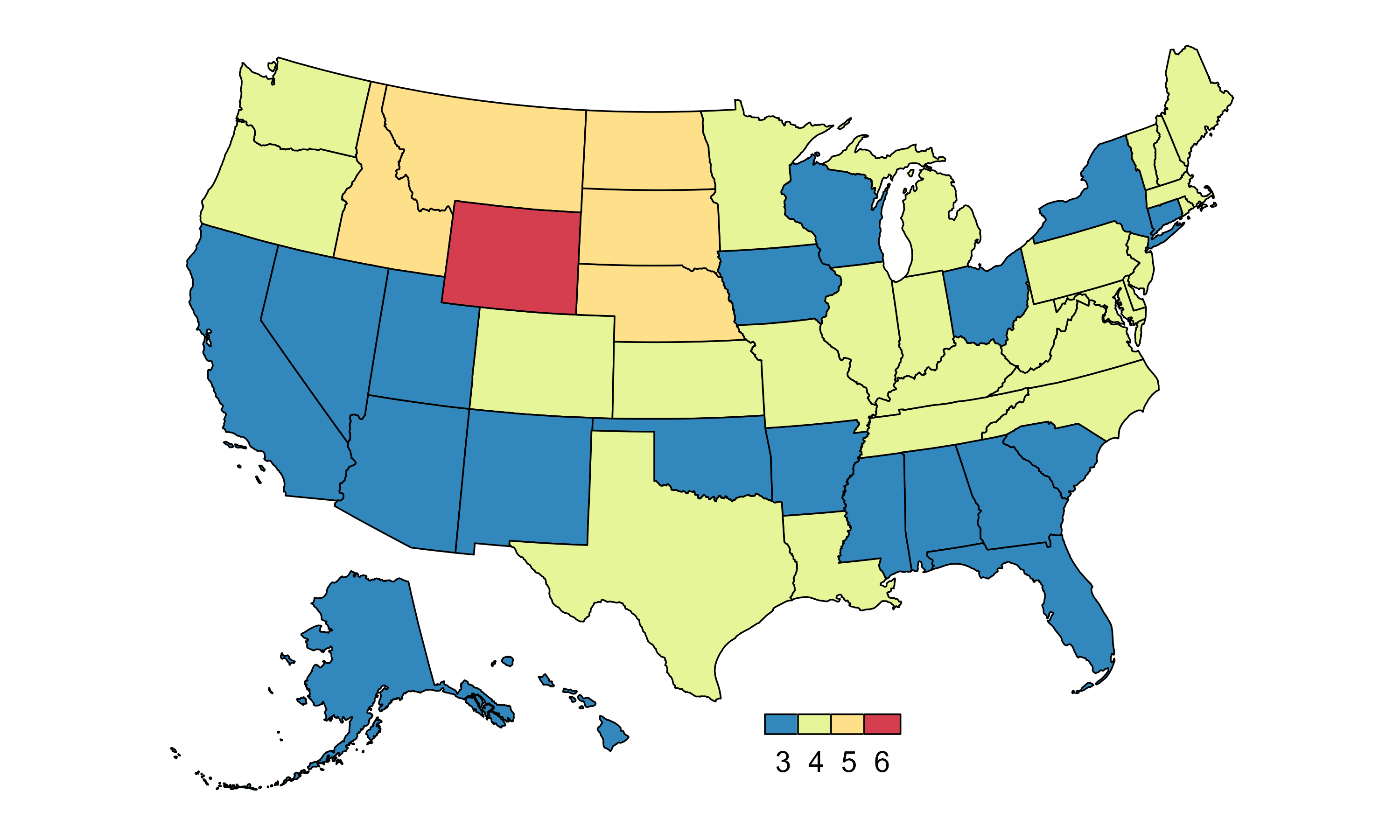}
    \caption{Size of state groupings, $|\mathscr{O}_s| \in \{3,4,5,6\}$.}
    \label{GIF:G3VSG4}
\end{figure}

Figure \ref{GIF:G3VSG4} visualizes the sizes of $|\mathscr{O}_s|$ across the U.S.. The ``default'' procedure is to group each state with 2 other most similar states. A state  $s \in \mathcal{S}$  ends up in a group of 4 if the latter are not geographically contiguous with it, that is a state is less similar to its neighbors then to other states that are further away. There is no particular pattern regarding regions where this occurs. Due to the sparse population of a cluster of states in Mountain West, these are in groups of 5 (6 for Wyoming) in order to achieve a total population of at least 3 million for each $\mathscr{O}_s$.

\subsection{State Mortality at Age 75}\label{app:mortality-75}

\begin{figure}[H]
\minipage{0.46\textwidth}
  \includegraphics[width=\linewidth]{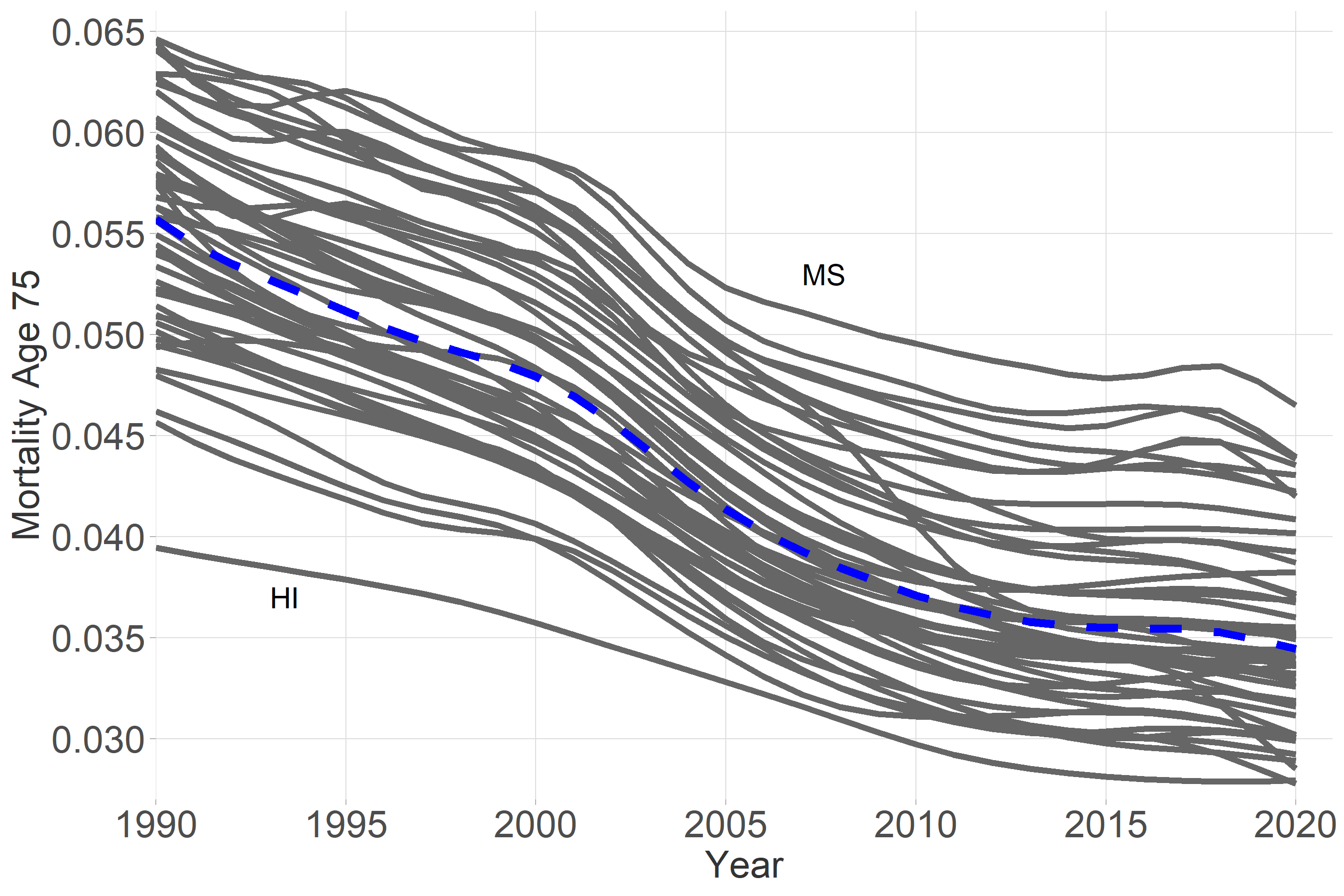} \\
  \hspace*{1in} Males Age 75  
\endminipage\hfill
\minipage{0.46\textwidth}
     \includegraphics[width=\linewidth]{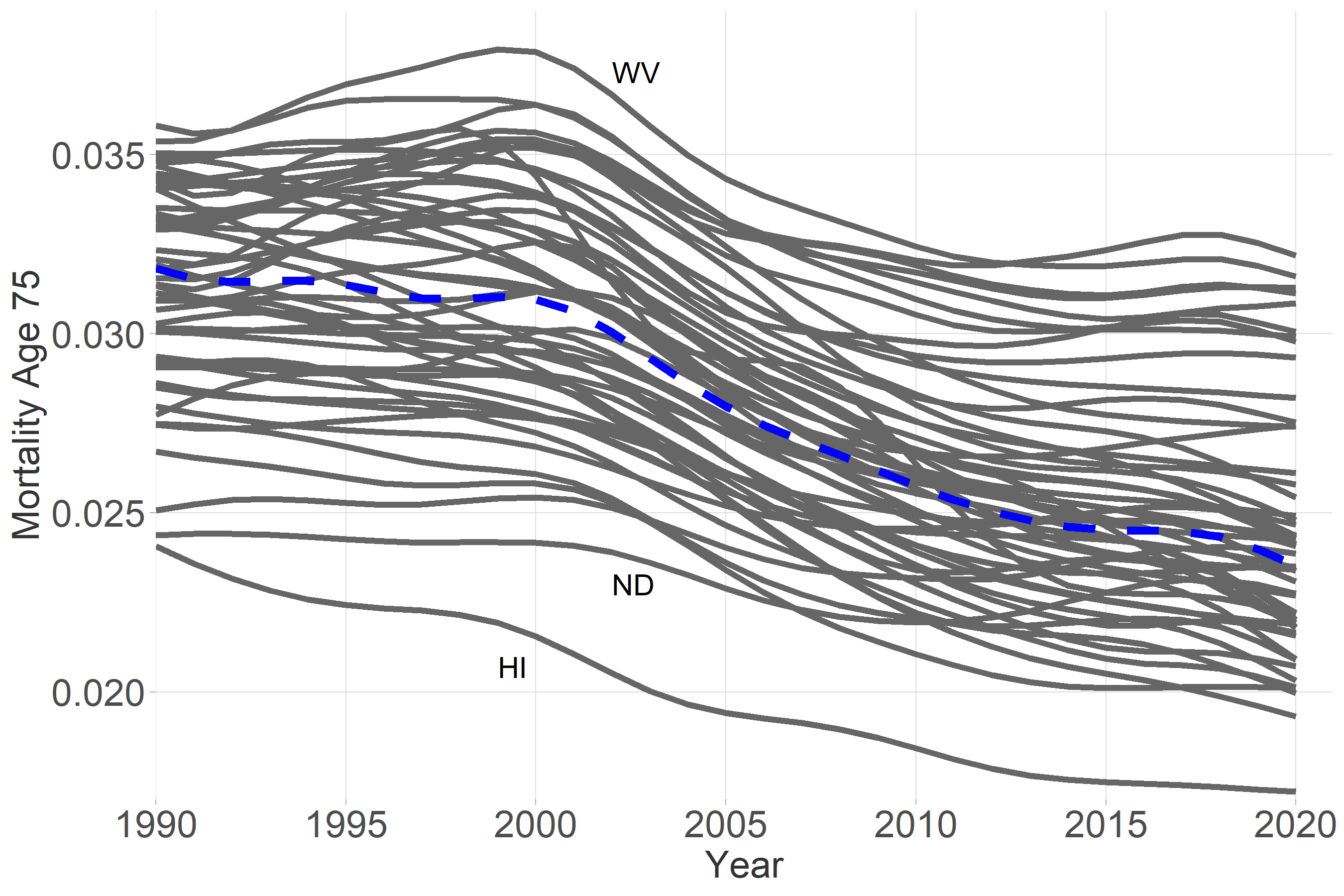}   \\
  \hspace*{1in} Females Age 75 
\endminipage
\caption{Multi-Output GP Mortality Rates for Age 75 Males (Left) and Females (Right) for Years 1990-2020. U.S.~national average shown in blue. Most states continued to experience mortality improvement at this Age as of 2020.} \label{MR-years1990-2020-70}
\end{figure}

\subsection{Supplementary Plots for Mortality Improvement Factors}\label{APPEND:matern-MI}

\begin{figure}[H] \hspace*{0.4in}
\minipage{0.39\textwidth}
  \includegraphics[trim=1.25in 0in 1in 0in,width=\linewidth]{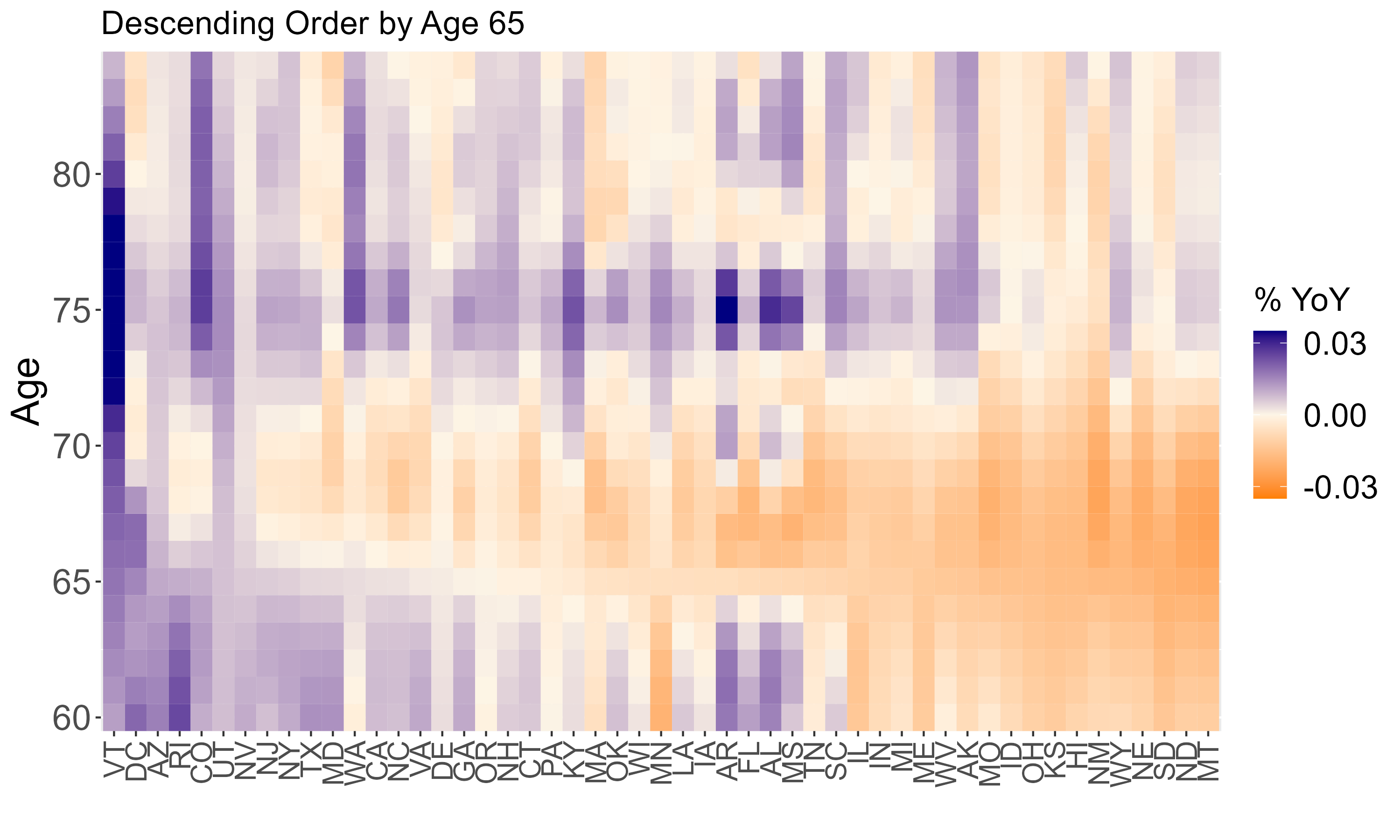} \\
   \hspace*{1in} Males
\endminipage\hspace*{0.7in}
\minipage{0.39\textwidth}
  \includegraphics[trim=1.25in 0in 1in 0in,width=\linewidth]{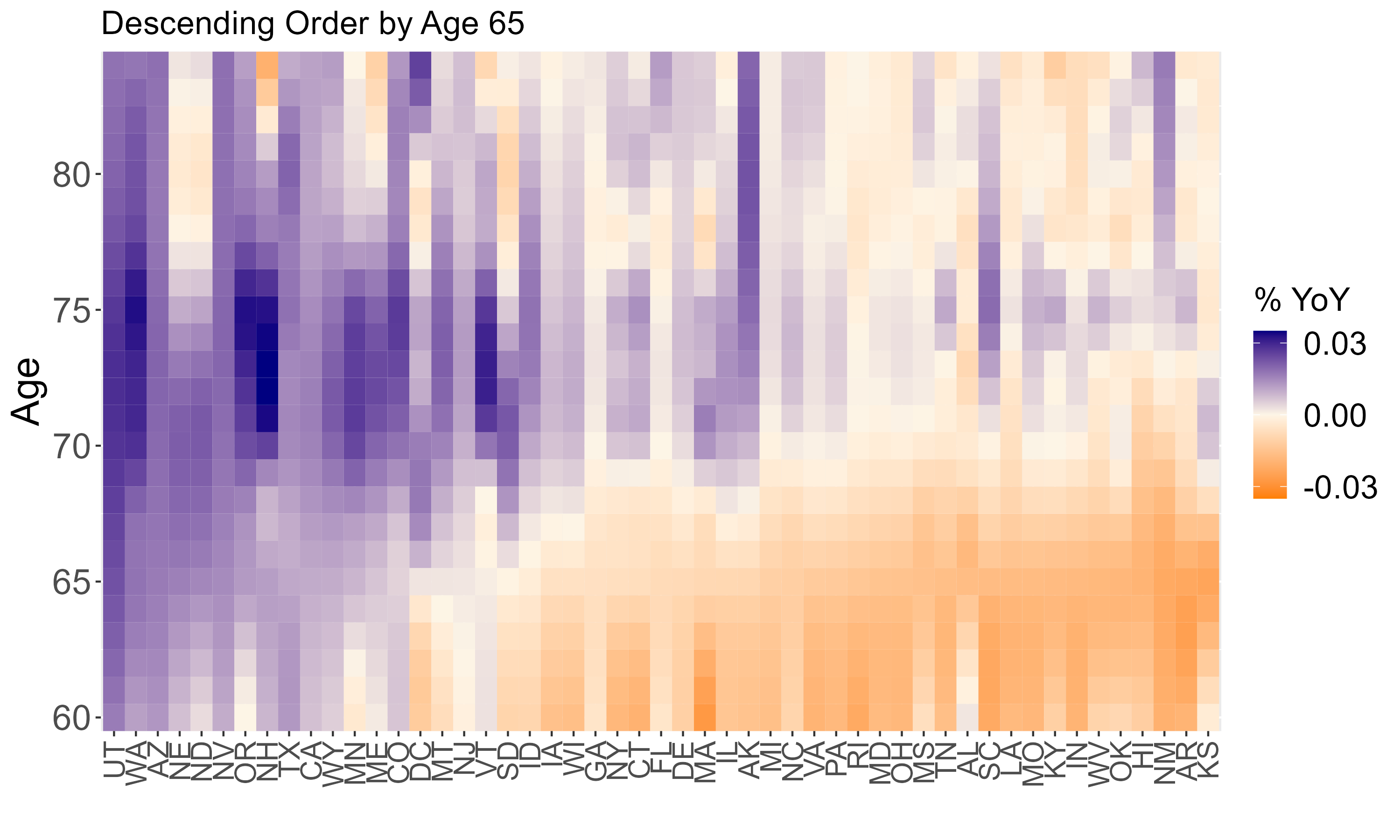} \\
  \hspace*{1in} Females
\endminipage
\caption{MOGP-generated Mortality Improvement Factors in 2020 using the M52 APC kernel \eqref{EQU:kernel}. Males (Left) and Females (Right). States are sorted by MI at Age 65.} \label{FIG:65-ir-hm}
\end{figure}

\begin{figure}[H]
\centering
\minipage{0.48\textwidth}
\includegraphics[trim=0.25in 0in 0.1in 0in,width=0.99\linewidth]{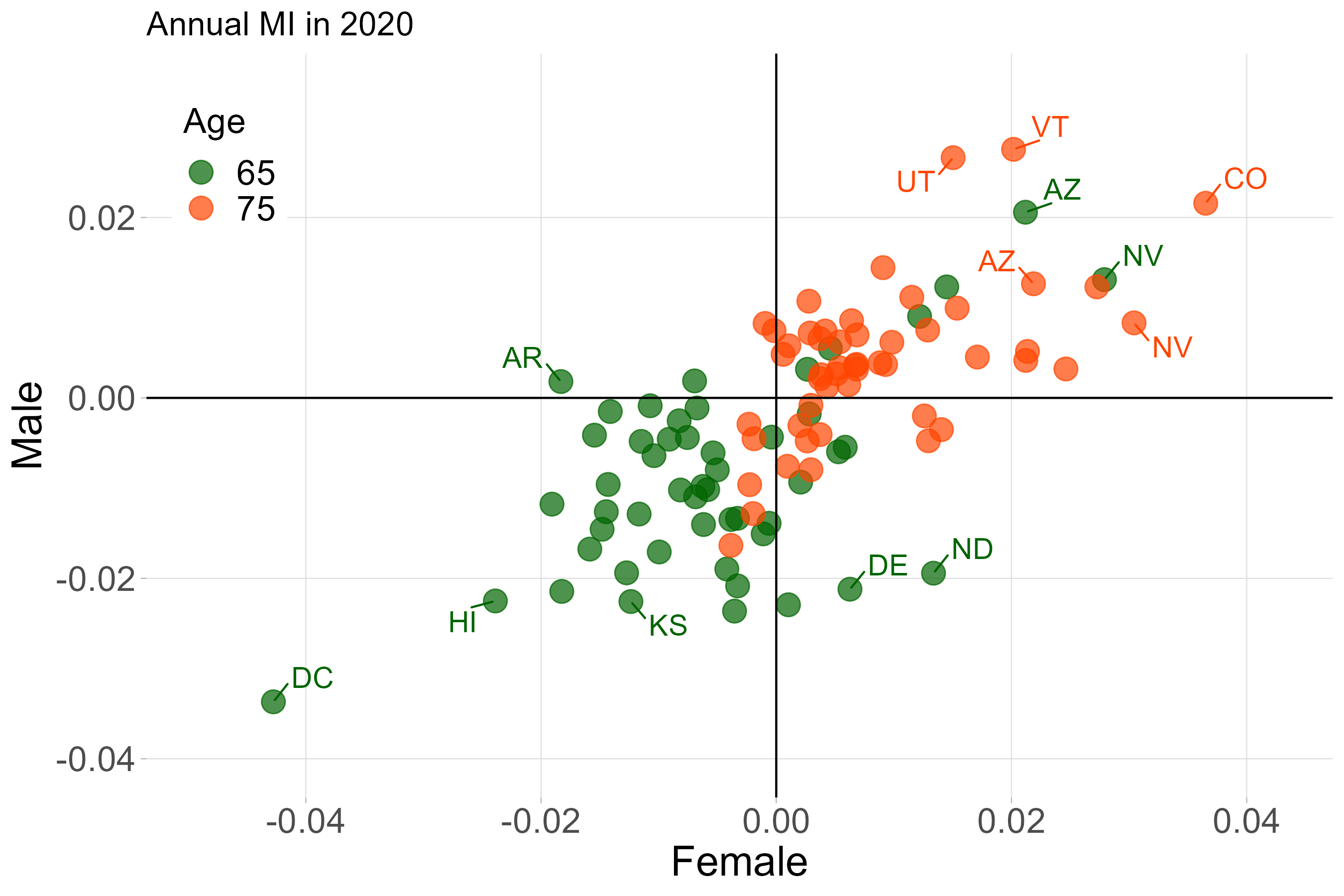} 
\endminipage\hspace*{0.2in}
\minipage{0.48\textwidth}
\includegraphics[trim=0.25in 0in 0.1in 0in,width=0.99\linewidth]{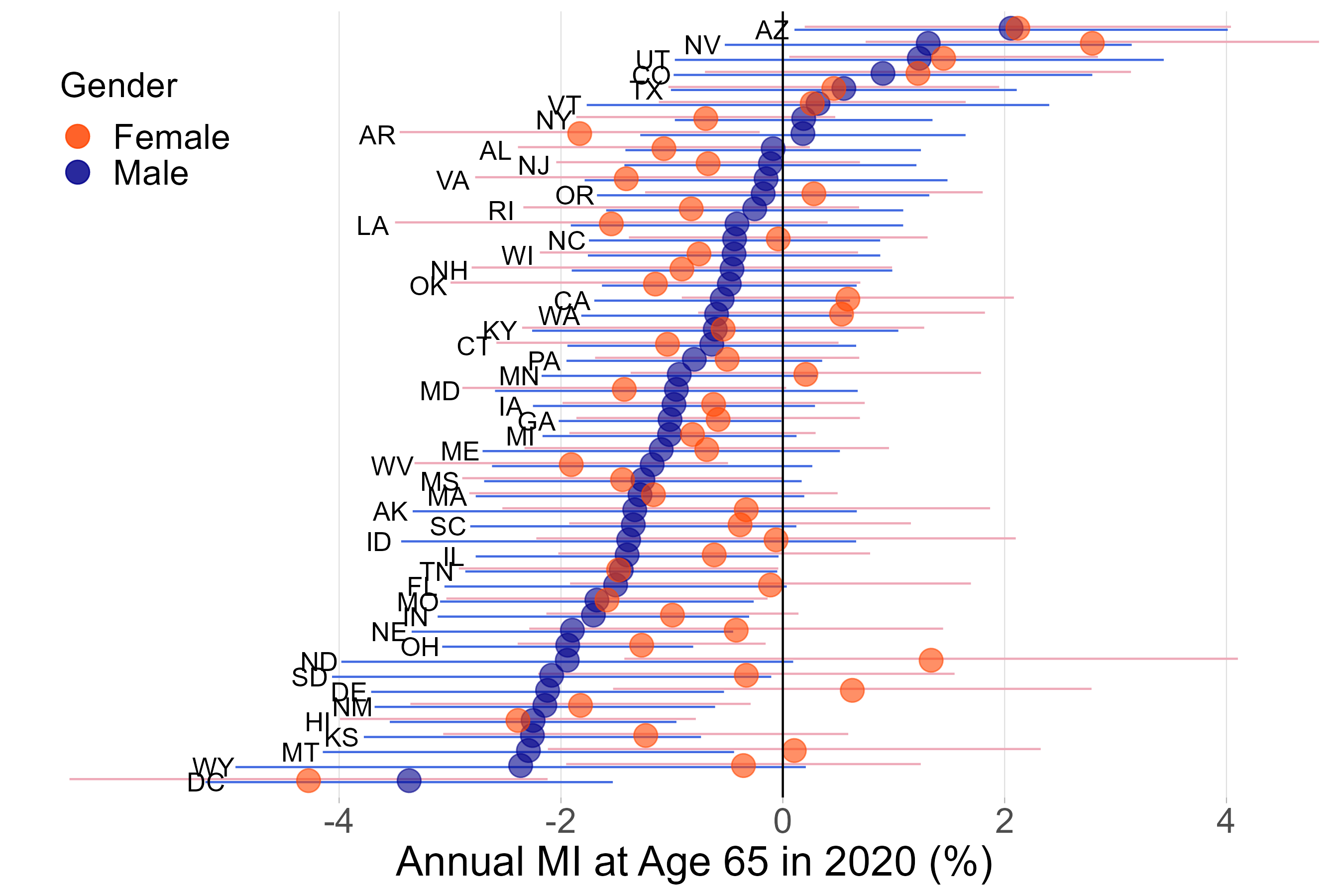} 
\endminipage
\caption{\emph{Left}: State-wise mortality improvement factors across genders at Ages 65 and 75. MIs are computed based on the SqExp MOGP-PCA model~\eqref{eq:se-mogp}.  \emph{Right:} Male vs Female MIs at Age 65, together with the respective 90\% posterior credible intervals. The states are sorted according to the Male MIs. 
\label{fig:mi-scatter} }
\end{figure}

\bibliographystyle{plain}
\bibliography{usmdb}

\end{document}